\documentclass[twocolumn,tighten,apj]{openjournal}
\usepackage{amsmath}
\usepackage{longtable}
\usepackage{xcolor}
\usepackage[utf8]{inputenc}
\usepackage[english]{babel}
\usepackage{hyperref}
\hypersetup{
    unicode,
    colorlinks=true,
    linkcolor=linkcolor,
    citecolor=linkcolor,
    filecolor=linkcolor,
    urlcolor=linkcolor,
}
\usepackage{color,colortbl}
\definecolor{linkcolor}{rgb}{0.0,0.3,0.5}
\usepackage[normalem]{ulem}
\usepackage{orcidlink}
\makeatletter
\newdimen\movetabledown
\newenvironment{longrotatetable}{%
  \clearpage
  \onecolumngrid
  \begingroup
  \LongTables
  \def\tablehead##1{%
    \@table@not@headedfalse%
    \kill
    \caption{\\%
      \@tablecaption}%
      \\
    \hline
    \hline\\[0.4ex]
    ##1\hskip\tabcolsep\\[1.2ex]
    \hline\\[0.4ex]
    \endfirsthead
    \caption[]{--- \emph{Continued}}\\[0.8ex]
    \hline
    \hline\\[0.4ex]
    ##1\hskip\tabcolsep\\[1.2ex]
    \hline\\[0.4ex]
    \endhead
    \hline
    \endfoot%
  }%
  \clearpage
  \def\LS@rot{%
    \setbox\@outputbox\vbox{%
      \vskip\movetabledown\hbox{\rotatebox{90}{\box\@outputbox}}%
    }%
  }%
  \let\LS@makecol\@makecol
  \let\LS@makefcolumn\@makefcolumn
  \vsize=\textwidth
  \hsize=\textheight
  \linewidth=\hsize
  \columnwidth=\hsize
  \@colroom=\vsize
  \textheight=\vsize
  \@colht=\vsize
  \def\@makecol{\LS@makecol\LS@rot}%
  \def\@makefcolumn##1{\LS@makefcolumn{##1}\LS@rot}%
}{%
  \clearpage
  \twocolumngrid
  \clearpage
  \endgroup
}
\makeatother

\newcommand{\auv}{A$_{\rm UV}$}
\newcommand{\auvd}{A$_{\rm UV,direct}$}
\newcommand{\av}{A$_{\rm V}$}
\newcommand{\cv}{C$_{\rm V}$}
\newcommand{\llp}{$\log(\lambda_{\rm peak})$}
\newcommand{\cuv}{C$_{\rm UV}$}
\newcommand{\kratio}{$\kappa_{\rm UV}/\kappa_{\rm FIR}$}
\newcommand{\sdust}{$\Sigma_{\rm dust}$}
\newcommand{\lir}{L$_{\rm IR}$}
\newcommand{\luv}{L$_{\rm UV}$}
\newcommand{\mdust}{M$_{\rm dust}$}
\newcommand{\mstar}{M$_\star$}
\newcommand{\fobs}{f$_{\rm obs}$}
\newcommand{\funobs}{f$_{\rm unobs}$}
\newcommand{\tdust}{T$_{\rm dust}$}

\def\ltsima{$\; \buildrel < \over \sim \;$}
\def\simlt{\lower.5ex\hbox{\ltsima}}
\def\gtsima{$\; \buildrel > \over \sim \;$}
\def\simgt{\lower.5ex\hbox{\gtsima}}

\defcitealias{kokorev24a}{K24}
\defcitealias{akins24a}{A24}
\newcommand {\etal}{et~al.}
\newcommand {\uJy}{$\mu$Jy}
\newcommand {\um}{$\mu$m}
\newcommand {\muv}{$M_{\rm UV}$}

\newcommand{\msun}{{\rm\,M$_\odot$}}
\newcommand{\sfr}{{\rm\,M$_\odot$\,yr$^{-1}$}}
\newcommand{\lsun}{{\rm\,L$_\odot$}}

\newcommand{\lpeak}{$\lambda_{\rm peak}$}

\begin{document}

\title{Dust in the Average Galaxy: Attenuation, Emission, and Opacity from $0<z<7$}

\author{Caitlin M. Casey\orcidlink{0000-0002-0930-6466}$^{1,2}$}
\noaffiliation

\author{Hollis B. Akins\orcidlink{0000-0003-3596-8794}$^{3,\ast}$}
\noaffiliation

\author{Andrew J. Battisti\orcidlink{0000-0003-4569-2285}$^{4,5}$}
\noaffiliation

\author{Jed McKinney\orcidlink{0000-0002-6149-8178}$^{3,\dag}$}
\noaffiliation

\author{Ezequiel Treister\orcidlink{0000-0001-7568-6412}$^{6}$}
\noaffiliation

\author{Jorge A. Zavala\orcidlink{0000-0002-7051-1100}$^{7}$}
\noaffiliation


\author{Hiddo Algera\orcidlink{0000-0002-4205-9567}$^{8}$}
\noaffiliation

\author{Manuel Aravena\orcidlink{0000-0002-6290-3198}$^{9,10}$}
\noaffiliation



\author{Yingjie Cheng\orcidlink{0000-0001-8551-071X}$^{11}$}
\noaffiliation


\author{Nicole E. Drakos\orcidlink{0000-0003-4761-2197}$^{12}$}
\noaffiliation


\author{Andreas L. Faisst\orcidlink{0000-0002-9382-9832}$^{13}$}
\noaffiliation

\author{Maximilien Franco\orcidlink{0000-0002-3560-8599}$^{14}$}
\noaffiliation

\author{Seiji Fujimoto\orcidlink{0000-0001-7201-5066}$^{15}$}
\noaffiliation


\author{Ghassem Gozaliasl\orcidlink{0000-0002-0236-919X}$^{16,17}$}
\noaffiliation


\author{Ali Hadi\orcidlink{0009-0003-3097-6733}$^{18}$}
\noaffiliation

\author{Santosh Harish\orcidlink{0000-0003-0129-2079}$^{19}$}
\noaffiliation

\author{Michaela Hirschmann\orcidlink{0000-0002-3301-3321}$^{20,21}$}
\noaffiliation


\author{Olivier Ilbert\orcidlink{0000-0002-7303-4397}$^{22}$}
\noaffiliation

\author{Kohei Inayoshi\orcidlink{0000-0001-9840-4959}$^{23}$}
\noaffiliation


\author{Jeyhan S. Kartaltepe\orcidlink{0000-0001-9187-3605}$^{24}$}
\noaffiliation

\author{Anton M. Koekemoer\orcidlink{0000-0002-6610-2048}$^{19}$}
\noaffiliation


\author{Claudia del P. Lagos\orcidlink{0000-0003-3021-8564}$^{4,2}$}
\noaffiliation

\author{Ronaldo Laishram\orcidlink{0000-0002-0322-6131}$^{25}$}
\noaffiliation

\author{Erini Lambrides\orcidlink{0000-0003-3216-7190}$^{26,\ddag}$}
\noaffiliation

\author{Daizhong Liu\orcidlink{0000-0001-9773-7479}$^{27}$}
\noaffiliation

\author{Arianna S. Long\orcidlink{0000-0002-7530-8857}$^{11}$}
\noaffiliation


\author{Georgios E. Magdis\orcidlink{0000-0002-4872-2294}$^{2,28,29}$}
\noaffiliation

\author{Sinclaire M. Manning\orcidlink{0000-0003-0415-0121}$^{7}$}
\noaffiliation

\author{Crystal L. Martin\orcidlink{0000-0001-9189-7818}$^{1}$}
\noaffiliation

\author{Felix Martinez III\orcidlink{0000-0002-9883-1413}$^{24}$}
\noaffiliation

\author{Richard Massey\orcidlink{0000-0002-6085-3780}$^{30}$}
\noaffiliation

\author{Jacqueline E. McCleary\orcidlink{0000-0002-9883-7460}$^{31}$}
\noaffiliation

\author{Henry Joy McCracken\orcidlink{0000-0002-9489-7765}$^{32}$}
\noaffiliation

\author{Lauro Moscardini\orcidlink{0000-0002-3473-6716}$^{33,34,35}$}
\noaffiliation


\author{Desika Narayanan\orcidlink{0000-0002-7064-4309}$^{36,2}$}
\noaffiliation


\author{Louise Paquereau$^{37}$}
\noaffiliation


\author{Jason Rhodes\orcidlink{0000-0002-4485-8549}$^{38}$}
\noaffiliation

\author{Brant E. Robertson\orcidlink{0000-0002-4271-0364}$^{39}$}
\noaffiliation


\author{Rasha M. Samir\orcidlink{0000-0003-2716-8332}$^{40}$}
\noaffiliation

\author{Claudia Scarlata\orcidlink{0000-0002-9136-8876}$^{41}$}
\noaffiliation

\author{Marko Shuntov\orcidlink{0000-0002-7087-0701}$^{2,29,42}$}
\noaffiliation

\author{Laura Sommovigo\orcidlink{0000-0002-2906-2200}$^{43}$}
\noaffiliation



\author{Aswin P. Vijayan\orcidlink{0000-0002-1905-4194}$^{44}$}
\noaffiliation


\author{Wuji Wang\orcidlink{0000-0002-7964-6749}$^{13}$}
\noaffiliation


\author{Can Xu\orcidlink{0000-0002-8437-6659}$^{45,46,47}$}
\noaffiliation


\author{Dhruv Zimmerman\orcidlink{0009-0008-7017-5742}$^{36}$}
\noaffiliation

\begin{abstract}

We present constraints on the dust emission and attenuation properties
of galaxies across $0\!<\!z\!<\!7$ using JWST imaging from the
COSMOS-Web Survey combined with deep far-infrared (FIR) and
(sub)millimeter data from {\it Spitzer}, {\it Herschel}, SCUBA-2,
NIKA2 and ALMA. We analyze over 500,000 galaxies to independently
constrain attenuation as inferred in the rest-frame
 ultraviolet (UV) and optical as well as dust emission
from stacked FIR spectral energy distributions (SEDs),
enabling a direct comparison between the two.  We find UV/optical
attenuation systematically underpredicts IR luminosity by a factor of
$\sim$3$\times$ at $0.5<z<7$ and up to an order of magnitude for
$M_\star>10^{10.5}$\,\msun.  We derive empirical relationships for the
effective attenuation (\auv\ and \av), dust temperature, fraction of
star formation that is unobscured, and dust-to-stellar mass ratio as
functions of redshift and stellar mass.
On galaxy-integrated scales, we separate the first order effect of
star/dust geometry from dust grain properties by combining constraints
on the IR SED, the UV SED, and the dust mass surface density.
Importantly, we measure over an order of magnitude decrease in \kratio
--- the ratio of dust mass absorption coefficients in the UV at
1600\AA\ and FIR at 500\um\ --- from $z\sim0$ to $z\sim7$.  A
depressed \kratio\ ratio is consistent with a deficit of small dust
grains, possibly attributable to the intense radiation fields of
high-$z$ star formation; indeed, we find a redshift-invariant inverse
relationship between \kratio\ and $\Sigma_{\rm SFR}$.  Assuming the
varying dust opacity is dominated by changes to small grains and not
large grains, we infer a modest evolution in the dust-to-stellar mass
ratio $\propto(1+z)^{1.36}$ with secondary stellar mass dependence,
$\propto M_\star^{-0.3}$. Most evolution in the dust-to-stellar ratio
is at $z<1$, the product of mild downward evolution in the dust-to-gas
ratio combined with steep evolution in the gas-to-stellar ratio.  The
significant evolution and dynamic range of \kratio\ and prevailing
disconnect between the UV/optical and FIR regimes emphasize that
direct dust constraints are irreplaceable for the majority of
star-forming galaxies at $z<7$, not just the most extreme
star-formers.
\end{abstract}

\begin{keywords}
    {Galaxies, Attenuation, Dust}
\end{keywords}

\maketitle

\section{Introduction}

Dust makes up a negligible fraction of the cosmic mass budget
($\simlt1$\%), and yet it has a profound impact on the perception of
starlight via attenuation \citep{salim20a,schneider24a}.  While dust
efficiently absorbs ultraviolet and optical photons and converts them
to thermal energy re-radiated at much longer wavelengths, our ability
to fully characterize interstellar dust beyond our immediate
neighborhood has been severely limited by observational hurdles and
degeneracies.  Dust extinction measurements are limited to individual
stellar sightlines in the Milky Way relative to localized measurements
of the hydrogen column density
\citep{reina73a,gorenstein75a,predehl95a,guver09a}.
Dust attenuation curves have been constrained for nearby galaxies and
increasingly larger samples out to $z\sim3$
\citep{reddy15a,shivaei20a,battisti22a}, but the inference relies on
assumptions -- empirical effective laws \citep{calzetti00a}, specific
star/dust geometries, or fixed grain populations -- that have not been
independently tested. Direct comparison between attenuation derived
from the UV/optical and absorption derived from the FIR remains rare.
Dust in emission has been seen predominantly in the rarest of bright
submillimeter galaxies \citep[e.g.][see reviews by \citealt*{blain02a}
  and \citealt*{casey14a}]{smail97a}, with relatively little
understanding as to how exceptional that dust emission is relative to
more `normal' galaxies at high redshift.

The framework through which we study dust in the distant universe is
thus anchored to nearby universe calibrations, with a limited range of
attenuation curves, dust mass absorption coefficients, and assumptions
about star/dust geometries.  We implicitly assume that these
calibrations apply at greater redshifts. If these assumptions are
systemically inaccurate, then much of what we infer regarding dust at
high redshift may not hold --- dust masses may be off by up to an
order of magnitude, and even stellar masses could be
under/over-estimated based on the application of an inappropriate
attenuation law \citep[e.g.][]{mckinney25a}.  Furthermore, dust is
built from metals and complex molecules, thus its total mass is also
inherently linked to the metallicity of the interstellar
  medium (ISM) and star formation histories in galaxies
\citep{draine07a,remy-ruyer14a,de-vis19a}.  Both metallicity and star
formation histories are believed to be quite different for the
`average' galaxy at $z=2-8$ than at $z=0$
\citep{maiolino08a,mannucci10a,zahid14a,jain26a}: more metal poor and
potentially burstier.  Observations of dust at these epochs may reveal
interesting details about the integrated enrichment and star formation
history. Further it may also reveal its future potential for star
formation as a key catalyst for gas cooling in molecular clouds.

Studying and calibrating the impact of dust is intrinsically difficult
beyond the local universe, given that extraordinary sensitivity is
required to understand the `normal' galaxy population.  In this paper
we place empirical constraints on galaxies' dust emission and
absorption, and the direct relationship between the two, out to
$z\sim7$ using the marriage of improved constraints from the James
Webb Space Telescope, JWST, and the Atacama Large Millimeter and
submillimeter Array, ALMA, combined with other FIR datasets, together
in the Cosmic Evolution Survey (COSMOS) field \citep{scoville07a}.  By pushing near-infrared
imaging more than 10$\times$ deeper than previous imaging at high
spatial resolution over large areas, JWST has revolutionized the
precision at which we characterize galaxies' photometric and
spectroscopic redshifts as well as the intrinsic shape of galaxies'
rest-frame optical SEDs \citep{shuntov25a}. This in turn also
dramatically improves constraints on their star formation histories
\citep{arango-toro25a} and stellar masses measurements. JWST
fundamentally pushes this characterization of the rest-frame
UV/optical for $\sim3\times$ more galaxies per unit area on the sky,
and for galaxies much fainter (1-2\,magnitudes) than previously
possible.  ALMA, along with other (sub)mm facilities, adds constraints
on galaxies' dust emission.  The wealth of ALMA data in JWST-imaged
areas has ballooned in recent years, such that many galaxies less
luminous than traditional dusty star-forming galaxies are now
routinely accessible.  Furthermore, large areas in the COSMOS field
now have deep ALMA continuum maps, enabling truly sensitive stacking
experiments propelled in sensitivity by the density of JWST-detected
galaxies on the sky.

This paper is organized as follows. \S~\ref{sec:framework} provides
a framework for understanding the goals of this work; it discusses the
physical relationship between attenuation and emission, and
discusses key unknowns.
\S~\ref{sec:data} presents the JWST, ALMA, and other submillimeter data
used in this work, focused in and around the COSMOS-Web JWST Survey
\citep{casey23a}.
\S~\ref{sec:methods} presents the methodology by which we constrain
the UV/optical and, independently, the dust SEDs for galaxies measured
as a function of stellar mass and redshift through a FIR stacking
analysis.
\S~\ref{sec:results} presents a number of key measurements of implied
physical characteristics, their evolution, and redshift dependence;
this includes properties derived from the UV/optical portion of the
SED, properties derived from the stacked dust SEDs in the FIR, and
properties that present physical measurements through a combination of
the two.
\S~\ref{sec:discussion} connects the unique constraints presented in
the previous section; we focus in particular on applying the framework
of \S~\ref{sec:framework} to separate out the impact of star/dust
geometry from dust grain properties and comment on the underlying
drivers of the evolution in the dust-to-stellar ratio.
\S~\ref{sec:conclusions} concludes.  Throughout, we presume a Kroupa
initial mass function \citep{kroupa01a} and a Planck cosmology
\citep{planck-collaboration20a}.  All appearances of $\log$ imply
$\log_{10}$ and $\ln$ implies a natural logarithm.

\section{Framework \&\ Approach}\label{sec:framework}

It is beneficial to revisit fundamental definitions linking galaxies'
dust attenuation to their emission and mass to clarify our aims in
this work.  The relationship between optical depth $\tau$, the dust
mass absorption coefficient $\kappa_{\lambda}$, path length $ds$, and
dust density $\rho$, is $d\tau_\lambda=-\kappa_{\lambda}\rho ds$.
From that, it follows that the magnitudes of attenuation
$A_{\lambda}$ and dust mass surface density, $\Sigma_{\rm dust}$,
relate via:
\begin{equation}
  A_{\rm \lambda,screen} = \frac{2.5}{\ln(10)} \kappa_{\lambda} \Sigma_{\rm dust}.
\label{eq:alambda1}
\end{equation}
This presumes a simple foreground screen of dust in front of a light
source (i.e. starlight).

The dust mass absorption coefficient $\kappa_\lambda$
\citep{weingartner01a,draine03a} encodes the complex absorption
properties of different dust grains, and the grain type and size
distribution, as a function of wavelength. It may also be broken into
two components corresponding to the intrinsic {\it absorptive}
properties of the dust ($\kappa_{\lambda,abs}$, which goes directly
into re-radiated thermal emission) and a scattered light component
($\kappa_{\lambda,scat}$, accounting for photons of wavelength
$\lambda$ that are not destroyed, but rather scattered), such that the
total $\kappa_\lambda = \kappa_{\lambda,abs} + \kappa_{\lambda,scat}$.
Milky Way dust has significant scattering for UV/optical photons such
that $\kappa_{\lambda,scat}/\kappa_{\lambda}\approx0.3-0.6$
\citep{cardelli89a}.  The scattering vs. absorption distinction is
crucial in the consideration of narrow sightlines over which
extinction is being measured (e.g. in the Milky Way); in the case of
integrated light from entire galaxies, $\kappa_\lambda$ is taken
entirely as an absorption factor due to all scattered light being
accounted for within the galaxy aperture, as discussed in
\citet{calzetti00a}.

Generalizing to a mixed star/dust geometry then can be written as:
\begin{equation}
  A_\lambda = \frac{2.5}{\ln(10)}\mathcal{G}_{\lambda}\kappa_{\lambda}\Sigma_{\rm dust}
\label{eq:alambda2}
\end{equation}
where the additional term, $\mathcal{G}_{\lambda}$, we introduce as a
wavelength-dependent factor that encodes the star/dust geometry.  It
may be defined as the ratio of the effective optical depth
($\tau_\lambda^{\rm eff}$, or optical depth of ``apparent
extinction'', \citealt{charlot00a}) to the physical optical depth, or
similarly the ratio of observed $A_\lambda$ (measured using standard
techniques in the rest-frame UV/optical) to directly inferred $A_{\rm
  \lambda, direct}$ (inferred by measures of dust mass surface
density, \sdust).  By its nature, $\mathcal{G}_\lambda$ is bounded
such that $0<\mathcal{G}_{\lambda}\le1$ \citep[the following works all
  explore the crucial role of geometry, though they do not explicitly
  use the convention we have outlined with
  $\mathcal{G}_\lambda$;][]{disney89a,byun94a,witt96a,witt00a,charlot00a,chevallard13a,narayanan18a,trayford20a,ferrara22a,qin24a,sommovigo26a}.
$\mathcal{G}_\lambda=1$ corresponds to a uniform foreground slab, and
$\mathcal{G}_\lambda\approx1/2$ for a uniform, optically thin
($\tau\ll1$) slab with stars mixed uniformly throughout.
A case where stars are completely decoupled from dust would correspond
to $\mathcal{G}_\lambda\approx0$ (i.e. a {\it background} dust
screen), but the reality of mixed star/dust geometries is that some
emergent starlight usually escapes near the surface, and
$\mathcal{G}_\lambda$ should not get very close to 0.  The impact of
$\mathcal{G}_\lambda$ is to {\it lessen} $A_\lambda$ for a given dust
mass surface density, and while $\mathcal{G}_\lambda$ can be worked
out analytically for some simple toy models (dust screen, dust slab,
exponential disks at given inclination, etc.), it will generally not
be a function that is easy to directly measure or derive.  For
example, it is understood that geometry also has a strong wavelength
dependence, such that a clumpy ISM with narrow escape channels for
starlight results in a {\it grayer} attenuation curve
\citep{natta84a,calzetti94a,witt00a,salim20a} or bluer rest-frame UV
slope \citep{goldader02a,howell10a,casey14b,narayanan18a} for fixed
dust mass. A clear illustration is given by \citet{granato00a}, who
find that radiative transfer through the age-dependent, two-phase
geometry of a model starburst produces an emergent attenuation curve
closely resembling the empirical \citet{calzetti00a} law, despite
adopting a grain mixture tuned to reproduce the Milky Way extinction
curve.  In the language of Eq.~\ref{eq:alambda2}, this is a case where
the wavelength dependence of $\mathcal{G}_\lambda$ dominates over that
of $\kappa_{\lambda}$ in setting the shape of the observed attenuation
curve.

What is known about $\kappa_\lambda$ from observational constraints?
The dust mass absorption coefficient has been empirically measured in
the Milky Way via the relation:
\begin{equation}
\frac{A_\lambda}{\Sigma_{\rm dust}} = \frac{A_\lambda/N_H}{(M_{\rm dust}/M_H)m_H} \mathcal{G}_\lambda = \frac{2.5}{\ln(10)}\mathcal{G}_\lambda\kappa_\lambda 
\label{eq:alambdasdust}
\end{equation}
where $A_\lambda/N_H$ is the attenuation per unit hydrogen column
density measured for individual stellar sightlines in the Milky Way
to be
\av/N$_H$\,$\approx$\,5.3$\times10^{-22}\,$mag\,cm$^{2}$\,/H \citep{bohlin78a,diplas94a,guver09a} in the V-band, where a dust-screen is
implicitly assumed ($\mathcal{G}_V=1$).
  \mdust/$M_H$ is the dust to
hydrogen gas ratio which has some metallicity dependence
\citep{draine07a,remy-ruyer14a};
for the Milky Way, solar metallicity, and a uniform dust screen
($\mathcal{G}_V=1$), this gives
$\kappa_{V}\,\approx\,3.2-3.3\times10^{4}\,$cm$^2$\,g$^{-1}$. 
This is
fairly well constrained to $\lesssim$10\%\ under these conditions.
For a general metallicity-dependent dust-to-gas ratio, and no
metallicity-dependent geometry, one would expect $\kappa_\lambda$ (in
the optical where $A_V/N_H$ is measured) to scale with metallicity following, e.g., Eq~9 of \citet{draine14a}.

At longer wavelengths, the total dust mass of a galaxy scales linearly
with the rest-frame flux density in the optically thin portion of the
spectrum \citep{hildebrand83a}:
\begin{equation}
M_{\rm dust} = \frac{S_{\nu}^{\rm rest} D_{\rm
    L}^2}{\kappa_{\nu}B_{\nu}(T_{\rm dust})}.
\label{eq:dustmass}
\end{equation}
Here $S_{\nu}^{\rm rest}$ is the rest-frame flux density of a given
dust reservoir emitting longward of $\ge$250\,\um\ (i.e. on the
Rayleigh-Jeans tail of dust blackbody emission), $D_{L}$ is the
luminosity distance, $\kappa_{\nu}$ is the dust mass absorption
coefficient at the frequency where the flux density is inferred
(i.e. in the FIR), $T_{\rm dust}$ is the mass-weighted average dust
temperature in the reservoir, and $B_{\nu}(T_{\rm dust})$ is the
Planck function evaluated at temperature $T_{\rm dust}$.  The
re-radiated spectrum per dust grain is given by\footnote{In practice,
the Planck function is computed in
erg\,s$^{-1}$\,cm$^{-2}$\,Hz$^{-1}$\,str$^{-1}$ and the use of $\nu$
in Equation~\ref{eq:dustmass} reflects the use of Janskys in the FIR
as typical convention.} $\kappa_\nu B_\nu(T_{\rm dust})$ and the
total dust luminosity set by $4\pi D^2_L S_\nu^{\rm rest}$; another
factor of 4$\pi$ arises from the conversion of $B_{\nu}$ from specific
intensity to a grain-normalized luminosity, and they cancel out.

The optically thin portion of the dust spectrum typically occurs at
rest-frame wavelengths longward of $\lambda\gtrsim250$\,\um.  For this
work, we will use rest-frame 500\um\ emission to anchor dust mass
calculations.  Here we adopt a dust mass absorption coefficient at
500\,\um, $\kappa_{\rm FIR}^{\rm fix}=2\,$cm$^{2}$\,g$^{-1}$ following
\citet{clark19a}, and we note that $\kappa_{\rm
  FIR}\propto\lambda^{-\beta}$ where $\beta=2$ throughout the FIR/mm
regime.  Throughout this paper we will refer to $\kappa_{\rm FIR}$ as
specifically evaluated at 500\um, with the \citet{clark19a} value
denoted $\kappa_{\rm FIR}^{\rm fix}$.
Importantly, we emphasize that dust mass coefficients in the FIR are
highly uncertain because the FIR lacks direct empirical constraints
unlike the UV/optical where attenuation along stellar sightlines can
be directly constrained.  A suite of foundational grain models
\citep{mathis77a} calculate $\kappa_{\lambda,abs}$ from the UV through
the submm \citep{draine84a}, and highlight the impact of dust grain
morphology --- fluffy aggregates vs. bare compact grains --- on
$\kappa_\lambda$ \citep{ossenkopf94a,henning96a}.  These models were
updated to include polycyclic aromatic hydrocarbons (PAHs) in
\citet{weingartner01a,li01a} and \citet{draine03a} with a
complementary analysis on grain shape in \citet{siebenmorgen14a}.
Given the lack of direct, empirical constraints on $\kappa_{\rm FIR}$,
we conservatively caution that dust masses cannot be known better than
within a factor of $\sim$2$\times$ following the 2$\times$ modeling
uncertainty on $\kappa_{\rm FIR}$.
The dust mass also has a dependence on flux density and presumed dust
temperature \tdust\ via the Planck Function, and there is also a
concern about applying single-temperature models to recover the total
dust mass, which is discussed in detail in \citet{sommovigo25a}.
However, despite the dependence on dust temperature and flux density
precision, the uncertainty of both ($\sim$20\%) are dwarfed by the
uncertainty of $\kappa_{\rm FIR}$.

Dust mass can then be translated to dust mass surface density with
knowledge of the dust emitting size effective radius, $R_{\rm e}$.  In
practice, resolved dust sizes are not available for large samples of
galaxies to high redshift, but for those that exist, FIR sizes and
morphologies draw parallel to trends observed in optical light: higher
redshift galaxies are smaller and FIR emission broadly appears
consistent with S\'{e}rsic profiles with $n=1$ \citep{hodge20a}.  In
this work, we will generalize to use galaxies' measured sizes from
JWST NIRCam imaging (rest-frame optical to near-infrared) as a good
proxy of their dust sizes for the purpose of calculating dust mass
surface densities.  We will assert in this work that $R_{e}^{\rm
  dust}\approx R_{e}^{\rm stars}/\sqrt{2}$.  This presumption is one
of convenience and necessitated by the limitation of current datasets;
it could be lifted in the future with improved constraints on that
relationship. The factor of $\sqrt{2}$ accommodates the fact that dust
reservoirs are often slightly more compact than their stellar
reservoirs and centrally peaked \citep{conselice14a,mosenkov19a}; while this
prefactor is not precisely constrained, the important component of
this is that $R_{e}^{\rm dust}$ does not evolve in a wildly different
manner than $R_{e}^{\rm stars}$, which is not just consistent with
recent observations but also theoretical expectation
\citep[e.g.][]{popping22a}.  It then follows that dust mass surface
density is then given by:
\begin{equation}
\Sigma_{\rm dust} = \frac{M_{\rm dust}}{\pi R_{e}^2} = \frac{S_{\rm 500}^{\rm rest}D_L^2}{\pi \kappa_{\rm FIR}B_{\rm 500}(T_{\rm dust}) R_e^2}
\label{eq:sigmadust}
\end{equation}
The dominant uncertainty on $\Sigma_{\rm dust}$ still comes from the
uncertainty on $\kappa_{\rm FIR}$ based on the critical size
assumption made here (where future observations could change the error budget).

Joining Equations~\ref{eq:alambda2} and \ref{eq:sigmadust} for attenuation at wavelength $\lambda$ in the UV/optical regime gives:
\begin{equation}
  A_\lambda = \frac{2.5}{\pi\ln(10)} \frac{S_{\rm 500}^{\rm rest}D_L^2}{B_{\rm 500}(T_{\rm dust}) R_e^2} \frac{\kappa_\lambda}{\kappa_{\rm FIR}}
\mathcal{G}_\lambda
\label{eq:alambda3}
\end{equation}
Here the observables from the FIR are $S_{\rm 500}^{\rm rest}$, from
an SED fit to observed-frame photometric constraints, $R_{e}$ from
high-resolution imaging, and $A_\lambda$ from a detailed SED fit to
the UV/optical SED.  The dominant uncertainty in this equation comes
from the product of $(\kappa_\lambda/\kappa_{\rm FIR})
\mathcal{G}_\lambda$ which we will set empirical limits on in this
work in the UV ($\lambda$=1600$\AA$).  A simplified form of
Eq.~\ref{eq:alambda3}, applied in the limit of a fixed geometry, was
used by \citet{casey24b} to estimate dust masses in little red dots
from their compact sizes and inferred attenuation; here we retain
$\mathcal{G_\lambda}$ and $\kappa_{\lambda}/\kappa_{\rm FIR}$ as
quantities to be constrained rather than assumed.

Following from Eq.~\ref{eq:alambda3}, it will be handy later in this
work to define the term
\begin{align}
  C_\lambda &\equiv \frac{2.5}{\ln(10)}\kappa_{\rm FIR}^{\rm fix}\Big(\frac{\kappa_\lambda}{\kappa_{\rm FIR}}\Big)\mathcal{G}_\lambda \left[ 10^{5}\,M_\odot\,{\rm kpc}^{-2}\right] \nonumber \\
  & = A_\lambda \left[ \frac{10^{5}\,M_\odot\,{\rm kpc}^{-2}}{\Sigma_{\rm dust}} \right]
\label{eq:clambda}
\end{align}
which represents the ratio of effective attenuation to dust mass
surface density. Here, $\kappa_{\rm FIR}^{\rm
  fix}$=\,$4.2\times10^{-10}$\,kpc$^2$\,\msun$^{-1}$, our fiducial
adopted value of the dust mass absorption coefficient at rest-frame
500\um\ in units of kpc$^2$\,\msun$^{-1}$. This renders $C_\lambda$
a unitless ratio of attenuation to dust mass surface density.
In the V-band, \citet{aniano12a} infer \cv\,=\,0.67, and
\citet{draine14a} find \cv\,=\,0.74, both based on Milky Way and dust
model calibrations assuming a foreground screen.

If the FIR SED of a galaxy can be observed, then the integrated IR
luminosity is known (\lir, typically integrated from rest-frame
8--1000\,\um, but dominated by emission right around
$\sim$100\,\um\ rest-frame), in addition to constraints on dust mass
\mdust\ from the Rayleigh-Jeans tail.  While \mdust\ depends on
$\kappa_{\rm FIR}$ which traces dust grain properties, {\it
  the IR luminosity is free of any such dependence.}  Through energy
balance, the IR tells us, objectively, how much light has ultimately
been absorbed in the rest-frame UV/optical and reprocessed by dust
grains.  \lir\ is simply equal to the difference between the intrinsic
UV (and optical) luminosity and the observed \luv.  A wealth of
literature exists on IRX\,$\equiv$\,\lir/\luv\ where \lir\ and
\luv\ are both {\it observed} quantities \citep[e.g.][]{meurer99a} and
its relationship to both \auv\ \citep{buat05a,hao11a,cortese08a} and
rest-frame UV slope
\citep{calzetti94a,reddy06a,gil-de-paz07a,takeuchi12a}.  We can thus
draw a direct line between IRX, this ratio of energy output in the IR
and UV, and the attenuation that {\it should} be seen in the optical
if a simple dust screen is assumed:
\begin{align}
A_{\mathrm{UV,direct}} &= 2.5 \log \left[ 1 + \frac{\mathrm{IRX}}{B^\prime} \right] \nonumber \\
&= 2.5 \log \left[
1 + \Big(\frac{C^\prime_{\rm TIR}}{C^\prime_{\rm FUV}}\Big)\frac{1 - f_{\mathrm{unobs}}}{B^\prime\,f_{\mathrm{unobs}}}
\right]
\label{eq:auvdirect}
\end{align}
The coefficient $B^\prime$ is the ratio of two bolometric correction
factors, BC(1600${\rm \AA}$)/BC(FIR), as discussed in
\citet{meurer99a} and \citet{mclure18a}.  $B^\prime$ is approximately
equal to one, dominated by the bolometric correction in the UV (where
1600\AA\ luminosity does not reflect the total UV light that is
absorbed by dust); later in \S~\ref{sec:auvcompare} we discuss
$B^\prime$ further and describe how it can be directly inferred for a
given UV/optical SED.  The second term in Eq.~\ref{eq:auvdirect}
follows from the first, where \funobs\ is the ratio of the star-formation rate (SFR)
that is unobscured relative to the total,
i.e. \funobs$\equiv$SFR$_{\rm UV}$/(SFR$_{\rm UV}$+SFR$_{\rm IR}$),
which is another convenient conceptual way of framing IRX.  The
coefficient, $C^\prime_{\rm TIR}/C^\prime_{\rm FUV}$, then represents
the ratio of SFR-to-luminosity scalings\footnote{$C$ is used in
\citet{kennicutt12a} to denote the coefficient between luminosity and
SFR, but here we use $C^\prime$ to be clear this is a different
quantity than is denoted in Eq.~\ref{eq:clambda}} between the total
infrared (TIR), and far ultraviolet, FUV, in \citet{kennicutt12a}: $C^\prime_{\rm
  TIR}/C^\prime_{\rm FUV}\approx$1.15.

We can then draw direct comparison between \auvd, the attenuation that
{\it should} be present if all of the dust were concentrated in a
screen, and \auv, the effective or `observed' attenuation in the UV.
The latter may be derived through detailed UV/optical SED fitting when
sufficient constraints are at hand to also constrain the star
formation history \citep[there is always some degeneracy between
  stellar age and attenuation causing redder SEDs in the UV/optical;
  see review on stellar population synthesis by][]{conroy13a}.  The
difference between \auvd\ and \auv\ is attributable to
geometry alone, such that:
\begin{equation}
  \mathcal{G}_{\rm UV} = \frac{A_{\rm UV}}{A_{\rm UV,direct}}
  \label{eq:geometry}
\end{equation}
This also follows from dividing Eq.~\ref{eq:alambda2} by
Eq.~\ref{eq:alambda1} in the UV (we will use 1600\AA\ to denote UV in
this work).  So while the geometry itself is complex, we can use the
rest-frame UV to directly parameterize $\mathcal{G}_{\rm UV}$ in a
global, integrated sense. Then we may combine Eq.~\ref{eq:alambda3}
with Eq.~\ref{eq:auvdirect} and Eq.~\ref{eq:geometry} and place
constraints on the galaxy-integrated ratio of dust mass absorption
coefficients, \kratio\ directly:
\begin{equation}
  \Big(\frac{\kappa_{\rm UV}}{\kappa_{\rm FIR}}\Big) = \pi \ln(10)
  \log\left[1+\frac{IRX}{B^\prime}\right] \frac{B_{\rm
      500}(T_{\rm dust}) R_e^2}{S_{\rm 500}^{\rm rest}D_L^2}
  \label{eq:kratio}
\end{equation}
Measurement of this ratio can provide some first steps towards a
greater understanding of the dust grain properties in the early
Universe through a relatively straightforward set of observations.
The observables\footnote{The framework we discuss takes for granted
that redshift, $z$, is reasonably well constrained.} here are \lir,
\luv, $S_{\rm 500}^{\rm rest}$, \tdust\ and $R_{e}$.  We also note
that the ratio of \lir\ and $S_{\rm 500}^{\rm rest}$ form a proxy for
\tdust, leaving the number of true independent, observational
constraints needed to four.  We note that, theoretically, \kratio\ is
a microphysical characteristic of dust grain properties, but here we
are calculating a quantity linked in meaning, but integrated on
macroscopic scales, which inherently folds in second-order effects of
ISM geometry, which are complex and not directly measured.  While this
ratio may still be uncertain within a factor of two or more (again,
from uncertainty on $\kappa_{\rm FIR}$), we can infer if it {\it
  evolves}, which can inform future models of dust grain physics in
the early Universe where no direct constraints on $\kappa_\lambda$
exist.

Our work aims to make such measurements, and simultaneously present
observations on the evolution and stellar-mass dependence of related
quantities: the magnitudes of attenuation in the UV and optical, the
effective attenuation law slope, \lir, \mdust\ and \tdust, the
fraction of unobscured star formation relative to total (\funobs), and
the dust-to-stellar mass ratio.  This paper focuses on a stacking
analysis, while a future work will focus on the specific
characteristics of galaxies that have direct detections in the IR.

\section{Data}\label{sec:data}

We use data from the COSMOS-Web Survey \citep[PIs: Kartaltepe
  \&\ Casey;][]{casey23a}.  The unique combination of depth and area
covered by JWST and ALMA datasets in recent years is particularly
notable in the COSMOS field, making it the only dataset uniquely
capable of this type of measurement: stacking many thousands of
galaxies across uniquely diverse (sub)mm datasets.

\subsection{JWST Data}

The JWST data in COSMOS, which forms the backbone of the dataset used
in this paper, is described in the imaging reduction papers presenting
NIRCam data \citep[a contiguous 0.54\,deg$^2$ in four
  filters;][]{franco25a} and MIRI data \citep[0.2\,deg$^2$ taken in
  parallel;][]{harish25a} and the COSMOS-Web ``COSMOS2025'' catalog
description paper \citep{shuntov25a}.  To briefly summarize these
data, NIRCam imaging in COSMOS-Web includes four filters (F115W,
F150W, F277W, and F444W) with 5$\sigma$ point source depth (in
0$\farcs$3 diameter apertures) roughly 27.5-28.2 AB.  MIRI includes
one filter (F770W) to a 5$\sigma$ point source depth of 25.5-26 AB.
The COSMOS2025 catalog was assembled using a NIRCam-only combined
$\chi^2_{+}$ detection image, hot$+$cold detection with Source
Extraction and Photometry \citep[SEP;][]{barbary16a}, 2D morphological
S\'{e}rsic model construction using SourceExtractor$++$ (SE$++$) on
NIRCam, and fixed model photometric extraction on 37 photometric bands
spanning ultraviolet through mid-infrared.  The catalog contains
784,016 sources spanning the 0.54\,deg$^2$ of NIRCam coverage.
Photometric redshifts are fit using LePhare
\citep{arnouts02a,ilbert06a} and a broad range of templates spanning a
wide range of attenuations, emission line strengths, and complex star
formation histories \citep[see \citealt{shuntov25a} for details on
  modifications made to LePhare for COSMOS-Web and ][ for more detail
  on CIGALE-derived physical properties of sources in COSMOS2025 and
  the consistency of those properties with LePhare]{arango-toro25a}.

Of the 0.54\,deg$^2$ area, 16\%\ of the area (0.086\,deg$^2$) is
contained within ground-based star masks; in other words, the
ground-based photometry in these regions is severely limited by
spatial confusion with bright foreground stars that are less
problematic in space-based imaging.  Space-based photometric
redshifts, including only {\it HST} and {\it JWST} photometry, are
generated for all objects in the catalog and these are used in place
of the full ground+space redshifts inside of these star masked regions
(see \citealt{shuntov25a} for discussion of the broad consistency
between ground+space and space-only photometric redshifts).

\subsection{(Sub)Millimeter Data}

COSMOS has a wealth of (sub)millimeter ([sub]mm) data from which we
can derive accurate dust emission constraints.  The most notable
datasets of competitive depth and unique wavelength coverage include
{\it Spitzer}/MIPS, {\it Herschel}/PACS+SPIRE, SCUBA-2, NIKA2, and ALMA
coverage.
{\it Spitzer} MIPS 24\um\ imaging \citep{sanders07a} covers the full
field to 1$\sigma$=16\,\uJy\ depth. {\it Herschel} PACS 100\um\ and
160\um\ imaging \citep{lutz11a} also cover the full field to 2.37\,mJy
and 4.75\,mJy 1$\sigma$ depth, respectively.  Similarly, {\it
  Herschel}/SPIRE imaging, at 250\um, 350\um, and
500\um\ \citep{oliver12a} is confusion-limited to a 1$\sigma$ rms of
5.8\,mJy, 6.3\,mJy and 6.8\,mJy respectively.  SCUBA-2 imaging in the
field is described by \citet{simpson19a}, which has 1$\sigma$ depth
ranging from 0.6--1.4\,mJy at 870\um\ (while the effective wavelength
of the filter is 870\um, the filter is often referred to as 850\um).
NIKA2 data in the field covers an area of $\sim$0.3\,deg$^2$
\citep{carvajal-bohorquez26a,bethermin26a} mostly overlapping the
COSMOS-Web area; maps at 1.3\,mm and 2\,mm have 1$\sigma$ depths of
0.32\,mJy and 0.091\,mJy, respectively, with corresponding spatial
resolutions of 11$''$ and 17$''$.

Finally, ALMA data, though not exhaustive or uniform, covers a large
number of sources at different frequencies and depths.  The two
roughly uniform blank-field survey datasets used in this work are the
Ex-MORA survey \citep[][ a continuation of the Mapping
    Obscuration to Reionization with ALMA, MORA, survey described in
  \citealt{casey21a} and \citealt{zavala21a}]{long26a}, a 2\,mm deep
field (1$\sigma\,\approx$\,90\,\uJy\ at $\sim$2.1\,mm) covering
0.16\,deg$^2$, and the CHAMPS survey \citep[Faisst et al., in prep,
  Martinez et al., in prep][]{zavala26a}, a 1.2\,mm mosaic
(1$\sigma\approx$0.14\,mJy) covering the 0.2\,deg$^2$ of JWST MIRI
imaging in the field from COSMOS-Web and PRIMER \citep{donnan24a}.

Beyond these blank-field efforts, we make use of the
A3COSMOS\footnote{See https://sites.google.com/view/a3cosmos} survey
\citep{liu19a,adscheid24a} to formulate a complete list of individual
dust detections in the field, combining data from the SCUBA-Dive
project \citep{mckinney24a} on the brightest submm galaxies in the
field, to CHAMPS, and Ex-MORA detections.  The A3COSMOS dataset (in
particular the 2025-03-12 data release) contributes most of the direct
detection sources: 725 out of a total of 1024 across the field.  In
the stacking analysis that follows we keep track of these individual
detections and their influence on the results but save analysis of
their characteristics to a future work.

\section{Methods}\label{sec:methods}

Here we present the derivation of physical parameters closely
associated with dust emission and absorption.  To do this, we treat
dust emission and dust absorption independently.  The former is
measured in the submm and the latter is measured from the rest-frame
UV/optical portion of the SED.
This is complementary to, though different than, common efforts that
use energy balance techniques to fit both portions of the spectrum
simultaneously.  Energy balance approaches directly translate absorbed
rest-frame UV/optical light to re-radiated FIR/millimeter emission
\citep{da-cunha08a,da-cunha15a,burgarella05a,noll09a,boquien19a}.
Such energy balance techniques, though insightful, rely on some key
assumptions we aim to break down empirically in this work: fixed (or
relatively inflexible) attenuation laws, fixed dust mass absorption
and emissivity assumptions, as well as fundamental presumptions made
regarding star/dust geometry.  A distinct class of models instead
solves the radiative transfer explicitly through a specified, often
age-dependent star/dust geometry, such that the emergent attenuation
law is a prediction rather than an input
\citep[e.g. GRASIL;][]{silva98a,granato00a}.  These avoid the
assumption of a fixed attenuation curve, but require the geometry
itself to be specified a priori -- which is the quantity we aim to
constrain empirically here. By not enforcing energy balance, we are
able to compare the inferred attenuation to the output \lir\ and
assess, under what conditions, the UV/optical SED accurately recovers
re-radiated dust emission.

In what follows we first describe the selection and pruning of
galaxies in the COSMOS-Web catalog for stacking in FIR/(sub)mm
datasets in \S~\ref{sec:stacking}.  Note that we explicitly aim to
exclude quiescent systems from our stacks for ease of physical
interpretation; at high masses and low redshifts, a significant
population of quiescent galaxies are known to have markedly different
dust characteristics than mass-matched galaxies with higher
star-formation rates, with quiescent galaxies showing both suppressed
dust emission and absorption \citep[e.g.][]{paspaliaris23a,chang26a}.
So it is in the interest of not mixing physically distinct populations
-- quenched systems and star forming systems -- that we restrict to
those thought to have dust warmed by the interstellar radiation field
generated from star formation.  

We then describe the IR SED fitting procedure to the stacked SEDs in
\S~\ref{sec:irsed}.  Finally, we bootstrap the COSMOS2025 catalog to
re-derive flexible attenuation law characteristics using priors based
on the COSMOS2025 photometric redshifts (\S~\ref{sec:bagpipes}).

\subsection{Stacking}\label{sec:stacking}

Of 784,016 sources in the COSMOS-Web catalog, only 1024 (0.1\%)
currently have direct detections in the FIR/(sub)mm.  Those individual
galaxies are a focus of a separate work.  We use submm stacking to
characterize the dust properties of a broader dynamic range of
`normal' galaxies here.  Submillimeter stacking is not a new
enterprise \citep[for
  example,][]{dole06a,pascale09a,penner11a,viero13a,bethermin15b,coppin15a,tacconi18a,inami20a,magnelli20a,euclid-collaboration26a}.
However, the limits of stacking results (in redshift, IR-luminosity,
and mass) are dependent both on the size of the sample being stacked,
and the precision with which its characteristics (like redshift, mass)
are known, and the depth and wavelength of the (sub)mm data used in
the stacking.  JWST has taken us into a new realm of stackable
samples: the on-sky density of galaxies has increased significantly,
both at higher redshifts and for lower mass galaxies.  JWST data has
also significantly improved galaxies' photometric redshifts, improving
the purity of stackable samples \citep{shuntov25a}.  Similarly, ALMA
has now built up significant (sub)mm datasets to competitive depths
such that stacking of yet-unreached mass regimes is now possible.
ALMA also has more long wavelength data ($>$1\,mm) than single-dish
facilities that have covered wide areas (like Herschel, SCUBA-2); the
long wavelengths, as well as ALMA-like depths at those wavelengths,
are crucial for measuring reliable dust masses out to high redshifts
(because they still probe the regime where the dust SED is
safely optically thin, longward of rest-frame $\gtrsim$250\,\um).

The FIR/(sub)millimeter data we stack in this analysis includes: {\it
  Spitzer} MIPS-24\um\ \citep{sanders07a}, {\it Herschel}/PACS
100\um\ and 160\um\ maps \citep{lutz11a}, {\it Herschel}/SPIRE 250\um,
350\um, and 500\um\ maps \citep{oliver12a}, SCUBA-2
850\um\ \citep{simpson19a}, NIKA2 1.3mm and 2mm
\citep{carvajal-bohorquez26a,bethermin26a}, CHAMPS 1.2mm (Martinez
\etal, Faisst \etal, in prep), and Ex-MORA 2mm \citep{long26a}.

We stack these datasets by stellar mass and photometric redshift as
presented in the COSMOS2025 catalog \citep{shuntov25a}.  We first
apply some basic filtering of the COSMOS2025 catalog to remove sources
of potential low quality.  Of the 784,016 sources in the catalog, we
select those with:
\begin{enumerate}
\vspace{-2mm}
\item {\tt mag\_model\_f277w} $\le$\ 28.2 \&
\vspace{-2mm}
\item {\tt warn\_flag} = 0 \&
\vspace{-2mm}
\item ($\log$({\tt chi\_star}/{\tt chi2\_best}) $>$ 1 {\tt OR}\\ $\log$({\tt radius\_sersic}$\times$3600) $>$ -2.3) \&\ 
\vspace{-2mm}
\item ({\tt flag\_star\_hsc} = 0 {\tt OR} $|${\tt zpdf\_med}-{\tt zpdf\_med\_space}$|$/(1+{\tt zpdf\_med}) $\le$ 0.15) \&\
\vspace{-2mm}
\item $|${\tt zpdf\_u68}-{\tt
    zpdf\_l68}$|$/(1+{\tt zpdf\_med}) $\le$ 3
\vspace{-2mm}
\end{enumerate}
The first criterion selects sources brighter than the nominal
5$\sigma$ point source detection limit in the deepest COSMOS-Web band;
the second criterion selects non-flagged sources \citep[see][for
  flagging details, mostly related to artifacts]{shuntov25a}. The
third criterion serves to exclude Milky Way stars from analysis:
unresolved point sources where the $\chi^2_{\rm stars}<10\chi^2_{\rm
  galaxy}$.  This is a bit more conservative than the stellar flag
included in the COSMOS-Web catalog public release to prevent marginal
or unclear sources from entering the stackable galaxy population.  The
fourth criterion requires a source lie outside the Hyper Suprime-Cam
(HSC) star mask region (where ground-based data is potentially faulty
due to bright star contamination) or, if within the HSC star mask
region, that there is close agreement between the redshift solutions
found with and without exclusion of the ground-based photometric
constraints.  The last criterion down-selects to sources without
significant bimodality in their redshift PDFs.  This filtering of the
catalog produces a set of 595,351``stackable'' sources (accounting for
79.6\%\ of the COSMOS2025 catalog).

We make two final cuts to the set of stackable sources.  The first cut
removes sources with suspiciously high stellar mass estimates for
their photometric redshift. To do this we calculate a mass threshold
above which sources should be regarded with suspicion for formation
within $\Lambda$CDM. We determine this redshift-dependent mass
threshold using the \citet{sheth99a} halo mass function and calculate
the most massive object expected in a given redshift bin (of width
$dz=0.5$) in the 0.54\,deg$^2$ area of the survey, assuming that
source has converted 30\%\ of its baryons to stars \citep[following
  the same logic as in][]{boylan-kolchin23a}.  The threshold of a
30\%\ conversion exceeds typical expectation for the efficiency of
star formation of, at most,
$\sim10$\%\ \citep[e.g.][]{evans09a,bigiel10a,thompson05a,ostriker11a}.
Objects with masses higher than this redshift-dependent maximum mass
limit are excluded from the stacks.  There are 142 such sources above
the limit, likely because their photometric redshifts are
overestimated; alternatively their stellar masses may be overestimated
due to significant non-stellar contributions to their SEDs.  Whatever
the origin, we make the conservative choice of cutting them out of the
stacking process.

\begin{figure}
\includegraphics[width=0.99\columnwidth]{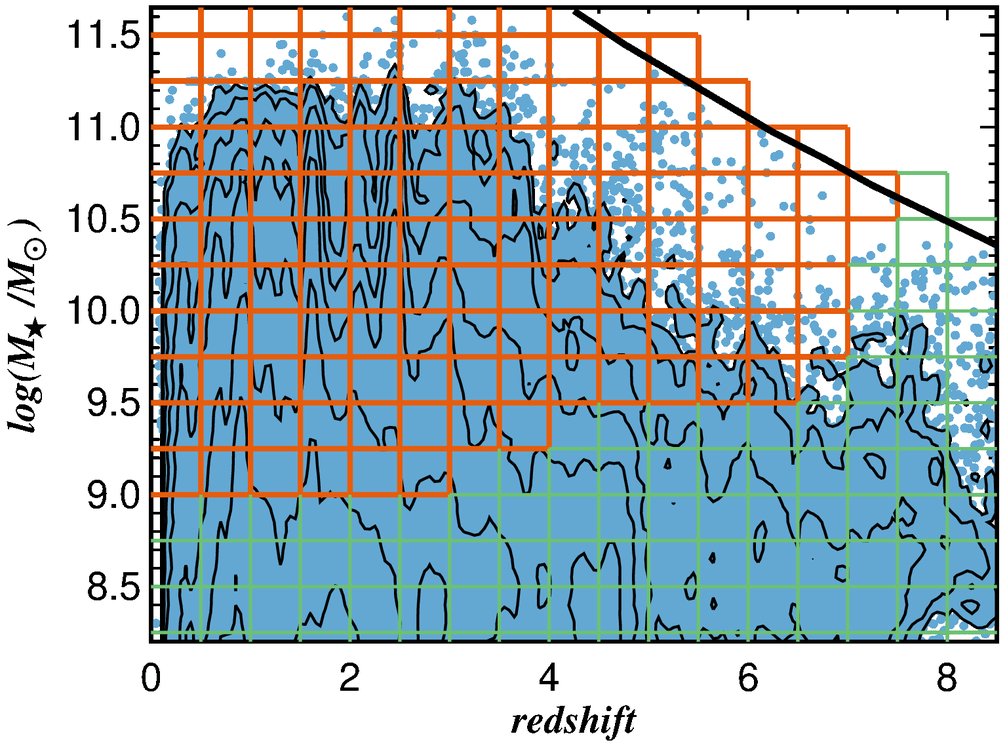}
\caption{The distribution of the stackable 501,656 sources from the
  COSMOS field used in this work.  Contours indicate concentrations of
  sources exceeding $>$5 per unit $\Delta z=0.05$ and
  $\Delta\log(M)=0.05$ (contours mark densities of 5, 10, 20, 50, 100,
  ... sources per same binning).  The bins used for stacking are shown
  with green and orange gridlines, every $\Delta z=0.5$ in redshift
  and $\Delta\log(M_\star/M_\odot)=0.25$.  Bins with converged dust
  SEDs (constraints for both dust emission and attenuation) are
  outlined in orange while bins with binned attenuation
  characteristics only (no derived dust emission) are shown in
  green. The thick black line marks the maximum allowable stellar mass
  we include in the stacking analysis: this redshift-dependent mass
  limit is determined by calculating the maximum halo mass found per
  survey volume and assuming that halo has a stellar baryon fraction
  of 30\%.}
\label{fig:stackables}
\end{figure}

One final, important cut is made to remove quiescent galaxies.
Galaxies at low redshifts and high stellar masses follow a bimodal
division where the dust characteristics of star-forming massive
galaxies are quite a bit different than for quiescent massive galaxies
\citep{paspaliaris23a}.  We expect dust emission in the former and not
the latter \citep{toni26a}.  We cut out quiescent systems using a SFR
cut with respect to the galaxy main sequence, or SFR-M$_\star$
relation.  For a given {\tt zpdf\_med} and {\tt mass\_med}
(i.e. $z_{\rm phot}$ and M$_\star$), we calculate the expected main
sequence SFR ``SFR$_{\rm MS}$'' using the \citet{speagle14a}
parameterization and excludes sources that are more than a dex lower
than the main sequence.  In other words, retained sources have
SFR$>$SFR$_{\rm MS}$/10; this includes all sources down to an order of
magnitude below the main sequence.  We explicitly test that the
adoption of the \citet{speagle14a} main sequence in this instance does
not bias our results unintentionally; we alternatively test the
\citet{popesso23a} main sequence formulation and find our stack
samples to differ by $\le$1\%. This reduces the total stackable
star-forming galaxy sample to 501,656 sources (64.0\%\ of COSMOS2025).
The distribution of sources in the redshift, mass plane is shown in
Figure~\ref{fig:stackables}, along with the mass and redshift bins
used in this work ($\Delta z=0.5$ and $\Delta
\log(M_\star/M_\odot)=0.25$).

We note that all stackable sources are covered by MIPS, PACS, SPIRE
and SCUBA-2 maps. There are 351,222 of those within the NIKA2 1.3\,mm
map footprint, and 348,503 in the NIKA2 2\,mm map footprint.
Similarly there are 209,320 sources covered in the CHAMPS map and
149,685 in the Ex-MORA map.  These numbers are reflective of the area
within COSMOS-Web covered by each survey.

Note that we do not explicitly {\it exclude} any additional special
category of source from this list of stackable sources.  Both X-ray
active galactic nuclei (AGN) and submillimeter luminous galaxies are
included as a result.  We have relatively few concerns about
`contamination' of IR flux by the former.  Luminous X-ray AGN do
correlate somewhat with IR luminous galaxies \citep{iwasawa11a},
however the wavelengths of light we are stacking predominantly
correspond to cold ISM dust.  Constraints in the mid-infrared, where
an AGN might boost the luminosity substantially, are very weak
relative to the longer-wavelength, cold dust constraints.  The primary
concern regarding AGN is their impact on a galaxy's SED-modeled
stellar mass \citep{ciesla15a,florez20a,buchner24a}, such that stellar
masses may be overestimated with inclusion of buried AGN luminosity in
the UV/optical SED fitting.  This is likely a persistent issue,
indeed, and the fix for it would involve a very careful analysis of
AGN content of the galaxies being stacked, which is beyond the scope
of this work.  We do note the number of direct-detection X-ray AGN in
the COSMOS2025 catalog is 1039, which is statistically insignificant
compared to the number of stacked sources (1039/501,656$\approx$0.2\%)
and preferentially only exist for the most massive subset of systems
(above $10^{10}$\,\msun) at $z<3.5$.

The incidence of AGN activity has been measured by stellar mass and
redshift for mass-selected samples spanning our parameter space, using
independent methodologies -- forward-modeling the X-ray luminosity
function through the stellar mass function \citep{georgakakis17a},
stacking X-ray data for NIR-selected galaxies \citep{aird18a}, and
semiparametric inference across many different surveys
\citep{yang18a,zou24a} -- and is consistently found to vary smoothly
and monotonically, with the specific accretion rate distribution
shifting steadily up with increasing redshift and stellar mass.  This
suggests that whatever systematic offset is imparted to stellar masses
by AGN contamination is likely to vary smoothly across the redshift
and mass range studied here, like it is in the cosmically-averaged
black hole accretion rate density and star-formation rate density
\citep{delvecchio14a,madau14a}.  We expect the effect of AGN to be
most significant in our most massive, highest-redshift bins where AGN
duty cycles are relatively higher; this may manifest most importantly
in the elevated dust temperatures seen at high mass and high-$z$.

The inclusion of IR luminous galaxies (e.g. DSFGs broadly defined) is
a potential contaminant some readers may worry about as it is possible
their IR flux densities are significantly elevated above the norm.  To
check the impact of the IR luminous population, we have run two
versions of our stacks: one with the 1024 direct detections explicitly
removed and one including them.  We find our stacking results, with
and without DSFGs, consistent within errors and attribute this to the
relative on-sky rarity of IR luminous galaxies compared to the size of
the samples used in the stacking (1024/501,656$\approx$0.2\%).  The
bins that are most substantially affected (where $\simgt$50\%\ of the
stackable sample is removed) are at $\log(M_\star/M_\odot)>11$, though
the resulting flux densities are consistent within (large) errors. For
the sake of simplicity in our measurements, and in the interest of
avoiding cherry-picking the stacked sample, we proceed in our analysis
without excluding IR luminous galaxies.

Our stacked flux densities are measured via a weighted
inverse-variance median based on the rms of the (sub)mm datasets (not
the quality of the redshift or stellar mass estimates). Because we are
not accounting for redshift uncertainties, we will point out that
there will be some level of `bin-smearing' in bins where the
photometric redshift quality is a bit worse.  This happens at the
low-mass and high-$z$ end of our stacks where $\sigma_{z}/(1+z)$ is of
order the bin size; the vast majority of bins have somewhat precise
photometric redshift uncertainties, $\sigma_{z}/(1+z)\approx0.03$, out
to $z\sim7$.
A full description of the stacking procedure, including a comparative
analysis of correcting flux densities for clustering bias, is detailed
in the Appendix.  We note that confusion noise, due to low spatial
resolution maps from {\it Herschel} or {\sc Scuba-2}, is greatly
mitigated when combining data during a stack as the maps are set to
have zero background, a key advantage that lets us push the depth of
these maps with sufficient statistics.  We provide stacked cutouts as
well as a full tabulation of the measured flux densities
(Table~\ref{tab:masterflux}) and derived SED characteristics
(Table~\ref{tab:mastersed}, derivation thereof described below) from
the stacked samples above stellar masses 10$^9$\,\msun\ from $0<z<7.5$
in the Appendix\footnote{Both tables are available online in fits
format at https://github.com/caitlinmcasey/duststacks .}.

\subsection{Fitting IR Dust SEDs}\label{sec:irsed}

Our IR SED fitting follows the methodology of MCIRSED described in
\citet{drew22a}, built on earlier SED methodology described in
\citet{casey12a}, which joins a single modified blackbody to a
mid-infrared power law and fits SEDs to photometry in a Bayesian
framework.  In this framework, non-detections are handled in the same
fashion as detections with flux density measurements and uncertainties
folded into the fit, where non-detections are statistically consistent
with zero flux density at a given wavelength. The model is quite
flexible and requires some choices. The free parameters of the dust
SED are the following, from best-constrained to least-constrained:
\begin{enumerate}
  \vspace{-1mm}
  \item the integrated IR luminosity \lir\ (which determines the
    normalization of the fit and is canonically integrated between
    8--1000\um),
  \vspace{-1mm}
  \item the luminosity-weighted cold dust temperature \tdust,
  \vspace{-1mm}
  \item the emissivity spectral index $\beta$ (which determines the
    slope of the Rayleigh-Jeans tail of dominant dust blackbody emission),
  \vspace{-1mm}
    \item the mid-infrared spectral slope $\alpha_{\rm MIR}$ (which determines
the relative proportion of warm dust to the dominant cold dust in the
SED), and
  \vspace{-1mm}
\item an assumption regarding the opacity of the dust (i.e. at
what wavelength the SED transitions from optically thin to thick,
$\lambda_0\equiv\lambda_{\tau=1}$).
  \vspace{-1mm}
\end{enumerate}
In only rare cases --- DSFGs with a wealth of IR photometric
constraints as well as spatially resolved IR maps --- can all of these
parameters be directly constrained.  In practice, most galaxies have a
tremendous dearth of information in this wavelength regime, so some
assumptions must be applied.  That is the case for the stacked SEDs.

After extensive testing and evaluation of the resulting SEDs, we
decide to uniformly fix the last parameter described above: the
wavelength at which opacity shifts from thick to thin ($\lambda_0$).
We use $\tau=1$ at $\lambda_0=$\,100\,\um, which is consistent with
existing constraints in the literature
\citep{draine06a,conley11a,greve12a,simpson17a}.  We note that the
choice of $\lambda_0$ impacts the mapping of the rest-frame peak
wavelength, $\lambda_{\rm peak}$, to the dust temperature,
\tdust\ (see Figure~20 of \citealt*{casey14a}), but it does not
significantly change the shape of the best-fit SED.  For example, a
similar optically thin SED fit would produce the same peak wavelength
$\lambda_{\rm peak}$ and same \lir, but a correspondingly lower dust
temperature.  Note that the observables really are $\lambda_{\rm
  peak}$ and \lir, and our inference of temperature is model
dependent.  The adopted opacity model, similarly, does not impact dust
mass, because the mass is inferred from the optically thin portion of
the dust SED, longward of rest-frame $\gtrsim$250\,\um.

Our constraints on the mid-infrared, governing $\alpha_{\rm IR}$, come
from {\it Spitzer} 24\um\ and {\it Herschel} PACS.  Over a large range
of low-redshift SEDs both {\it Spitzer} and PACS measurements have
high signal-to-noise ratio (SNR), so we allow $\alpha_{\rm IR}$ to
vary as long as there are two or more photometric constraints
$>3\sigma$ significant short-ward of rest-frame 200\,\um.  When those
constraints are not present, we adopt $\alpha_{\rm IR}=4.5$
translating to a moderately steep drop off of the Wien's side of the
dust blackbody.  We note this is somewhat steeper than the empirically
measured $\alpha_{\rm IR}\approx2-2.5$ for local (U)LIRGs
\citep{u12a,casey12a}, but consistent with many other high-$z$ DSFGs
\citep{kirkpatrick12a,casey14a} and our own measurements in bins where
it is explicitly measured: $\langle \alpha_{\rm
  IR}\rangle=4.61^{+0.35}_{-0.22}$.

Our constraints on the Rayleigh-Jeans tail are quite good below
$z\sim4$ allowing for a direct fit of the emissivity spectral index,
$\beta$.  If there are more than two photometric constraints
$>3\sigma$ significant at rest-frame wavelengths longward of 300\,\um,
we allow a free fit for $\beta$.  We otherwise fix the value to
$\beta=2$ \citep{hildebrand83a,dunne01a,da-cunha08a}.  This fixed
value is consistent with the bins where $\beta$ was directly fit,
$\langle\beta\rangle=1.98^{+0.24}_{-0.18}$, as well as other recent
literature constraints \citep{casey21a,cooper22a,long26a}.
While one might be concerned about a comparison of fits with free
$\alpha_{\rm IR}$ and $\beta$ to those with fixed values, we note that
the dynamic range of both parameters impact the measured \lir\ by less
than 20\%.

\begin{figure*}
  \includegraphics[width=0.99\textwidth]{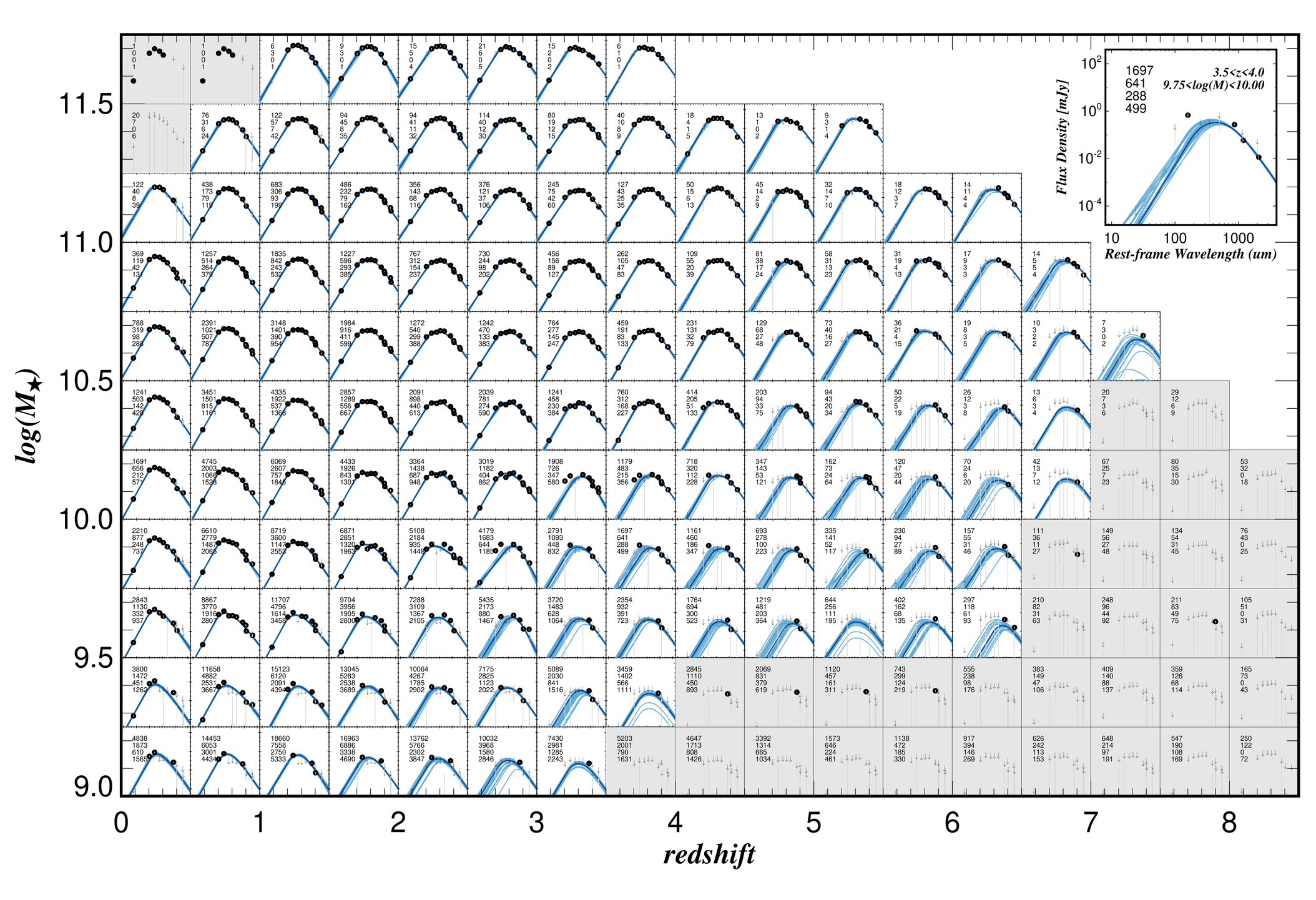}
  \caption{The best-fit dust SEDs to stacked (sub)millimeter
    photometry for stellar mass and redshift selected samples.  Each
    panel shows a schematic of the dust SED (blue, with light blue
    SEDs sampling the uncertainty) superimposed on the stacked
    photometry (black with gray error bars).  Upper limits
    ($<3\sigma$) are shown as downward facing arrows at their
    3$\sigma$ upper limits (error bars are still shown in light gray).
    Panels with gray backgrounds have photometric constraints of too
    poor quality to fit a converged dust SED (fewer than two points
    above 3$\sigma$ significance).  The numbers in the upper left of
    each panel indicate the number of sources stacked across (1)
    Spitzer, Herschel and SCUBA-2, (2) NIKA2, (3) CHAMPS, and (4)
    Ex-MORA (the latter three cover smaller areas than the full
    field). Inset to the top right is one example panel in finer
    detail to show the wavelength and flux density axis, which is
    fixed for every subpanel.}
\label{fig:masterseds}
\end{figure*}

Photometry is given an additional 10\%\ flux calibration uncertainty
and individual stack measurements are capped at a maximum SNR of 10 in
order to minimize the statistical power of individual high SNR
photometric measurements (e.g. 24\um).  \lir\ and \tdust\ are left as
free parameters for all fits.  Dust mass is not a parameter fit for
explicitly, but inferred directly within the SED fitting and
calculated using Eq.~\ref{eq:dustmass} for each set of SED parameters.
All SED fitting is corrected for cosmic microwave background (CMB) heating following the
prescriptions provided in \citet{da-cunha13a}. A gallery of resulting
SED fits over all stacked bins is shown in
Figure~\ref{fig:masterseds}.

\begin{figure*}
  \centering
  \includegraphics[width=1.8\columnwidth]{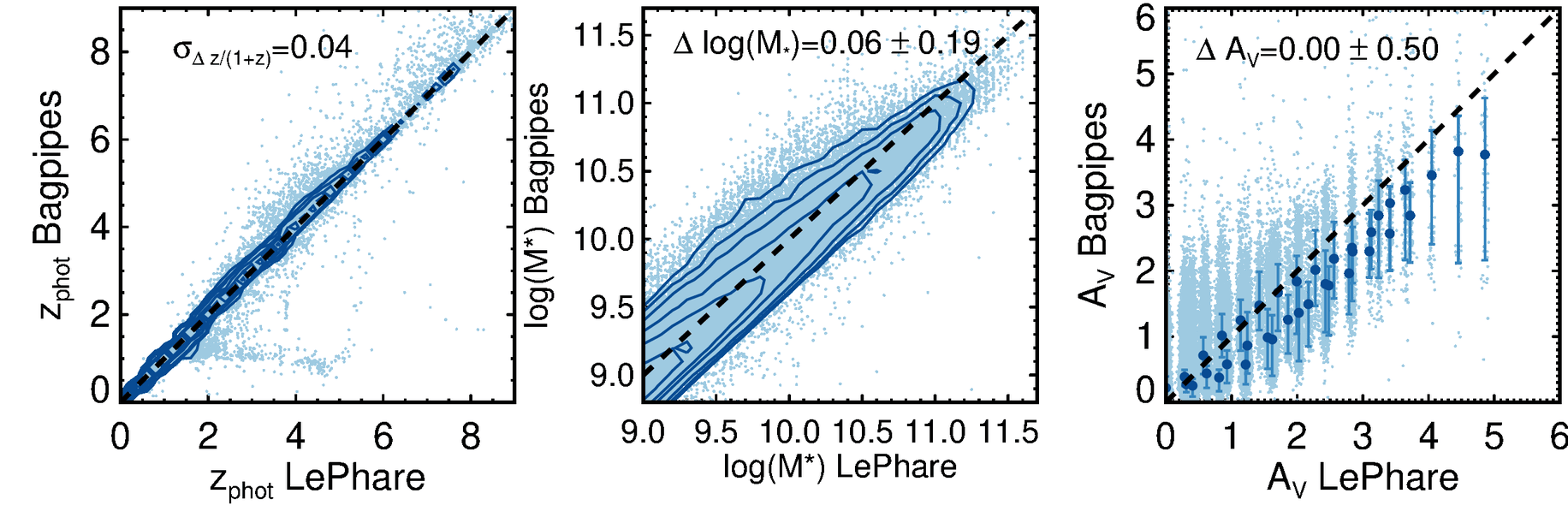}
  \caption{A comparison of derived properties from {\sc LePhare}
    \citep{shuntov25a} and our {\sc bagpipes}-derived quantities based
    on the same model-based photometry.  At left, {\sc LePhare}
    photometric redshifts and their uncertainties were used as input
    to {\sc bagpipes}; the resulting agreement is tight with
    $\sigma_{\Delta z/(1+z)}=0.04$.  Middle, the stellar masses agree
    within uncertainties.  At right, the derived \av\ from {\sc
      bagpipes} agrees with the (quantized) values of \av\ inferred
    from more rigid {\sc LePhare} SED fits; for visual clarity, we
    have perturbed {\sc LePhare} \av\ values about their fixed values
    to show the breadth of the underlying population.}
  \label{fig:bagpipes-compare}
\end{figure*}

\subsection{Derivation of Attenuation Characteristics}\label{sec:bagpipes}

The primary purpose of the COSMOS2025 catalog is to measure
photometric redshifts and, second, to measure basic physical
characteristics like rest-frame UV magnitude and stellar mass.
Photometric redshift fitting requires a delicate balance of increased
free parameters dictating the galaxy template SEDs against obtaining a
realistically constrained redshift. The COSMOS2025 approach, more
thoroughly described in \citet{shuntov25a}, uses more free parameters
than the simplest photometric redshift techniques like {\sc EaZY}
\citep{brammer08a}.  This is done to provide the most realistic
estimates on photometric redshift uncertainties; with too few free
parameters, the uncertainty is often under-estimated, and with too
many it will be overestimated.

A crucial piece of photometric redshift fitting is applying an
attenuation law to galaxy templates.  So as not to introduce too many
free parameters into the fits, COSMOS2025 adopts three possible rigid
attenuation laws -- the \citet{calzetti00a} starburst attenuation law,
the \citet{arnouts13a} attenuation law and the \citet{salim18a}
attenuation law -- with a range of $E(B-V)$ spanning 0--1.2 in bins of
0.1 magnitudes.  This roughly translates to magnitudes of
attenuation $0<A_{\rm V}<5$, with a majority of sources ($\sim$99\%)
fit to $A_{\rm V}<2.5$.  Most sources (51\%) are fit to the
\citet{calzetti00a} attenuation law, while the remainder are evenly
split between the \citet{arnouts13a} and \citet{salim18a} laws.  This
range of choices -- quantized choices of $E(B-V)$ and attenuation law
-- is well suited for the broad application of fitting photometric
redshifts.  However, in this work we are digging deeper into the
attenuation characteristics of galaxies as measured from the UV/optical and
to do that, we need some more flexible, direct fits to the primary
parameters governing attenuation.

To constrain sources' attenuation characteristics, we fit detailed
SEDs to all COSMOS-Web galaxies' fiducial model-based photometry using {\sc
  bagpipes} \citep{carnall18a}.  We use the BPASS stellar model grids
\citep{eldridge09a}.  A Gaussian prior is placed on redshift centered
on the median redshift from {\sc LePhare} ({\tt zpdf\_med}) with
standard deviation set to the maximum value of the upper or lower
68$^{th}$ percentiles with hard limits at $\pm$3$\sigma$ (or $0<z<20$,
whichever is more restrictive).  A non-parametric star formation
history with a continuity prior is used (a $t$-distribution with
degrees of freedom $\nu$=2 and scale = 0.3); twelve bins are used at
$z<4$, nine bins are used at $4<z<8$, and seven bins are used at
$z>8$.  The four most recent bins have fixed age intervals of 0-10,
10-30, 30-100 and 100-300\,Myrs, and the remainder are spaced
uniformly in $\log({\rm time})$ from $>$300\,Myr to $z=20$.  A uniform
prior on metallicity is placed between 0--2.5$\times$ solar, and
nebular emission is fit with uniform prior between $-4<\log U<-1$.

\begin{figure*}
  \centering
  \includegraphics[width=0.99\columnwidth]{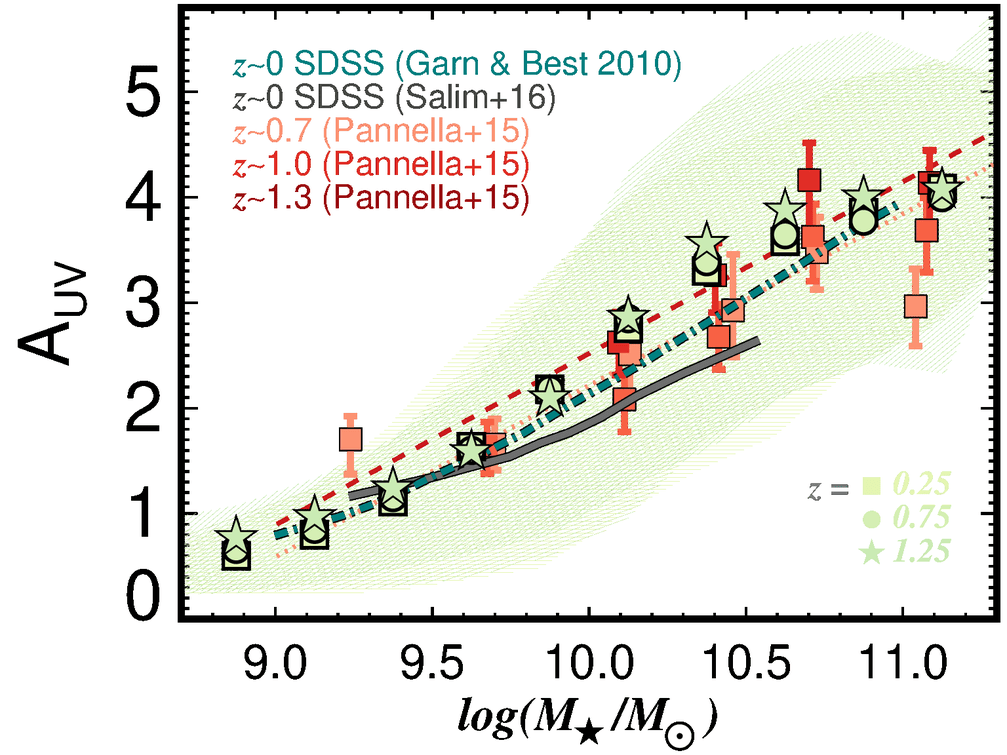}
  \includegraphics[width=0.99\columnwidth]{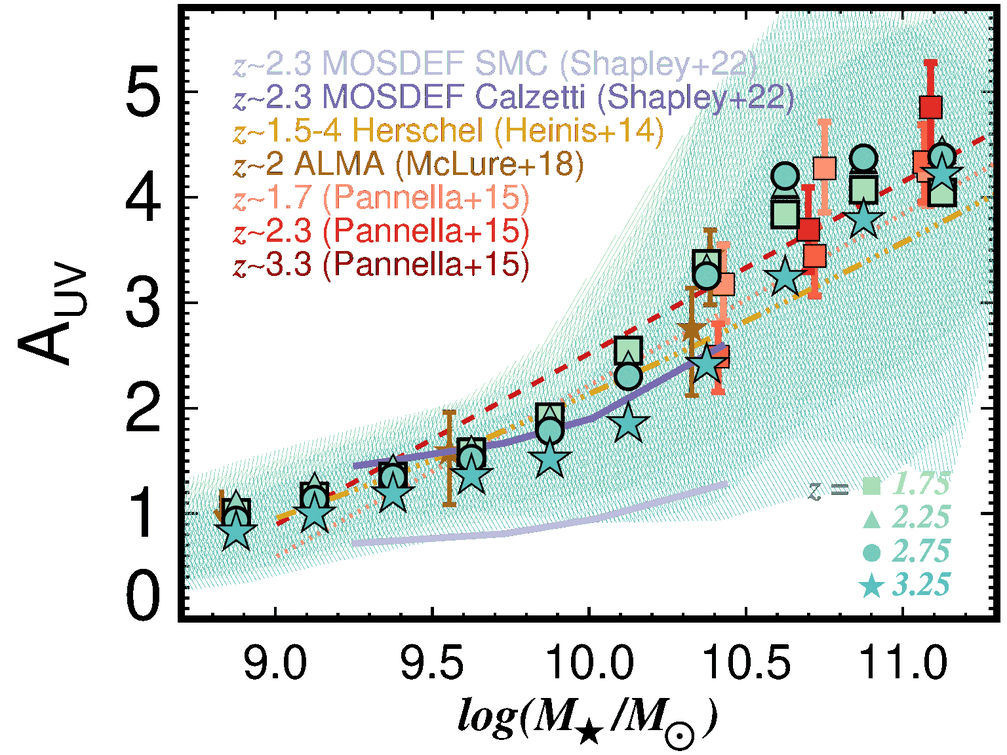}
  \caption{A comparison of our binned \auv-\mstar\ relation (shades of
    green/cyan) compared to literature compilations spanning $0<z<1.5$
    (left) and $1.5<z<3.5$ (right).  Our points represent the median
    \auv\ per mass and redshift bin with shaded regions showing the
    inner 68\%\ confidence interval on the spread of measured \auv in the corresponding
    bin; the errors on the median are of order the size of each data
    point. At $z\sim0$ we compare to \citet{garn10a} and
    \citet{salim16a}; while aligned at low masses, we find higher
    \auv\ around 10$^{10.5}$\,\mstar\ (albeit we are only presenting
    binned \auv\ for the star-forming subset and not the quiescent
    systems).  Our comparisons with {\it Herschel} measurements at
    $z\approx0.7-4$ \citep{heinis14a,pannella15a} and ALMA $z\sim2$
    \citep{mclure18a} are largely consistent within uncertainties.
    When comparing to MOSDEF measurements anchored to the Balmer
    decrement \citep{shapley22a}, here translated to \auv, our
    attenuations are consistent assuming a Calzetti attenuation curve
    (and less consistent for the steeper SMC attenuation curve); we
    discuss trends in the best-fit attenuation curves more in
    \S~\ref{sec:flexiblecurve}.}
  \label{fig:avmass1}
\end{figure*}

A flexible attenuation law is modeled with a \citet{salim18a}
parameterization originally presented in \citet{noll09a}, with uniform
prior on \av\ between $0<$\av$<6$, uniform prior on $\delta$, the
power-law deviation from a Calzetti dust law, between $-0.5<\delta<0.1$
(where $\delta=-0.4$ corresponds roughly to the Small
  Magellanic Cloud, SMC, attenuation curve), and a uniform prior on
the 2175\AA\ bump strength between $0<B<3$.  To reproduce the
attenuation curve for a given $\delta$ and $B$, the Calzetti
attenuation law, $k^\prime_{\rm Cal}(\lambda)$, \citep[specifically
  equation 4 of][]{calzetti00a} is manipulated via equations 3, 4 and
9 in \citet{salim18a}.

We save the posterior median and 68$^{th}$-percentile confidence
intervals on redshift, stellar mass, \av, \auv, $\delta$ (deviation
from Calzetti dust law), $B$ (bump strength), \muv, and $\beta$
(rest-frame UV slope).  \auv\ and \muv\ are evaluated at rest-frame
1600\AA.  We note that, given this is a photometrically constrained
sample with some redshift uncertainty, attenuation characteristics are
uncertain on a source-by-source basis: $\delta$ is only well
constrained for sources with well-sampled rest-frame UV and optical,
while bump strength, $B$ is not well constrained in a majority of
cases; however we still keep it as a free parameter in an effort to
realistically inflate the uncertainties on other attenuation-related
quantities.

We verify that {\sc bagpipes} output is consistent with {\sc
  LePhare}-derived characteristics, as shown in
Figure~\ref{fig:bagpipes-compare}. The posterior on derived {\sc
  bagpipes} redshift differs by $\sigma_{\Delta z/(1+z)}=0.04$, of
order the uncertainty on the photometric redshift, without any
systematic offset.  The derived {\sc bagpipes} stellar masses are
0.06\,dex lower than those from {\sc LePhare}, slightly less than the
average uncertainty on the {\sc LePhare} stellar mass, of 0.07\,dex.
Further, we also drew comparison to the stellar masses derived in
\citet{arango-toro25a} using {\sc cigale} and find similar
consistency.  While the {\sc LePhare} attenuation fits are quantized,
they too are broadly consistent with the attenuation fit in {\sc
  bagpipes} with a slight skew towards higher magnitudes of
attenuation with {\sc bagpipes}.

\section{Results}\label{sec:results}

Here we summarize a number of comparisons between physically derived
galaxy properties.  In \S~\ref{sec:auvmstar} and
\S~\ref{sec:flexiblecurve} we first analyze the results of attenuation
characteristics, derived from UV/optical SEDs alone, {\it independent}
from dust SEDs. We look at the attenuation - stellar mass relation and
the implied evolution of inferred attenuation characteristics.  Next
we analyze galaxies' FIR SED characteristics, measured {\it
  independently} from UV/optical SEDs.  In \S~\ref{sec:lirmdust} we
show the evolution of \lir\ and \mdust, and in \S~\ref{sec:tdust} we
show the evolution of dust temperatures.  \S~\ref{sec:alphabeta}
summarizes the measured mid-infrared slope and emissivity spectral
index from the dust SED. Finally, we measure quantities that draw both
on the results of UV/optical SED fitting and FIR SED fitting; in
\S~\ref{sec:funobsmstar} we analyze the fraction of star formation
that is unobscured as a function of stellar mass and redshift, and in
\S~\ref{sec:DTS} we present the mass and redshift dependence of
galaxies' dust-to-stellar (DTS) ratios.  Throughout this paper we use
a uniform color scheme to show different mass bins evolving with
redshift (drawn from the magma color palette) and redshift bins as a
function of stellar mass (viridis color palette).

\subsection{Attenuation - Mass Relation}\label{sec:auvmstar}

The relationship between stellar mass and globally-averaged
attenuation has been established in the literature
\citep{devour16a,bogdanoska20a,qin24a}.  The relation exists between
stellar mass and \auv\ as well as \av, where the difference is driven
by the underlying nature of the attenuation curve
(\auv$\approx$2.5\av\ for a Calzetti attenuation curve).  We first
anchor our measurements to other literature measurements of
\auv-\mstar\ and proceed to interpret trends we measure across the
full range of mass and redshift accessible in our analysis.  We
emphasize that we are discussing the effective or observed attenuation
here, derived from constraints in the UV/optical only.

Figure~\ref{fig:avmass1} presents a comparison of our derived $A_{\rm
  UV}-M_\star$ relation against literature measurements from
9$<\log(M_\star/M_\odot)<11$ and $0<z<3$.  Plotted are the median
values of \auv\ and 68\%\ spread on \auv\ values in each bin.  At
intermediate redshifts ($z\sim1-3$), literature measurements are
broadly consistent with our findings within uncertainties; these
literature measurements draw specifically on {\it Herschel}
\citep{heinis14a,pannella15a} and ALMA-observed \citep{mclure18a}
samples of galaxies, and sources with direct Balmer decrement
measurements \citep{shapley22a}.  The primary discrepancy is at the
lowest redshifts ($z\lesssim1$) where we infer systematically higher
median attenuation, \auv$\,\approx3.5-4$ for galaxies with stellar mass
$\sim10^{10.5}$\,\msun\ compared to local measurements of
\auv$\,\approx2.5-3$ in the same mass range \citep{garn10a,salim16a}.
While the median attenuations differ, we note that all literature
measures fall well within the inner 68\%\ confidence interval of
sample measurements. Note that constraints on \auv\ in the lowest
redshift bin, $z=0.25$, in our work will be somewhat more uncertain
than both the literature constraints at similar redshifts or our work
at higher redshifts because we have not explicitly included GALEX
photometric constraints in the COSMOS2025 catalog.  Our measurements
of \auv-\mstar\ suggest little evolution across this mass and redshift
range, and broad consistency with the literature, especially at
$1<z<3$.

\begin{figure*}
  \centering
  \includegraphics[width=0.99\columnwidth]{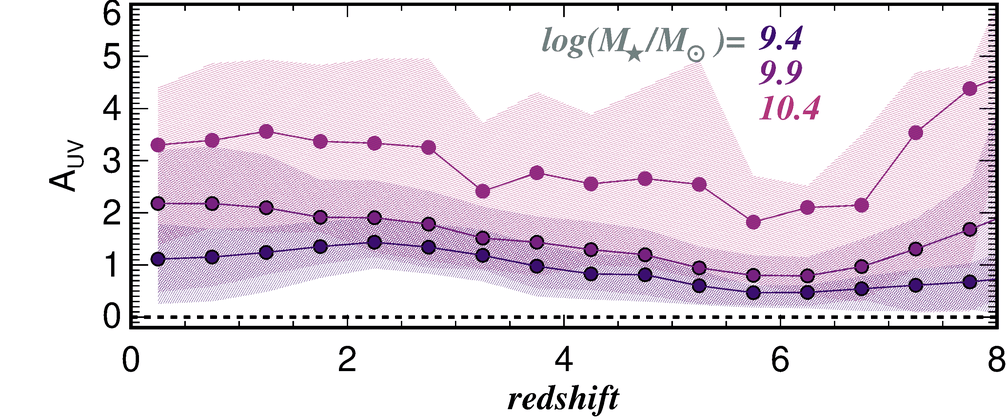}\includegraphics[width=0.99\columnwidth]{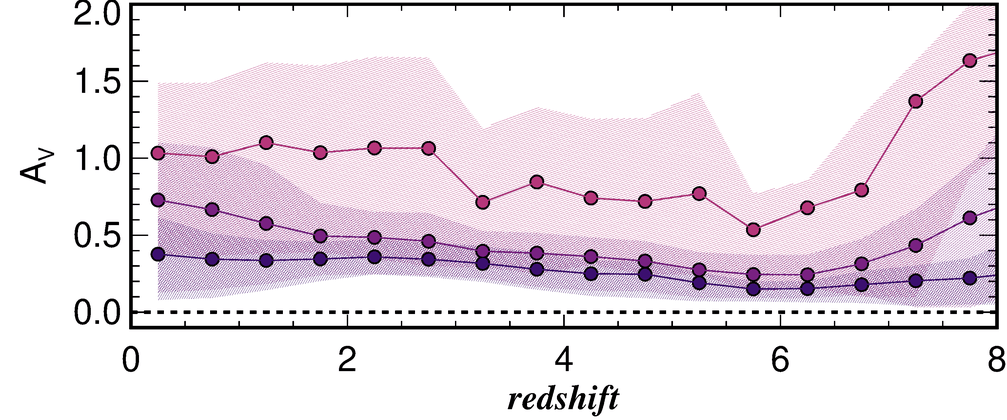}
  \includegraphics[width=0.99\columnwidth]{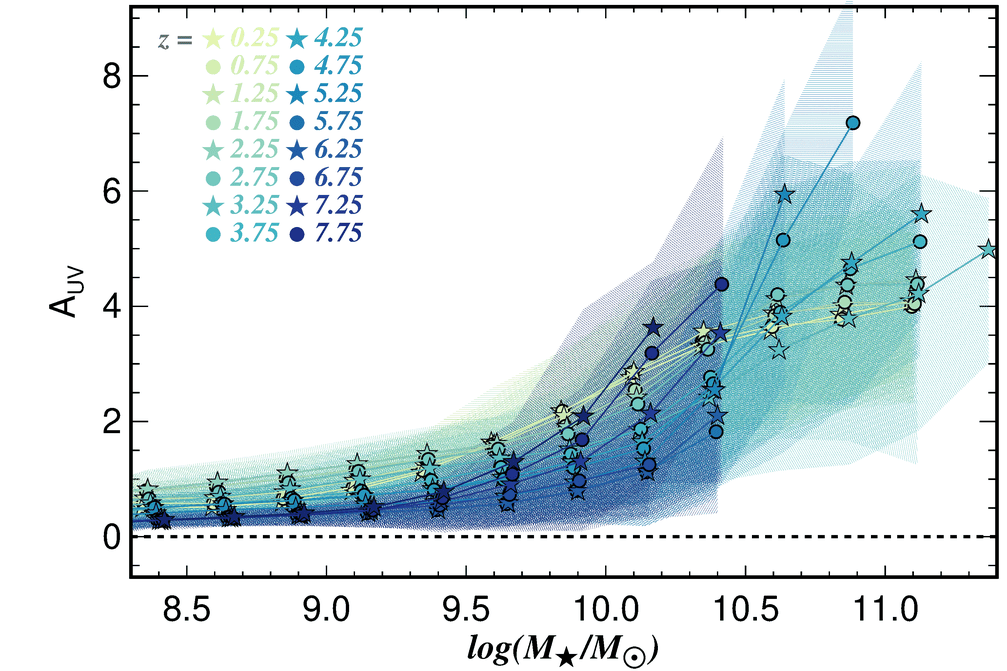}\includegraphics[width=0.99\columnwidth]{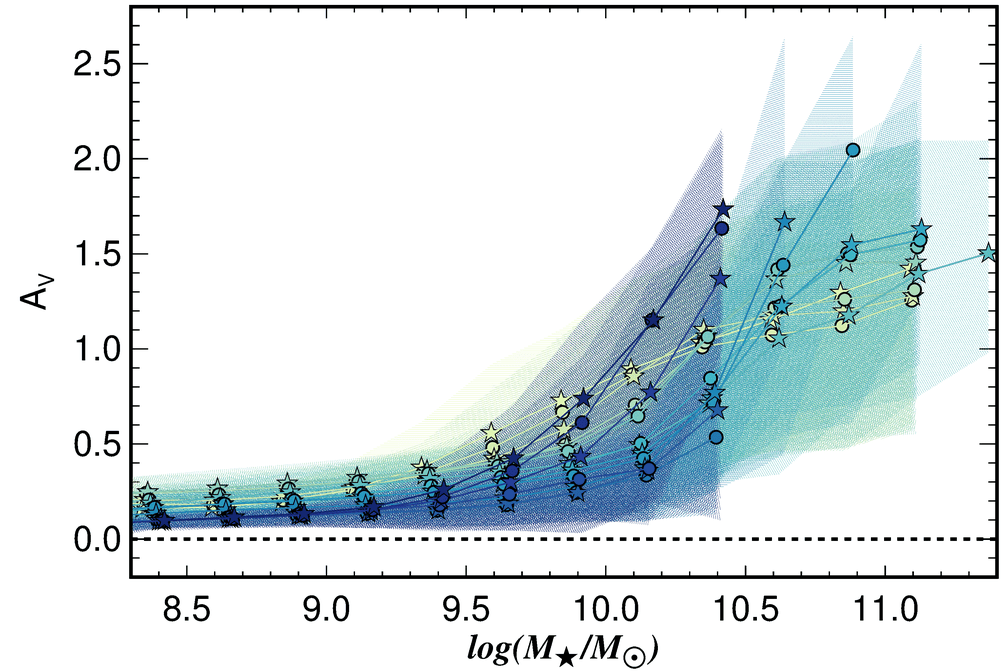}
  \caption{Our mass- and redshift-binned \auv-\mstar\ relation (left)
    and \av-\mstar relation (right) from $0<z<8$.  Note bins shown on
    Figure~\ref{fig:avmass1} at $z<3.5$ are the same as shown here.
    The shaded regions are the 68\%\ confidence interval for each
    binned interval.  Only bins containing more than 50 sources are
    shown.  We see mild evolution in the \auv-\mstar\ (\av-\mstar)
    relation with redshift, made more clear on the top-left and
    top-right panels at three fixed mass bins covered across the full
    range of our data.  Across all masses, we generally see a decrease
    in the average \auv\ (\av) with redshift to $z\sim6$; at higher
    redshifts, there is then an increase.  This inflection point
    coincides with the end of reionization but requires further
    investigation as to its origins. Two figures in the appendix
    (Fig.~\ref{fig:panels_auvmstar} \& \ref{fig:panels_auvz}), show
    the breakdown of this relation against our fitted relation
    (Eq.~\ref{eq:avmstar}) in more detail.}
  \label{fig:avmass2}
\end{figure*}

Given the relatively good agreement, we
present the extended attenuation - stellar mass relationships in fixed
redshift bins out to $z\sim8$ in Figure~\ref{fig:avmass2}.

  While there was little evidence for evolution in \auv-\mstar\ out to
  $z<3$ \citep{shapley22a}, the data in Figure~\ref{fig:avmass2}
  suggest some slight downward evolution in \auv\ at fixed stellar
  mass with increasing redshift out to $z\sim6$.  The same trend is
  seen in \av\ at fixed stellar mass as a function of increasing
  redshift.  The lower \auv\ and \av\ is most striking in the mass
  range $9.5<\log(M_\star/M_\odot)<10.5$ from $0<z<6$.  Higher stellar
  mass bins show little evolutionary trend, suggesting that
  non-quenched massive galaxies tend to be quite obscured regardless
  of redshift.

Both \av-\mstar\ and \auv-\mstar\ resemble logistic growth
functions with a monotonic decline in \auv\ and \av\ at fixed stellar
mass with redshift (out to $z\sim6$).  Thus we fit our binned data to
a simplified model describing the average \av\ expected at a given
stellar mass and redshift:
\begin{equation}
\label{eq:avmstar}
\begin{aligned}
A_V(M,z) &= I(M) + S(M)\,\bigl(1+z\bigr), \\[4pt]
I(M) &= \frac{A_{\max}}{1+\exp\!\left[-k\,(M-M_0)\right]}, \\[4pt]
S(M) &= -\exp\!\left(b_0 + b_1\,M\right).
\end{aligned}
\end{equation}
where $M\equiv\log(M_\star/M_\odot)$.  We similarly fit this same
model to \auv$(M,z)$ and report the derived parameters for both
relations in our appendix, Table~\ref{tab:allrelations}.  We also plot
the fitted relation split into individual redshift panels in the
appendix.

At redshifts higher than $z\sim6$, we see an interesting inflection in
the data, such that \auv\ and \av\ at fixed stellar mass increase with
redshift, where the minimum attenuation is around $z\approx6$.  The
sample scatter on binned \auv\ and \av\ are significant, but the
increase from $z\sim6$ to $z\sim8$ is seen in every mass bin between
$10^{9}-10^{10.5}$\,\msun.  The reason for this reversal is unclear.
$z\sim6$ is the end of reionization, before which the intergalactic
medium (IGM) is more optically thick; though our model fitting with
{\sc Bagpipes} lacks a direct implementation of the IGM damping wing
\citep{miralda-escude98b} or significant nebular 2-photon continuum
contributions \citep{katz24a}, both of which could artificially redden
the spectra of galaxies without dust.  Another possibility is that
galaxies' ISM at earlier times may undergo some type of transition
around $z\sim6$, related to their compact sizes, star/dust geometry,
and/or sources of early dust. We will leave a further investigation of
this phenomenon -- whether physical or an artifact of our SED modeling
-- to a future work.

\begin{figure}
\includegraphics[width=0.99\columnwidth]{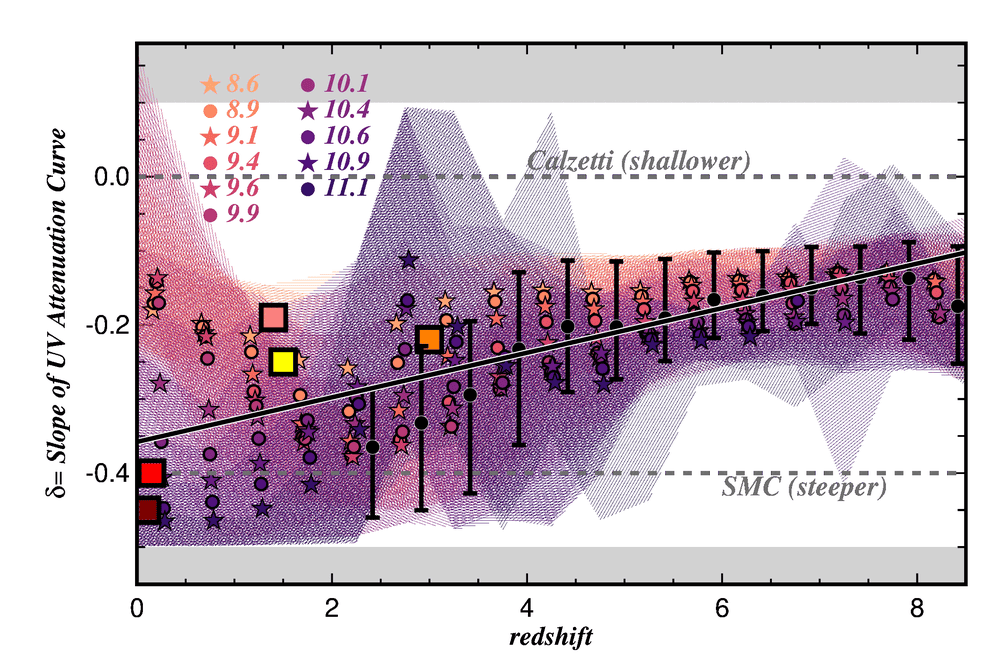}
\caption{The redshift evolution of $\delta$, the parameter describing
  the rest-frame UV slope of the attenuation curve.
  $\delta$\ describes the deviation from a Calzetti attenuation curve,
  where $\delta$=0 recovers Calzetti, and a negative
  $\delta$\ produces a steeper attenuation curve.  Gray regions denote
  the bounds of modeled $\delta$ allowed in our fits, ranging from
  $-0.5<\delta<0.1$. The SMC attenuation curve roughly follows
  $\delta$=-0.4, which is consistent with the average attenuation
  curve recovered by \citet[][red square]{salim18a} and \citet[][dark
    red square]{conroy10a}.  Higher-$z$ literature measurements come
  from \citet{buat12a} (yellow square), \citet[][pink
    square]{kriek13a} and \citet[][orange square]{reddy16a}. We show
  the evolution of $\delta$\ binned by stellar mass ($\Delta\log
  M=0.25$ bins) with colored points and stars, and the average
  $\delta$ across all masses as open black circles between $2<z<8$
  (shaded regions show the inner 68\%\ confidence
  interval of all measured $\delta$ per mass bin).  The best-fit
  relationship (black line) is presented in the text in
  Eq~\ref{eq:z_delta}.}
\label{fig:z_delta}
\end{figure}

\subsection{Evolution in the Attenuation Curve}\label{sec:flexiblecurve}

\citet*{salim20a} provide an overview of recent (pre-JWST) results
pertaining to the derivation of the attenuation curve, both for low
redshift and high redshift samples.  To briefly summarize the
consensus view, the literature clearly demonstrates that galaxies'
attenuation curves are diverse; there is significant evidence that
some star-forming galaxies in the local volume have attenuation curves
similar to (if not steeper than) the SMC
\citep{conroy10a,leja17a,salim18a} and significantly steeper than the
Calzetti attenuation law \citep[][see also
  \citealt{battisti16a}]{calzetti00a}, though it is perhaps stellar
mass and metallicity dependent in the local volume.  Similarly, work
beyond $z>0.5$ from the pre-JWST era suggests steeper-than-Calzetti
attenuation laws may be common, but not quite as steep as work on
local star-forming galaxies.

Figure~\ref{fig:z_delta} highlights our primary finding, that
$\delta$, the deviation of the attenuation law slope from Calzetti,
averaged over mass-matched bins shows redshift dependence from $z=2$
to $z=8$, with especially pronounced evolution seen over $2<z<4$.  We
do not find that $\delta$ has a significant stellar mass dependence
though slight mass variance is seen in Figure~\ref{fig:z_delta} at
fixed redshift, but we caution that lower mass sources are fainter and
their attenuation curve slopes are less well constrained
\citep[cf.][]{vijayan24a}.
Specifically we find that the average attenuation law 
steepens with decreasing redshift such that:
\begin{equation}
  \langle\delta(z)\rangle=(-0.36\pm0.07) + (0.030\pm0.014) z
  \label{eq:z_delta}
\end{equation}
This is explicitly fit to all sources above stellar masses
$10^{9}$\,\msun\ and $2<z<8$ where $\delta$ is best constrained.  Our
results, though of marginal significance (the redshift evolution is
$\sim$2.1$\sigma$ significant) are consistent with the prior
measurements of the attenuation law slope from both $z<0.5$
\citep{conroy10a,salim18a} and $z\sim1-3$ samples
\citep{buat12a,kriek13a,reddy16a}, though our data show significant
scatter at $z<2$.  We attribute the scatter at low-$z$ to a limitation
of our data: specifically, the GALEX data was not included in the
COSMOS2025 catalog, and thus measurement of our low-$z$ attenuation
curves are done from extrapolation from the $u$-band blueward.  At
$z>2$ our photometric data cover the entirety of the rest-frame UV,
and correspondingly, the distribution in measured slopes narrows.

We note that our results here are complementary though less
constrained than spectroscopic analyses
\citep{markov25a,markov25b,shivaei25a}. We also note that constraints
from Balmer decrements \citep{calzetti00a,reddy15a,battisti22a} can
differ from SED-fitting methods and favor a steepening of attenuation
law with increasing redshift.

\begin{figure*}
\centering
\includegraphics[width=0.99\columnwidth]{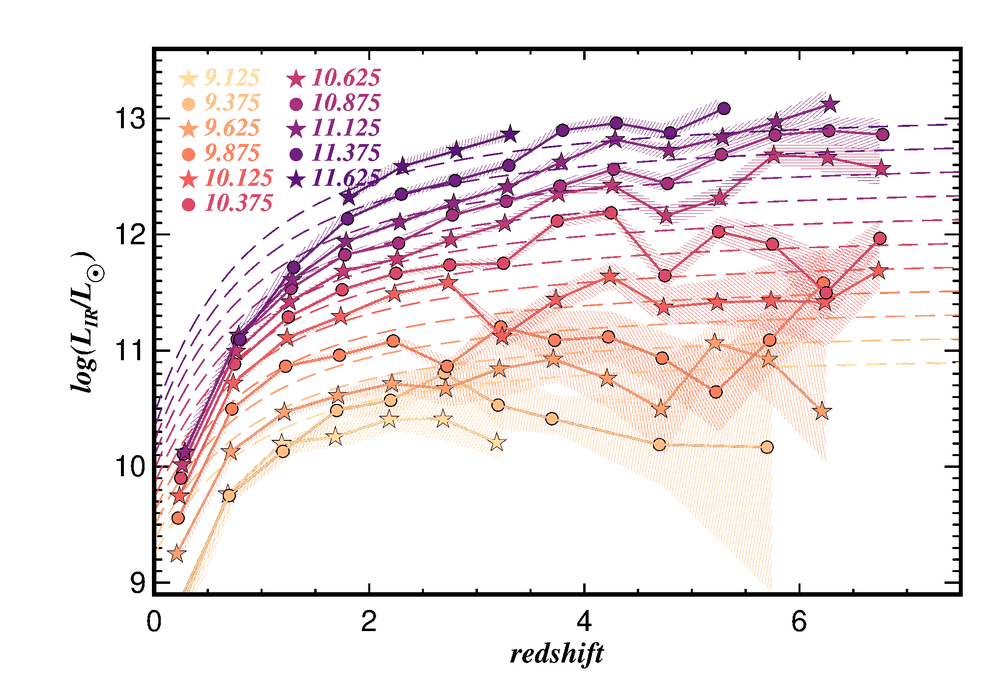}
\includegraphics[width=0.99\columnwidth]{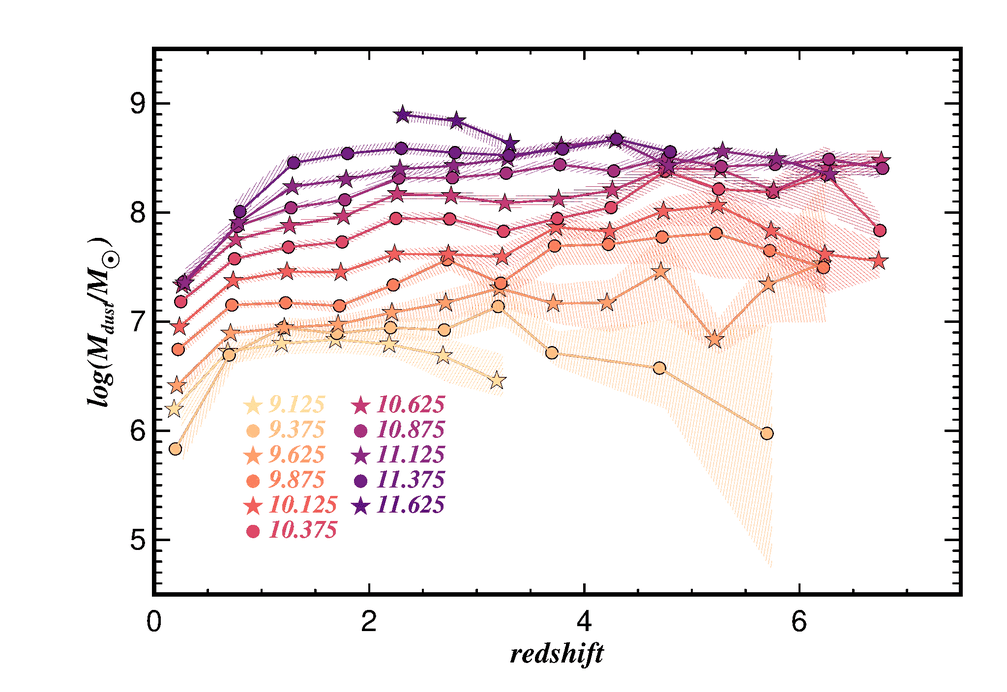}
\caption{The redshift evolution of \lir\ (left) and \mdust\ (right) 
  in fixed stellar mass bins.  Dashed lines at left denote the total
  IR luminosity that would be expected if 100\%\ of star-formation
  were obscured, and the SFR is taken from the main-sequence of star
  formation, or the \mstar-SFR relation \citep{speagle14a}.  Between
  $0<z<2$, we see a 100$\times$ increase in the average
  \lir\ and 5-10$\times$ increase in \mdust\ per mass bin above
  $10^{10}$\,\msun.  Beyond $z\sim2$, evolution is more subtle; at
  high \mstar, there is a factor of 10$\times$ higher \lir\ at
  $z\sim6$ than at $z\sim2$, while masses below
  $10^{10.5}$\,\msun\ see no evolution or a dropoff in \lir.  Dust
  masses in the highest mass galaxies show no evolution, while there
  is mild evidence suggesting a threefold increase in dust masses for
  intermediate masses.}
\label{fig:lirmdustz}
\end{figure*}

\subsection{\lir\ and \mdust}\label{sec:lirmdust}

Pivoting in focus to the FIR SEDs, we present the derived IR
luminosities and dust masses of our stacked SEDs in
Figure~\ref{fig:lirmdustz} split into stellar mass bins to highlight
their redshift evolution.  We observe a few general trends in the
SEDs: at $z<1$, IR luminosities are exceedingly low, with median
\lir$\lesssim$10$^{10}$\,\lsun\ even for galaxies around masses
$\sim10^{11.5}$\,\msun.  IR luminosities are a factor of
$\sim$100$\times$ brighter around $z\sim2$, the peak of the cosmic
star formation rate density, across all mass bins for which we have
measured SEDs.  This sharp evolution in galaxies' IR luminosities is
well documented in the literature going back to some of the first
measurements of the IR luminosity function from {\it Spitzer} and {\it
  Herschel} and implications for the star-formation rate density
\citep[e.g.][]{le-floch05a,caputi07a,casey12b,gruppioni13a}.  Two
phenomena conspire to produce the great abundance of IR-luminous
galaxies at $z\sim2$: the evolution of the galaxy SFR-\mstar\ relation
such that galaxies' SFRs at $z\sim2$ are much higher likely driven by
increased gas surface densities at $z\sim2$, and in parallel, the
observed star formation is dominated by dust-reradiated emission and
not direct UV light at high stellar masses, above
\mstar$>$10$^{9.5}$\,\msun, which we will discuss more in
\S~\ref{sec:funobsmstar}.

\begin{figure*}
\centering
\includegraphics[width=1.6\columnwidth]{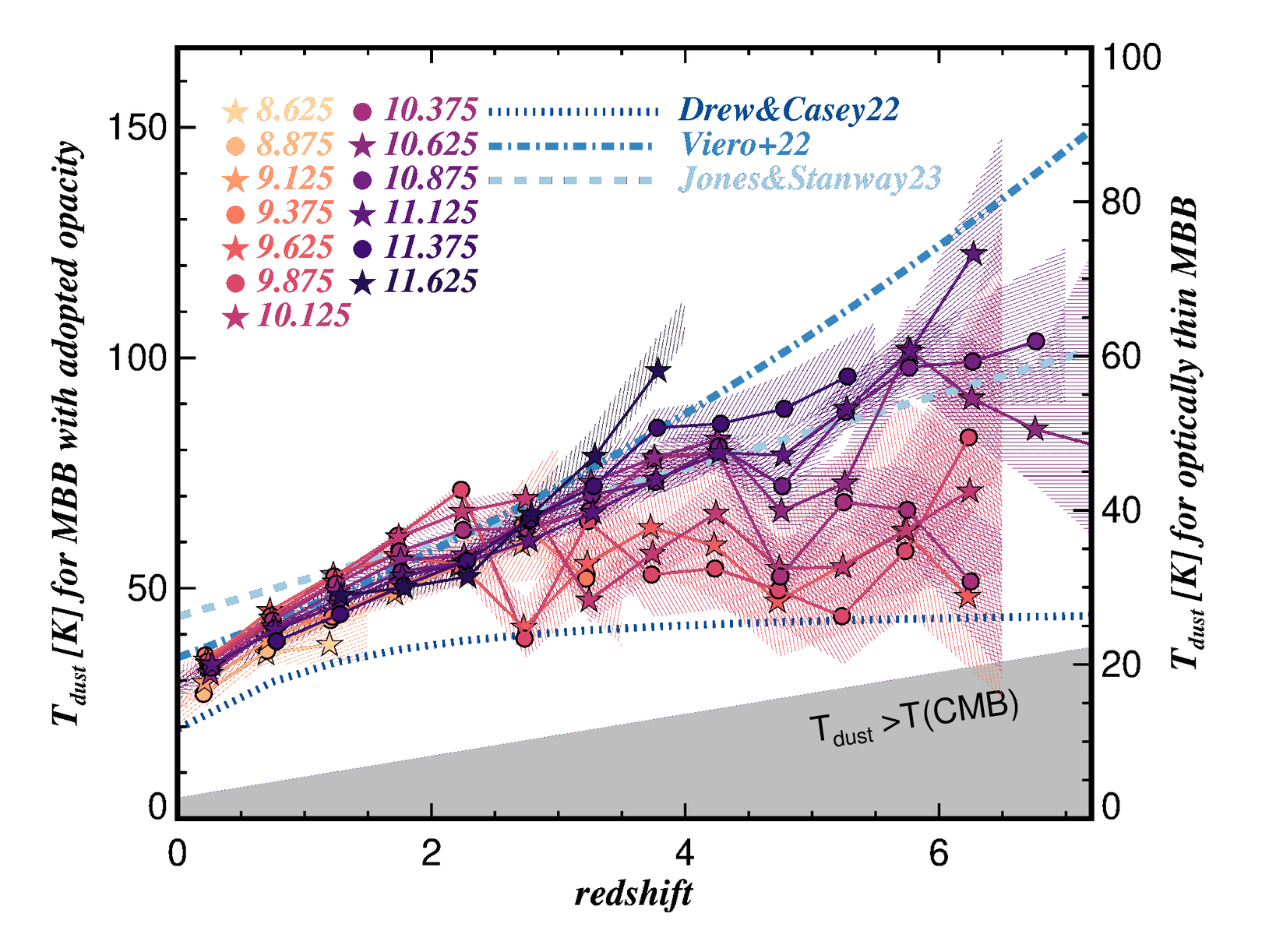}
\caption{The redshift evolution of dust temperature derived for
  stacked SEDs in fixed stellar mass bins ranging
  $8.6<\log(M_\star/M_\odot)<11.6$.  The floor on dust temperature is
  set by the temperature of the CMB which evolves as $T_{\rm
    CMB}=2.73(1+z)$ (gray filled region). Three literature relations
  are overplotted.  From \citet{drew22a} we show the expectation of
  the evolution of \tdust\ at fixed stellar mass given the evolution
  in the SFR-\mstar\ relation, where SFR scales to \lir, and
  \tdust\ has a direct relationship with \lir; the
  \citeauthor{drew22a} work was calibrated against $0<z<2$
  samples. \citet{viero22a} uses {\it Herschel} data to infer
  \tdust\ for stacked samples of galaxies drawn from COSMOS2020 and
  find a steep \tdust($z$) dependence. Last we show the empirical
  relation presented in \citet{jones23a} which uses a meta-analysis to
  address the \tdust($z$) dependence observed in various literature
  datasets.  Our data are shown from low-mass bins (yellow) through
  high-mass bins (dark blue) superimposed.  Below masses
  $\log(M_\star)\sim$10.4, we see relatively little evolution in
  \tdust\ beyond $z\sim2$ while higher mass bins show a marked
  increase in \tdust\ from $z\sim2$ to
  $z\sim6$. Figure~\ref{fig:panels_llp} in the Appendix show this
  evolution in more detail against the derived relation given in
  Eq.~\ref{eq:tdustfit}.}
\label{fig:tdust}
\end{figure*}

At $z>2$, extending out to $z\sim7$ (the limit of our stacked SEDs),
IR luminosities at fixed stellar masses appear flat to 0$^{th}$ order.
However, a more detailed look sees some higher order variation.  At
low stellar masses $<10^{9.5}$\,\msun, IR luminosities at $z>4$ are
markedly lower than they are at $z\sim2$ where they peak (though not
as low as at $z\sim0$).  Galaxies of intermediate masses,
$9.5\lesssim\log(M_\star/M_\odot)\lesssim10.5$, do appear to have non-evolving
IR luminosities.  Galaxies in the highest mass bins,
$>10^{10.5}$\,\msun, appear to have {\it higher} IR luminosities at
$z\sim4-6$ than at $z\sim2$.  Note that, if galaxies' IR luminosities
were to track exactly with their SFRs as given by the evolution of
SFR-\mstar\ \citep[using the parameterization of][]{speagle14a}, they
would rise by 0.3\,dex from $z\sim2$ to $z\sim4$ and then rise another
0.3\,dex out to $z\sim8$, following the dashed lines on the left panel
of Figure~\ref{fig:lirmdustz}.  

Dust masses tell a slightly different narrative.  The caveat of the
dust mass here is that the dominant source of uncertainty is
$\kappa_{\rm FIR}$, and here we apply a fixed value and assert no
redshift evolution, a topic we return to in the discussion.  With
fixed $\kappa_{\rm FIR}$ in mind, galaxies at $z\sim0$ have less dust
than at $z\sim2$.  The drop in the recent universe is less severe than
galaxies' IR luminosities, about a factor of 5-10$\times$.  Beyond
$z>2$, the highest stellar mass systems have a non-evolving average
dust mass.  At intermediate masses, dust masses at $z\sim5-6$ are
slightly larger than at $z\sim2$ (by 3-5$\times$) and lower beyond
$z>6$, though $z>6$ constraints are limited to two mass bins and
increasingly uncertain SEDs.  At lower stellar masses there may be a
significant inflection, such that galaxies with
$<10^{9.5}$\,\msun\ may have lower dust masses at earlier epochs than
they do at $z\sim2$.

We will return to a discussion of these trends in \lir\ and \mdust\ in
the context of potential predictive power from UV/optical-derived
characteristics in the discussion, in particular
\S~\ref{sec:auvcompare}.

\subsection{Evolution of Dust Temperature}\label{sec:tdust}

Galaxies' dust temperatures have been shown in a number of works to
evolve with redshift
\citep{bethermin15a,schreiber18a,bouwens20a,viero22a,drew22a,jones23a}.
Most commonly this evolution is traced with galaxies directly detected
in the IR, for the obvious reason that those are the sources for which
it can be measured, though recent stacking results also
  show increasing SED dust temperatures to $z\sim3$
  \citep{euclid-collaboration26a}.  In all works, the actual
observable is the rest-frame peak wavelength of the dust SED,
$\lambda_{\rm peak}$, but as discussed in \S~\ref{sec:irsed}, the
mapping to a dust temperature is opacity-model dependent, so works
that assume different IR SED methodology do not have directly
comparable dust temperatures.  In this work we explicitly constrain
$\lambda_{\rm peak}$ and convert that to a representative dust
temperature using:
\begin{equation}
\log(T_{\rm dust}/K)\approx 3.756 - 1.048 \log(\lambda_{\rm peak}/\mu\!m)
\label{eq:tdustlpeak}
\end{equation}
which falls out directly based on the presumption of an optically thick SED short-ward of $\lambda_0\sim$100\,\um\ as described in
\S~\ref{sec:irsed}; this equation is also explicitly described in
\citet{burnham21a}. We report dust temperatures from elsewhere in the
literature at face value as they are reported in those works.
Overall, existing studies have concluded that galaxies' dust SEDs 
increase in temperature with increasing redshift.

\begin{figure}
  \includegraphics[width=0.99\columnwidth]{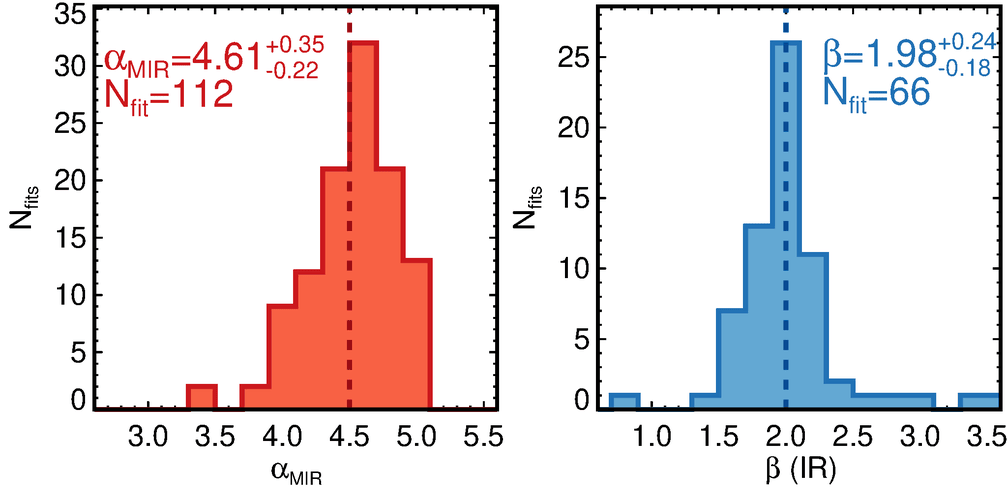}
  \caption{The distribution in the mid-infrared slope, $\alpha_{\rm
      MIR}$, and emissivity spectral index, $\beta$, for
    redshift-stellar mass bins where sufficient photometric
    constraints allow for direct fitting.  The vertical lines mark the
    adopted values for stacked SED bins where direct fits of
    $\alpha_{\rm MIR}$ and $\beta$ are not possible.}
\label{fig:alphabeta}
\end{figure}

Figure~\ref{fig:tdust} shows the evolution in measured dust
temperatures of our stacks, binned by stellar mass out to $z\sim7$.
We note a few key trends: from $0<z<2$, dust temperatures for all mass
bins evolve similarly, with the hottest SEDs at higher-$z$, ranging
between 50--70\,K (25--40\,K) at $z\sim2$ for the adopted opacity
model (for optically thin SEDs). In other words, at a given redshift
$z\lesssim$2, \tdust\ appears to have {\it no} stellar mass
dependence.  At higher redshifts, the lower mass bins
$<10^{10}$\,\msun\ have constant temperatures out to $z\sim6$, hovering
around 50\,K (30\,K for optically thin), while galaxies of higher
stellar masses continue to evolve upward, reaching average
temperatures of $\sim$90\,K (50\,K, optically thin) at $z\sim5$ above
$>10^{10.5}$\,\msun.

Three curves from the literature are overplotted on
Figure~\ref{fig:tdust}.  \citet*{drew22a} is a comprehensive analysis
of galaxies' dust temperatures at $z<2$ that deduces a fundamental
relationship between \lir\ and \tdust, such that more luminous
galaxies are hotter. They explore the possible redshift evolution of
\tdust\ due to galaxies' increasing IR luminosities at higher
redshifts, and that relation is shown in Figure~\ref{fig:tdust}; its
evolution marks the change in SFR of a main sequence galaxy, which
sees the most evolution from $0<z<2$ and evolves shallowly at $z>2$.
We also overplot the relation derived in \citet{viero22a} which drew
from stacked IR SEDs out to $z\sim10$ using COSMOS2020;
\citeauthor{viero22a} fit a polynomial to their \tdust($z$) data
finding a steep evolution.  Many of their higher mass, higher redshift
bins are somewhat overpopulated; in other words, the COSMOS2020
catalog, at face value, has many galaxies at high stellar masses and
high redshifts ($z>7$), some masses unphysically large.  These bins in
the \citeauthor{viero22a} analysis may skew towards unusually hot
temperatures either from contamination from low-$z$ sources (where
redshift and \tdust\ have a degeneracy, such that low-$z$ `cold' SEDs
could be mistaken as high-$z$ `hot' SEDs) or galaxies with especially
strong AGN boosting their stellar masses above a reasonable threshold
(and an AGN can also serve to heat ISM dust).

The last literature
relationship we show is the meta-analysis presented in
\citet*{jones23a}, that empirically draw a linear increase in dust
temperatures from $0<z<8$, which is between the extremes set by
\citet*{drew22a} and \citet{viero22a} works.  The \citet*{jones23a}
fits the evolution we see in \tdust($z$) fairly well, especially for
stellar masses $\sim10^{11}$\,\msun.  While \citeauthor{jones23a} does
not explicitly address the stellar mass range of their analysis, we
can infer they are likely high mass systems from the existence of
direct dust detections.
We note that the \tdust($z$) trends recovered here are properties of a
stellar-mass selected sample. For luminosity-limited samples of
870\um-selected galaxies, \citet{dudzeviciute20a} find no evolution in
dust temperature at fixed \lir\ across $z\approx1.5-4$, indicating
that much of the apparent temperature evolution has a complex
relationship with the SFR-\mstar\ relation and galaxies' size-mass
relation, which we elaborate on in \S~\ref{sec:tdustdiscuss}.

We directly fit all of our \lpeak\ data to a relation of the form:
\begin{equation}
  \lambda_{\rm peak}(z,M) = a + b(M-10) + c\log(1+z)
  \label{eq:tdustfit}
\end{equation}
where $M\equiv\log(M/M_\odot)$.  We infer coefficients of
$a=2.213\pm0.012$, $b=-0.035\pm0.007$, and $c=-0.53\pm0.02$.  This
relation, as well as the \llp-\tdust\ relationship, is given in our
appendix, Table~\ref{tab:allrelations}.  We discuss the physical
driver of dust temperature and its relation to other measured
quantities more in the discussion, in \S~\ref{sec:tdustdiscuss}.

\begin{figure*}
  \centering
  \includegraphics[width=1.99\columnwidth]{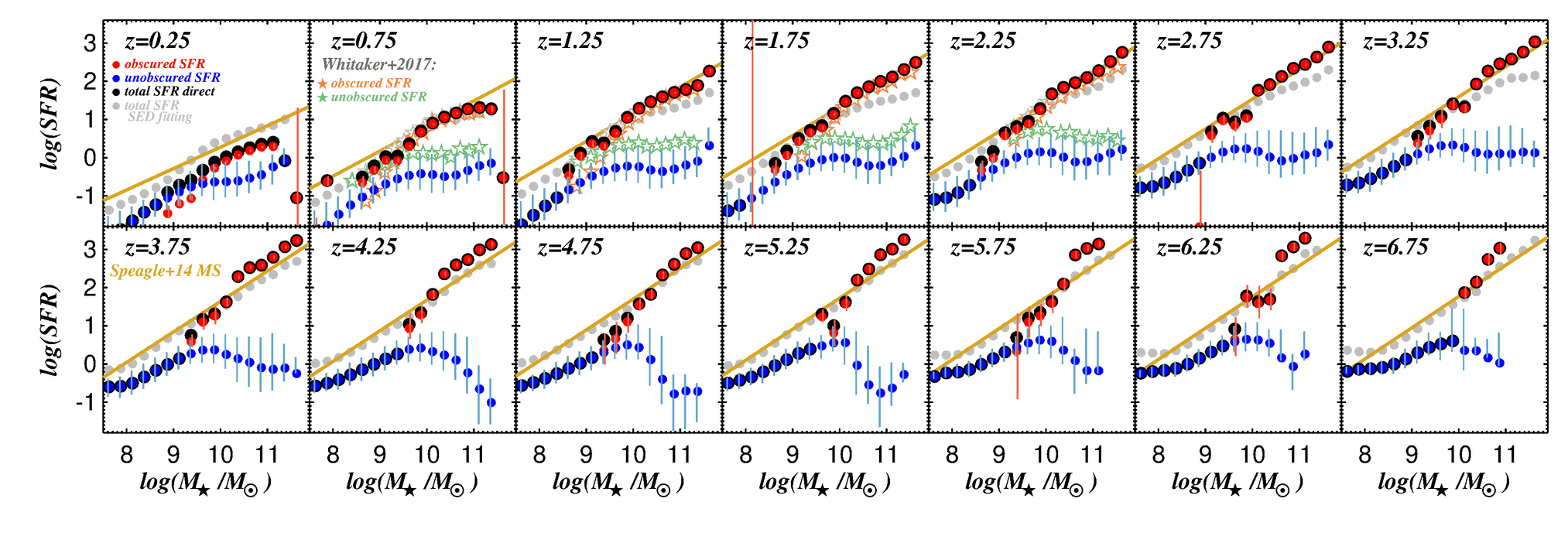}
  \caption{The star-formation rate, stellar mass relation (SFR-\mstar)
    shown in $dz=0.5$ bins broken down into the UV (blue) and IR (red)
    components.  The coaddition of the UV and IR together are shown as
    black points, while the average SFR taken directly from the
    COSMOS2025 catalog LePhare SED fits are shown as gray points.  In
    the four panels $0.5<z<2.5$ we overplot the measurements of
    \citet{whitaker17a} which used galaxies from the 3D-HST survey as
    well as 24\um-extrapolated IR constraints on SFR.  There is a
    significant discrepancy (as much as an order of magnitude) in
    total SFR reported in the COSMOS2025 catalog for massive galaxies
    ($>10^{10}$\,\msun) between $0<z<2.5$ and those derived via direct
    coaddition of UV and IR components.  }
  \label{fig:sfms}
\end{figure*}
\begin{figure*}
\centering
\includegraphics[width=0.99\columnwidth]{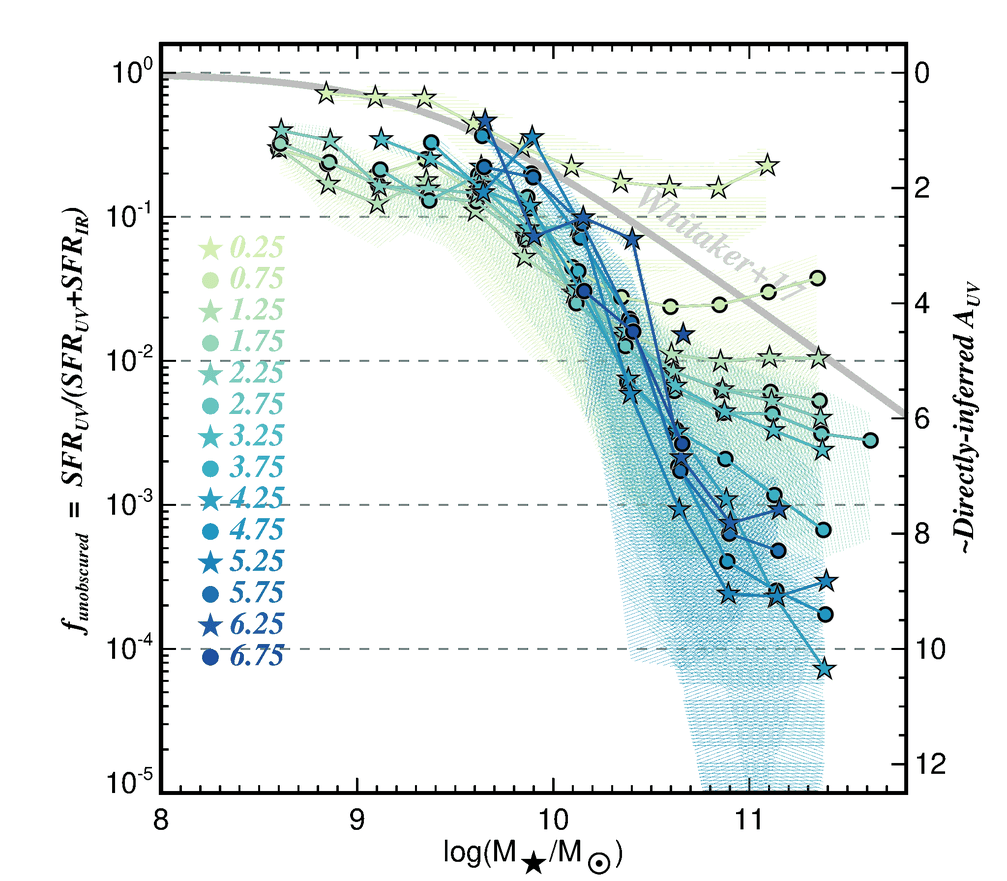}
\includegraphics[width=0.99\columnwidth]{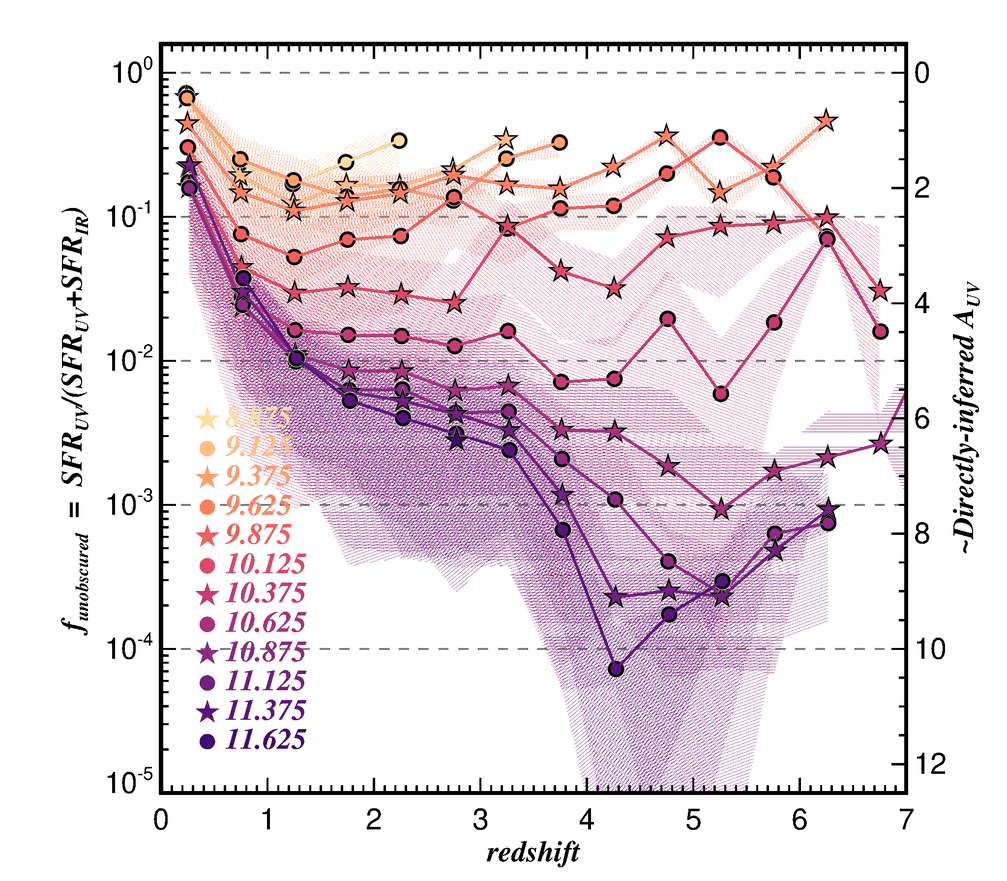}
\caption{The stellar mass and redshift dependence of \funobs.  Broadly
  consistent with prior findings at $z<3$ from \citet{whitaker17a}, we
  find a very strong stellar mass dependence in \funobs, such that
  $>$90\%\ of star formation is obscured above a stellar mass of
  $10^{10}$\,\msun\ at all redshifts beyond $z\sim0.5$.  The
  diminishingly low fraction of unobscured star formation that is seen
  at high masses is even smaller at $z\sim4-6$ than it is at $z\sim2$.
  Masses below $\sim10^{10}$\,\msun\ show relatively little evolution
  in \funobs\ between $2<z<6$, with slight indication of {\it higher}
  \funobs\ at $z>4$ compared to $z\sim2$.  Overall, we find that more
  star formation is obscured relative to \citet{whitaker17a} at fixed
  stellar mass and we attribute the difference to improved constraints
  on the IR SED in this work. Figure~\ref{fig:panels_funobsmstar} and
  \ref{fig:panels_funobsz} show a more detailed breakdown of this
  figure in redshift and stellar mass against the empirical fit
  described in Eq~\ref{eq:funobs} as well as simulations expectations
  from \citet{zimmerman24a}. }
\label{fig:funobs}
\end{figure*}

\subsection{Mid-IR Slope and Emissivity Spectral Index}\label{sec:alphabeta}

Two parameters of the IR SEDs that are less well constrained than
\lir\ and \tdust\ are the mid-infrared power law slope, $\alpha_{\rm
  MIR}$, and the emissivity spectral index, $\beta$.  $\alpha_{\rm
  MIR}$ reflects the slope of the dust emission on the Wien side of
the dust blackbody and $\beta$ reflects the emission on the
Rayleigh-Jeans tail of the blackbody.  Only when sufficient SED
constraints are on-hand do we explicitly fit for one or both.
Figure~\ref{fig:alphabeta} show the distribution of both for stacked
bins where the parameter is fit.  Within the range of available data,
we see no clear evolutionary or mass-dependent trend.  The median of
both distributions, $\alpha_{\rm MIR}=4.61^{+0.35}_{-0.22}$ and
$\beta=1.98^{+0.24}_{-0.18}$, motivate the choice of the fixed values
$\alpha_{\rm MIR}\equiv4.5$ and $\beta\equiv2$ for SEDs without
sufficient SNR on either end of the blackbody peak.

\subsection{\funobs - Mass Relation}\label{sec:funobsmstar}

The relationship between the unobscured fraction of star formation and
stellar mass gives some intuitive sense as to the dominant sources of
both UV and IR emission and is complementary to \auv($M$,$z$)
discussed in \S~\ref{sec:auvmstar}, though most closely a probe of the
`direct' attenuation as given in Eq~\ref{eq:auvdirect}.  The
unobscured fraction of star formation, \funobs, is equal to the
star-formation rate probed in the UV over the total (i.e. $f_{\rm
  unobs}=SFR_{\rm UV}/(SFR_{\rm IR} + SFR_{\rm UV})$).  Conversely,
the obscured fraction, $f_{\rm obs}\equiv SFR_{\rm IR}/(SFR_{\rm IR} +
SFR_{\rm UV})$, is $=1-f_{\rm unobs}$.  For the remainder of this
work, we focus on \funobs\ rather than \fobs\ given the dynamic range
of \funobs\ seen in the data.  Measurements of galaxies' star
formation rates from SED fitting should theoretically be sensitive to
SFR$_{\rm total}$, accounting for both unobscured and obscured
components.  However, they typically do this by extrapolating the
measurement of dust from \av\ with a fixed set of IR templates, and in
some cases implements a full energy-balance technique to fit out to
submillimeter wavelengths \citep{da-cunha15a,boquien19a}.  The
COSMOS2025 catalog does not fit the IR portion of the SED for lack of
constraints for the vast majority of sources which are not directly
detected at long wavelengths.  For the purposes of this paper
where we are intentionally decoupling the two, our {\sc Bagpipes}
fitting from \S~\ref{sec:bagpipes} also does not fit the IR.

To calculate \funobs\ for our mass and $z$ bins we need both SFR$_{\rm
  UV}$ and SFR$_{\rm IR}$. First we remeasure rest-frame UV magnitudes
for all individual sources from {\sc bagpipes}; this $M_{\rm UV}$ is
very close to the $M_{\rm NUV}$ circulated in the COSMOS2025 catalog,
but evaluated at 1600\AA\ rather than in the NUV filter.  That $M_{\rm
  UV}$ is then converted to $L_{\rm UV}$ using the
\citet*{kennicutt12a} FUV scaling.  SFR$_{\rm IR}$ is calculated from
\lir\ using the \citeauthor*{kennicutt12a} TIR scaling.
For each bin, we keep track of the
full distribution in $M_{\rm UV}$ and the posterior distribution on
\lir\ for the same bin to derive realistic, representative
uncertainties on \funobs\ at a given stellar mass and redshift.

Figure~\ref{fig:sfms} shows the breakdown between IR and UV
components of the star-forming main sequence (or SFR-\mstar\ relation)
relative to the total SFR calculated over the same mass and redshift
bins using COSMOS2025.  We note that the coaddition of UV and IR
components of the SFR agrees very well with the \citet{speagle14a}
meta-analysis of the main sequence parameterization from $0<z<6$
(shown as gold lines on Fig.~\ref{fig:sfms}) as well as the prior
measurements out to $z\sim2.5$ in \citet{whitaker17a}.

Figure~\ref{fig:funobs} shows the derived \funobs-\mstar\ relation in
redshift bins spanning $0<z<7$ as well as the redshift evolution of
\funobs\ in fixed mass bins.
We overplot the relation as measured by \citet{whitaker17a} in gray,
who found no evidence for an evolution from $0.5<z<3$; their analysis
used data from the 3D-HST survey \citep{whitaker14a,skelton14a} as
well as 24\um-extrapolated IR star-formation rates.  Our stacked SEDs
suggest a lower \funobs\ at fixed stellar mass than found by
\citet{whitaker17a}, though we largely attribute the offset to
the different treatment of the IR SED, primarily our direct fits vs. a
24\um-extrapolation, as well as a discrepancy in UV SFR tracer, NUV
($\approx$2300\AA) vs. FUV ($\approx1600\AA$).  We can reproduce the
\citet{whitaker17a} relation if we use a similar approach as described
therein.

The main takeaway from Figure~\ref{fig:funobs} is that all galaxies
seem heavily obscured, even at relatively low stellar masses at all
redshifts, with exception of the lowest redshift bin $z<0.5$ (we exclude
the $z=0.25$ bin as an outlier).  There is a steep mass dependence to
\funobs\ at all redshifts, with some subtle redshift evolution on
\funobs\ for values $<1\%$ at \mstar$>10^{10.5}$\,\msun.  

We fit these data to a generalized logistic growth function such that:
\begin{equation}
\log(f_{\rm unobs}) = \mathcal{F}_0 - \log\!\left[1 + 10^{\,\alpha(z)\,(M - M_t(z))}\right]
\label{eq:funobs}
\end{equation}

\noindent again where $M \equiv \log(M_\star/M_\odot)$.  We first
fit this function to each redshift bin and note some broad
evolutionary trends, such that $\mathcal{F}_0$ is constant and $\alpha$ and M$_t$ can be described as
smooth functions of $z$ within measurement error:
\begin{equation}
  \alpha(z) = \alpha_0 + \alpha_{1}\,\log(1+z)
\end{equation}
\begin{equation}
M_t(z) = M_{t,0} + M_{t,1}\,\log(1+z)
\end{equation}
\noindent Given the smooth evolution observed across a broad range of
redshifts, we fit the five parameters ($\mathcal{F}_0$, $\alpha_0$,
$\alpha_1$, $M_{t,0}$, $M_{t,1}$) to the measured \funobs($M$,$z$)
data from $0.5<z<7$ and present the derived values in
our appendix, Table~\ref{tab:allrelations}.

The first notable observation is that $\mathcal{F}_0$ tells us that
even at low stellar masses, only $\sim$38$\pm$5\%\ of the
star-formation is probed directly by the rest-frame UV on average.
The second notable effect is that the stellar mass at which the
obscured component overwhelms the unobscured component is between
$9<\log(M_\star/M_\odot)<10$, closer to 10$^9$\,\msun\ at $z\sim2$ and
evolving to higher masses with increasing redshift.  The third
observation is that the drop off with mass is {\it steep} and steepens
with increasing redshift, with $\alpha\sim1$ at $z\sim1$ steepening to
$\alpha\sim3$ at higher redshifts.  We note that \citet{whitaker17a}
observed no evidence for redshift evolution of \fobs-\mstar. Without
the extended redshift lever arm available in this work, our results
over the same redshift range $0.5<z<2.5$ would not be well-enough
constrained to claim redshift evolution either.  This is similar to
the observed trends in \fobs-\mstar\ seen in the {\sc simba}
cosmological hydrodynamic simulations, where redshift evolution is
seen over $0<z<6$ but is subtle at $z<3$ \citep{zimmerman24a}; the
\citet{zimmerman24a} results, translated to \funobs\ from \fobs, are
shown in the full panel version of the \funobs-\mstar\ relation in the
Appendix.  We will return to a detailed comparison of \funobs\ and
\auv\ measures further in the discussion.

\subsection{Dust to Stellar Mass Ratio}\label{sec:DTS}

Dust is the byproduct of star formation, thus it follows that the
total dust mass of star-forming galaxies should be fundamentally
linked to their stellar masses.  This relation may be further
complicated by growth of dust in the ISM or, alternatively, dust
destruction by supernovae and feedback, the rates of which depend on a
complex mix of physical conditions: density, temperature, clumpiness,
metallicity, and strength/hardness of the radiation field
\citep{galliano18a}.  The dust-to-gas ratio (DTG) is more often
studied,  as well as the gas-to-stellar ratio (GTS), due
to gas's direct link to the star formation process, and dust's key
role in cooling gas and thus catalyzing star formation.  However, the
lion's share of gas measurements at high-$z$, in practice, actually
measure dust and presume a DTG rather than provide an independent
anchor by which to assess the relative relationship
\citep{scoville16a}; this is because dust continuum observations are
much more time efficient than CO observations.

Measurements of the dust-to-stellar ratio (DTS) is less broadly
discussed in the literature but has advantages in that the two
quantities -- dust and stars -- are measured from vastly different
parts of a galaxy's spectrum.  Most literature samples with
measurements of the DTS ratio are either local samples, or limited
high-$z$ samples. For example, directly detected IR-luminous sources,
like local ULIRGs \citep[e.g.][]{da-cunha10a}, and {\it
  Herschel}-selected galaxies presented in the HATLAS survey
\citep{dunne11a} or PEP and HerMES surveys \citep{santini14a}.  Beyond
this, a volume-limited sample of local galaxies with IR constraints
exists from the {\it Herschel} Reference Survey (HRS) and is presented
in \citet{cortese12a} and \citet{andreani18a} and from DustPedia
presented in \citet{galliano21a}.

\begin{figure*}
\centering
\includegraphics[width=0.99\columnwidth]{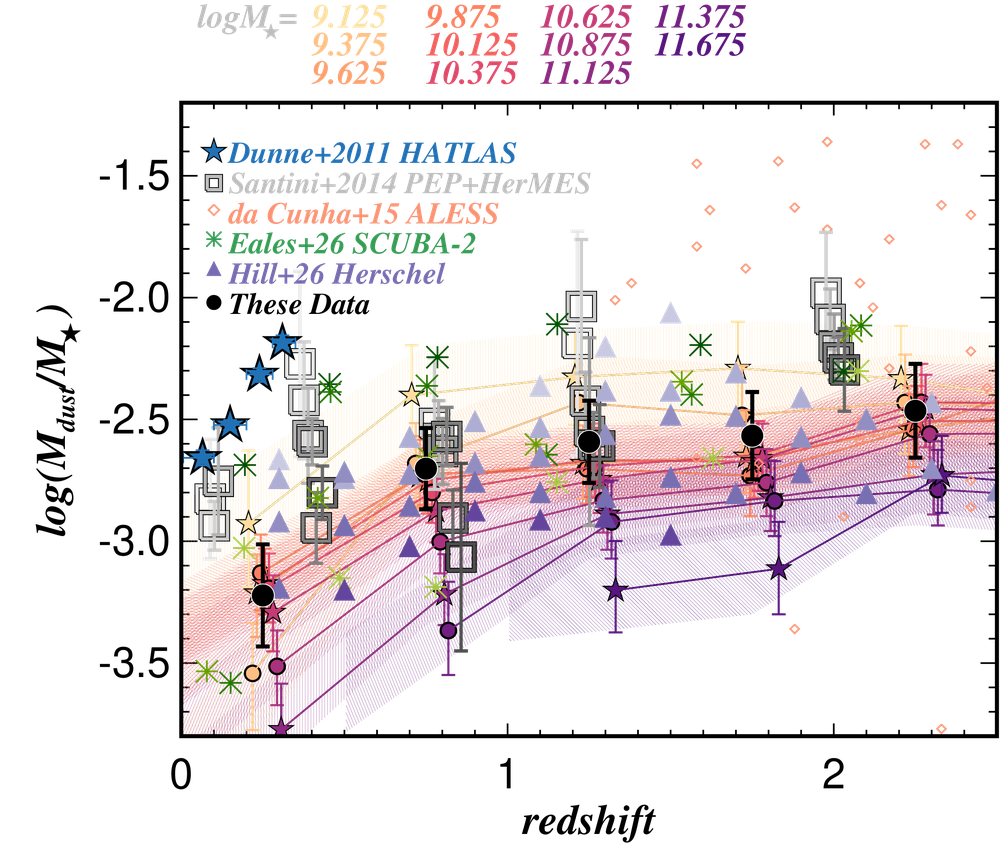}
\includegraphics[width=0.99\columnwidth]{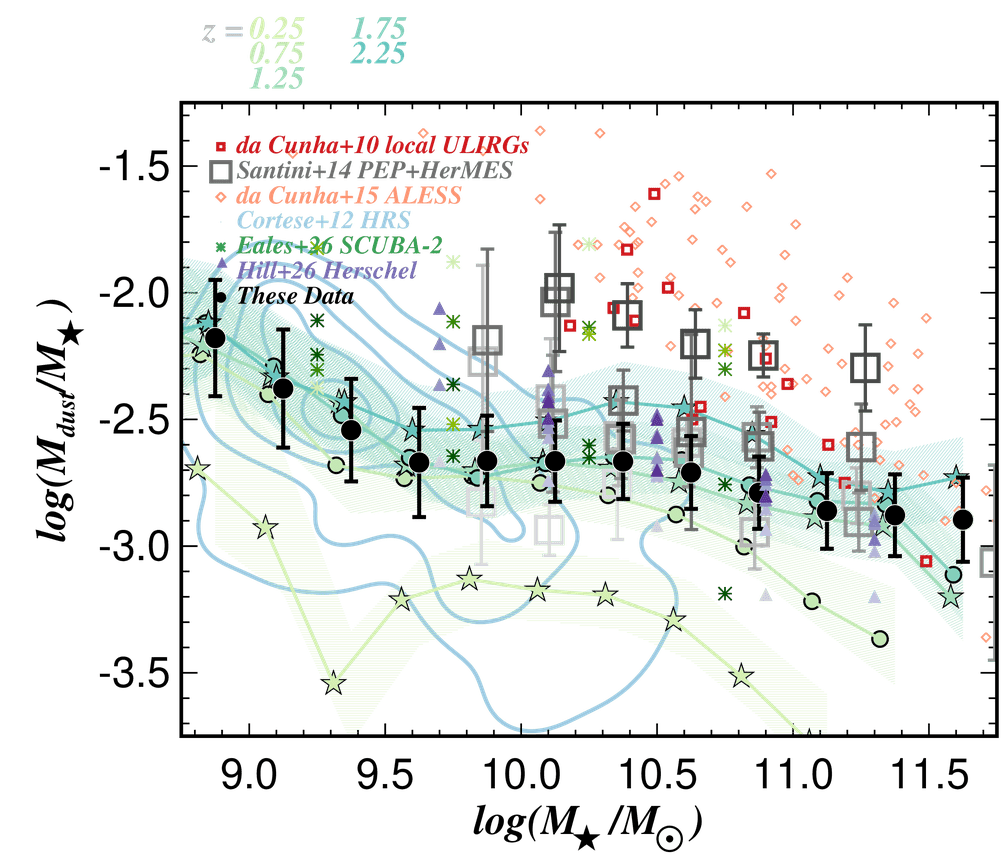}
\caption{A literature comparison of the dust-to-stellar ratio, DTS,
  with our data.  The redshift evolution is shown at left and stellar
  mass dependence at right. HATLAS-detected galaxies show strong
  evolution in the DTS from $z=0$ to $z\sim0.5$ \citep[blue
    stars;][]{dunne11a}.  Work from the PEP+HerMES surveys is shown in
  gray boxes \citep{santini14a}; lighter gray boxes correspond to
  lower stellar mass bins.  We also show individually detected dusty
  star-forming galaxies from ALESS \citep{da-cunha15a} in light red
  diamonds. Recent SCUBA-2 stacking from \citet{eales26a} and {\it
    Euclid-Herschel} stacking from Hill
  \etal\ \citep{euclid-collaboration26a} is shown in green asterisks
  and purple triangles, respectively. Though significant scatter in
  literature DTS measurements exist, our data agree with the steep
  redshift evolution seen at $z<0.5$ by \citet{dunne11a}, and
  shallower increase out to $z\sim2$ found by \citet{santini14a}.  At
  right, we see a net decrease in the DTS at high stellar masses.  The
  \citet{cortese12a} compilation of dust masses from the Herschel
  Reference Survey \citep[see also][]{andreani18a} are shown in
  contours broadly aligns with our measurement of the
  DTS-\mstar\ relation at $1<z<2$ rather than at $z=0$, though this is
  likely due to different adopted conventions for deriving dust mass.
  In comparison we note that both local ULIRGs \citep{da-cunha10a} and
  high-$z$ DSFGs \citep{da-cunha15a} are a bit more dust-rich, likely
  by construction given their selection as dust-luminous sources.}
\label{fig:mdustmstar_lit}
\end{figure*}

\begin{figure*}
\centering
\includegraphics[width=0.99\columnwidth]{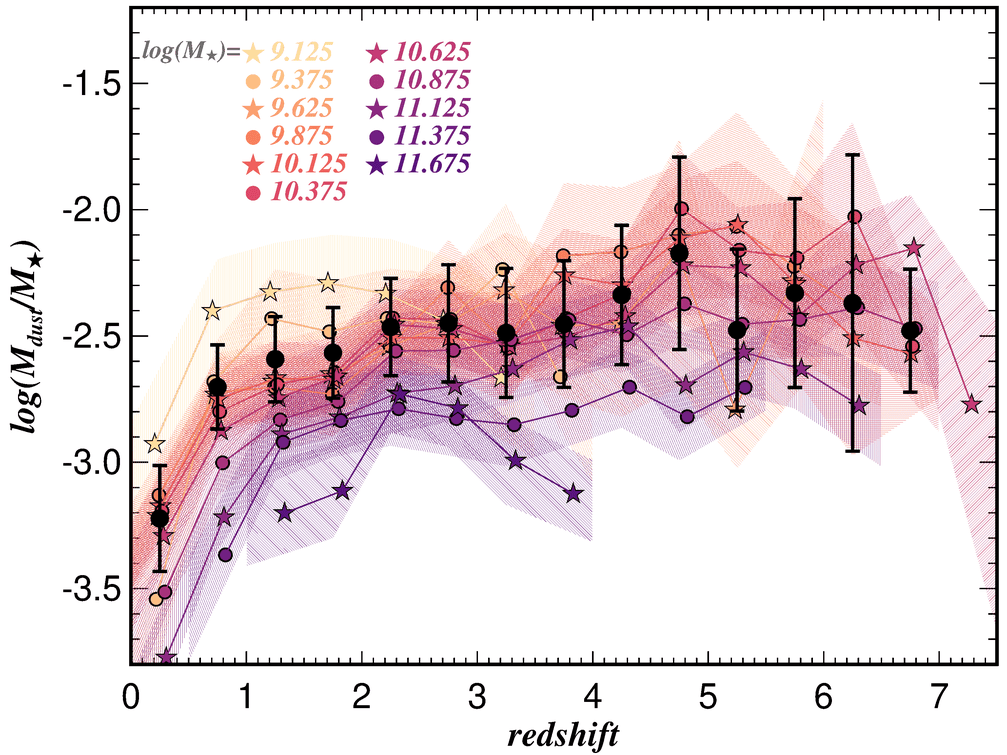}
\includegraphics[width=0.99\columnwidth]{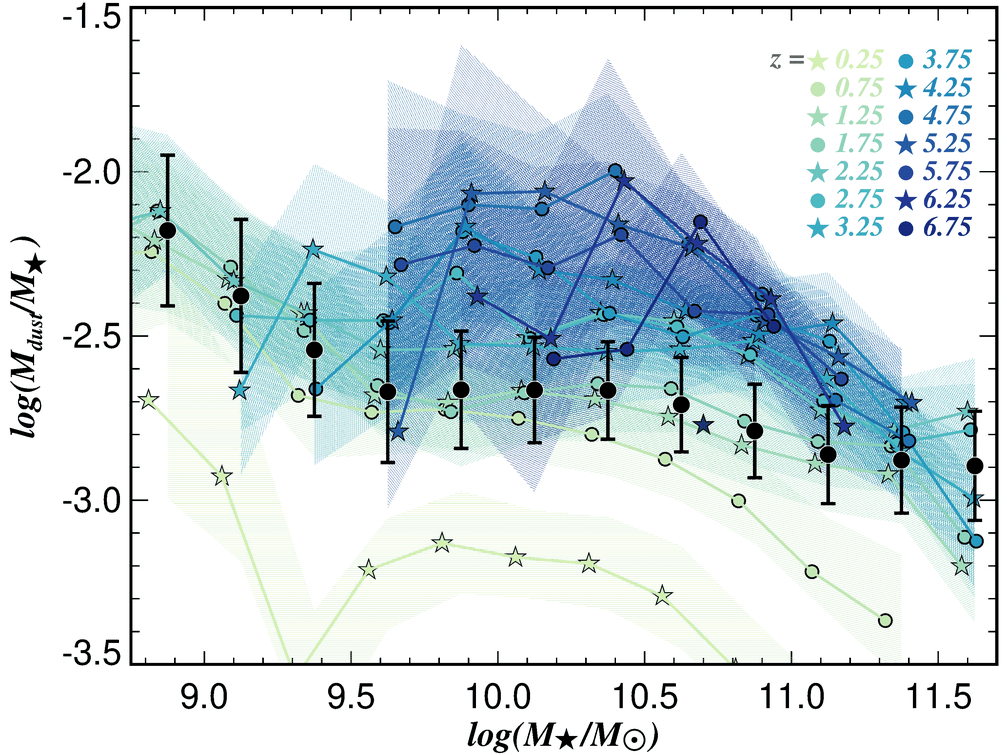}
  \caption{The redshift (left) and stellar mass (right) dependence of
    the dust-to-stellar ratio, DTS, measured via stacking.  The
    redshift evolution of the DTS is predominantly flat over all
    individual mass bins (light green to purple color) as well as
    averaged over all masses (black points).  The most significant
    variance is seen in the lowest redshift bin, $z<0.5$. In contrast,
    the stellar mass dependence of the DTS is strong and persists to
    high redshifts. We fit a redshift, and mass-dependent functional
    form to the DTS as given in Eq.~\ref{eq:dts1}.
    Figures~\ref{fig:panels_dtsmstar} and \ref{fig:panels_dtsz} in the
    appendix show the relation broken down relative to the
    Eq.~\ref{eq:dts1} fit.}
  \label{fig:mdustmstar}
\end{figure*}

We compare the inferred dust-to-stellar ratios of our stacked dust
SEDs to literature measurements at $z<2.5$ in
Figure~\ref{fig:mdustmstar_lit} and the full constraints of our data
out to $z\sim7$ are shown in Figure~\ref{fig:mdustmstar}.  In the
literature comparison, we note that both \citet{dunne11a} and
\citet{santini14a} observe a steep rise in the DTS from $z=0$ to
$z\sim0.5$.  Though our stacks are only sampled in $\Delta z=0.5$
bins, we note the same rise from $z=0.25$ to $z=0.75$.  Like
\citet{santini14a}, we see this rapid evolution in the DTS at low
redshift level out towards higher redshifts ($z\sim2$), and both
datasets exhibit a lower DTS at higher stellar masses.  The trend of a
lower DTS at higher masses is also seen in the nearby HRS sample
\citep[][note the DustPedia sample occupies very similar
  space]{cortese12a,andreani18a}.  Galaxies explicitly selected to be
dust-luminous \citep[ULIRGs and DSFGs;][]{da-cunha10a,da-cunha15a}
have higher dust-to-stellar ratios than our stacks suggest is typical
for galaxies of their same stellar mass.  We note that we broadly
observe the same trends in the DTS ratio with redshift and mass, but
the normalization of the DTS is often offset.  For example, the $z=0$
HRS sample seems to show a consistent DTS-\mstar\ relation as we
measure at $1<z<2$, at much higher DTS than our redshift-matched
sample.  Similarly the {\it Herschel} samples have consistently higher
DTS than we measure over the same mass and redshift ranges.  We
attribute the differences to the methodology of deriving the dust mass
itself (dust mass absorption coefficient assumed, which carries
significant uncertainty, and to a lesser extent dust temperature).

Looking in more detail at Figure~\ref{fig:mdustmstar}, we draw a few
conclusions over a wide dynamic range of mass and redshift. The DTS
ratio falls by nearly an order of magnitude between $z\sim3$ and
$z\sim0$, most of that evolution at $z<0.5$.  Also all mass bins show
no significant (or very shallow) evolution toward higher redshifts
($3<z<7$).  It is notable that no significant drop in the DTS is seen
even beyond $z>6$ where we have sufficient sensitivity to draw
measurements from our stacks.  Then where constraints on lower-mass
\mstar$<10^{9.5}\,$\msun\ galaxies exist ($z<3$), their DTS ratios are
markedly higher than at higher stellar masses.  The mass and redshift
dependence seen in Figure~\ref{fig:mdustmstar} is well described via a
power law, such that
\begin{equation}
  \log(M_{\rm dust}/M_\star)(z,M) = A + \eta_{0}(M-10) + \eta_{1}\log(1+z)
\label{eq:dts1}
\end{equation}
again where $M\equiv\log(M_\star/M_\odot)$.  We fit the parameters of
the function as $A=-3.237\pm0.036$, $\eta_{0}=-0.296\pm0.023$ and
$\eta_{1}=1.364\pm0.066$. These parameters are also summarized in
Table~\ref{tab:allrelations}. We will expand on how these constraints
on the DTS inform our understanding of the changing evolution of
galaxies' dust content, and relationship to gas and stars, further in
the discussion, \S~\ref{sec:dts_discuss}.

\subsection{$z<0.5$ measurements as an outlier}\label{sec:lowz}

The lowest redshift bin is a persistent outlier in this analysis. Why?
We have noted higher average \av\ (and \auv) than literature
measurements and low \lir\ and \mdust, as well as higher \funobs. The
low \lir\ and \mdust\ are in agreement with prior literature
\citep[e.g.][]{le-floch05a,santini14a}.  However, the slightly higher
average attenuation, particularly around $10^{10}\,$\msun, is
discrepant with the literature.  To form a clear understanding of
potential systematics in our dataset, we dug deeper into the cause of
this low-redshift discrepancy.  The relevant population are largely
Milky Way analogs at $z\lesssim0.5$: spirals with some lenticulars and
early type galaxies.

The \citet{salim16a}, henceforth S16, derivation of the local
\auv-\mstar\ relation is anchored on the GALEX-SDSS-WISE legacy
dataset, drawn from SDSS spectroscopically confirmed $z<0.3$ galaxies
in the local volume.  \citet{strauss02a} describes this sample,
selected to be brighter than $r<17.8$; they note that the color
distribution of the population is rather narrow about
$|g-r|\approx0.5$, which is consistent with photometry from COSMOS2025
in the HSC $g$ and $r$ filters for the matched population.  Forty-six
of the GALEX-SDSS-WISE sources sit in the COSMOS-Web field so we use
that sample to directly compare derived \av, \auv, and \mstar.
The predicted
masses are 0.21$\pm$0.38\,dex lower for {\sc bagpipes}-derived SEDs
than found in S16. Despite this, there is a persistent discrepancy
between predicted attenuations: \av=0.20$\pm$0.09 (S16) vs our
\av=1.36$\pm$0.53 and \auv=1.63$\pm$0.58 (S16) vs our
\auv=5.07$\pm$2.52.

One might suspect that differences in photometry or SED fitting
techniques cause the discrepancy in inferred dust attenuation with the
literature.  SED fitting in S16 is performed with CIGALE with a
similar suite of flexible (though parametric) star formation histories
as used here for {\sc bagpipes}. The photometry for the overlapping
sample is broadly consistent, whether modeled or based on Kron
apertures \citep{kron80a}, though one key difference is extreme ends
of wavelength coverage for either set of photometry.  Our fits include
much longer rest-frame wavelengths (i.e. F444W and F770W) and omit the
rest-frame UV (from GALEX).  These two differences are significant --
in the case of S16 it results in fits with lower SFR and older stellar
populations than we have fit.  This leads our models to have a higher
\av\ and SFR (and thus predicted \lir) than may hold.

Last, we emphasize our initial down-selection to our 'stackable'
sample, where we excluded quiescent galaxies.  S16 does not exclude
quiescent systems, and so the fact that they find lower \av\ at fixed
stellar mass correlate, in part, with real differences in either
sample.  For these reasons we have excluded the lowest redshift bin
from a number of our fitted relations and we issue caution in
particular on the interpretation of UV/optical-inferred properties of
the lowest redshift stacks in this work, leaving a more careful
approach to the low-$z$ subsample to a future work.

\section{Discussion}\label{sec:discussion}

The measurements in this paper --- thanks in particular to the
remarkable sensitivity of both JWST and ALMA --- present a new view on
the average dust absorption and emission characteristics of galaxies
over the majority of cosmic time for typical galaxies, down to stellar
masses of $\sim10^{9}$\,\msun\ and out to $z\sim7-8$.  Such progress
is enabled by the rich density of galaxies detected by JWST imaging,
and the increased precision with which we can characterize their
redshifts, masses, sizes, and star formation characteristics.  It is
also enabled by the decades' long accumulation of exquisitely deep
FIR/(sub)mm imaging in fields like COSMOS; taken in aggregate, from
{\it Spitzer}, {\it Herschel}, SCUBA-2 and ALMA, average galaxies'
dust SEDs can finally be well understood to much lower masses and
higher redshifts than previously accessible.

This discussion focuses on the relationship between quantities
presented in \S~\ref{sec:results}, the physical interpretation of
those quantities and relationship to ongoing discussions elsewhere in
the literature.  Specifically we touch on the UV/optical's ability to
capture dust attenuation and emission (\S~\ref{sec:auvcompare}), the
relationship between \av\ and \sdust\ (\S~\ref{sec:avsdust}), a
discussion of a possible grain size distribution evolution
(\S~\ref{sec:kratio}), dust temperature evolution
(\S~\ref{sec:tdustdiscuss}), and inference from the dust-to-stellar
ratio of galaxies (\S~\ref{sec:dts_discuss}).

\subsection{Can rest-frame UV/optical SEDs accurately measure attenuation, or infer \lir?}\label{sec:auvcompare}

A wealth of literature on DSFGs has made clear that the rest-frame
optical imaging or spectra for the most dust-rich galaxies has zero
constraining power on our understanding of the {\it actual}
attenuation present, because the ISM in such galaxies is optically
thick through the near-IR. \citet{swinbank04a} and \citet{chapman05a}
demonstrate that far-infrared star formation rates measured in the
submm are factors of 10-120$\times$ higher than predicted from the
UV/optical.  In other words, the most extreme star-forming galaxies
simply look like typical Lyman-break galaxies in the UV/optical
\citep{adelberger00a}; they may not even be particularly red
\citep{smail04a,mckinney24a}.  So, while DSFGs are ``catastrophic
failures'' of the predictive power of the UV/optical, what about the
full range of stellar masses and redshifts probed in this work?

\begin{figure*}
\centering
\includegraphics[width=0.99\columnwidth]{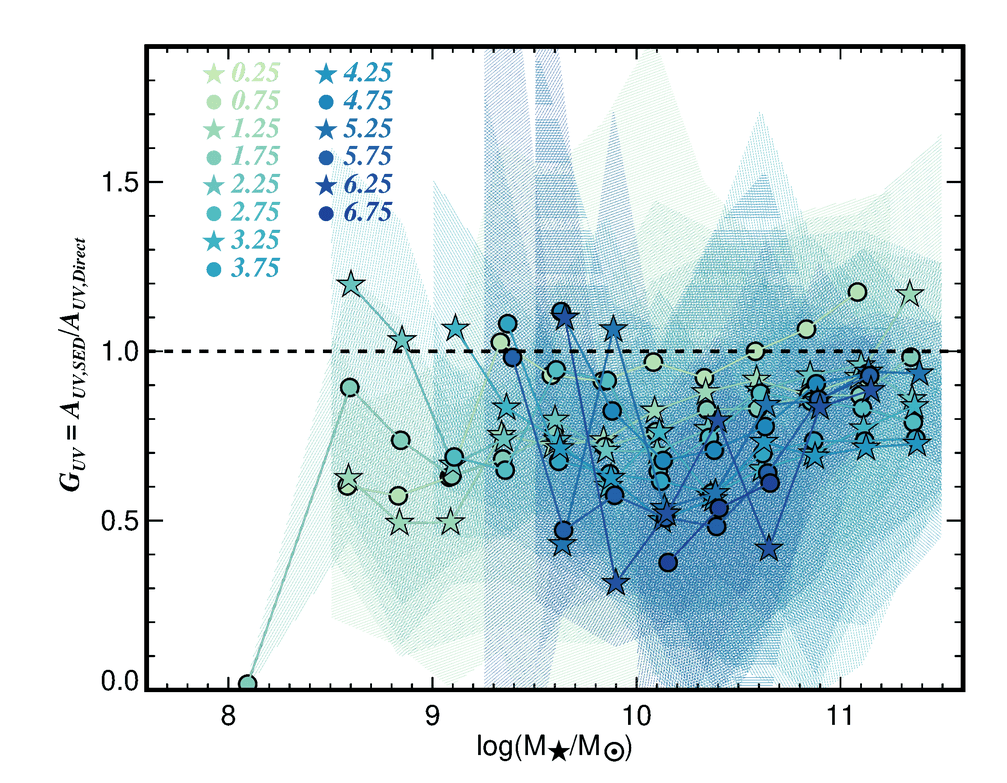}
\includegraphics[width=0.99\columnwidth]{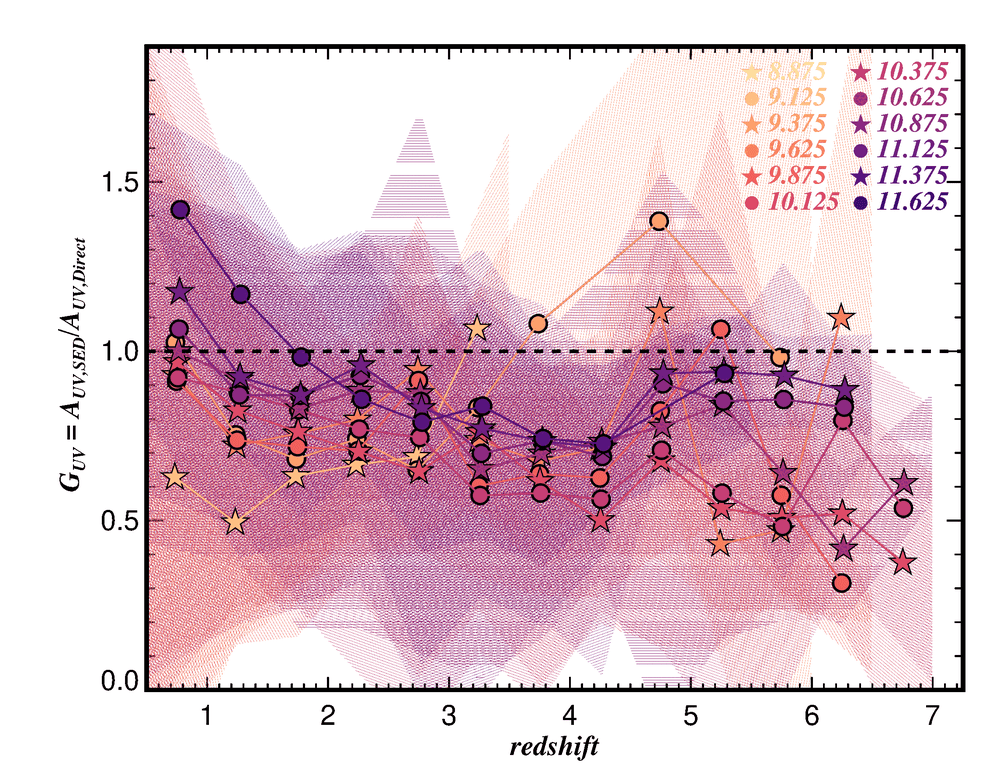}\\
\includegraphics[width=0.99\columnwidth]{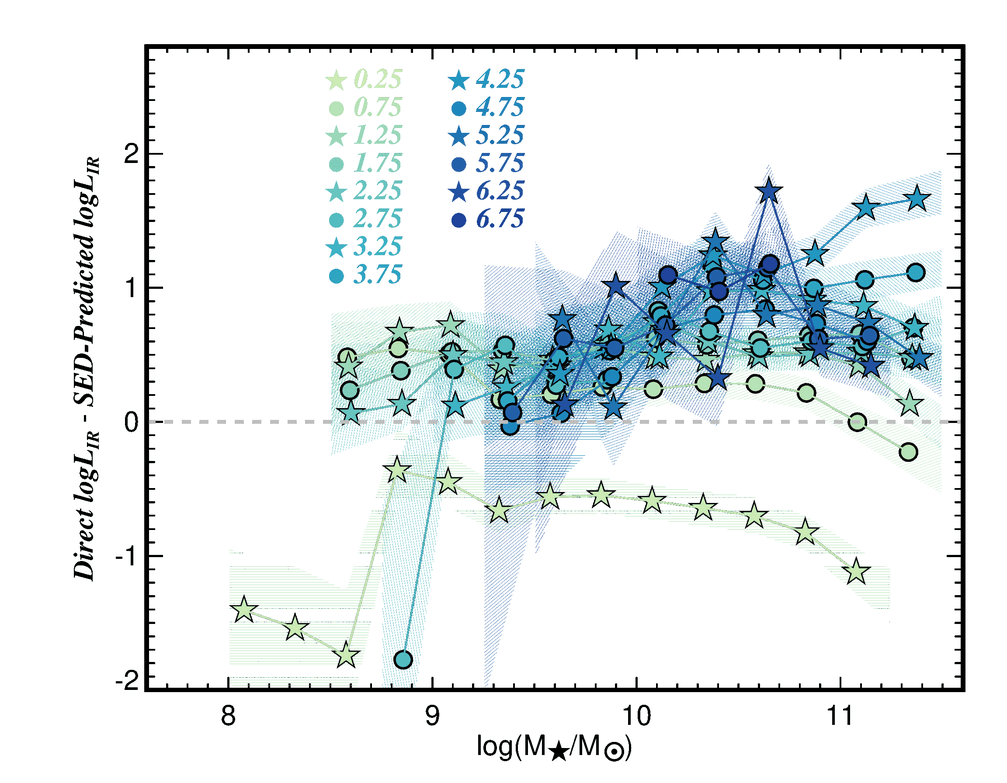}
\includegraphics[width=0.99\columnwidth]{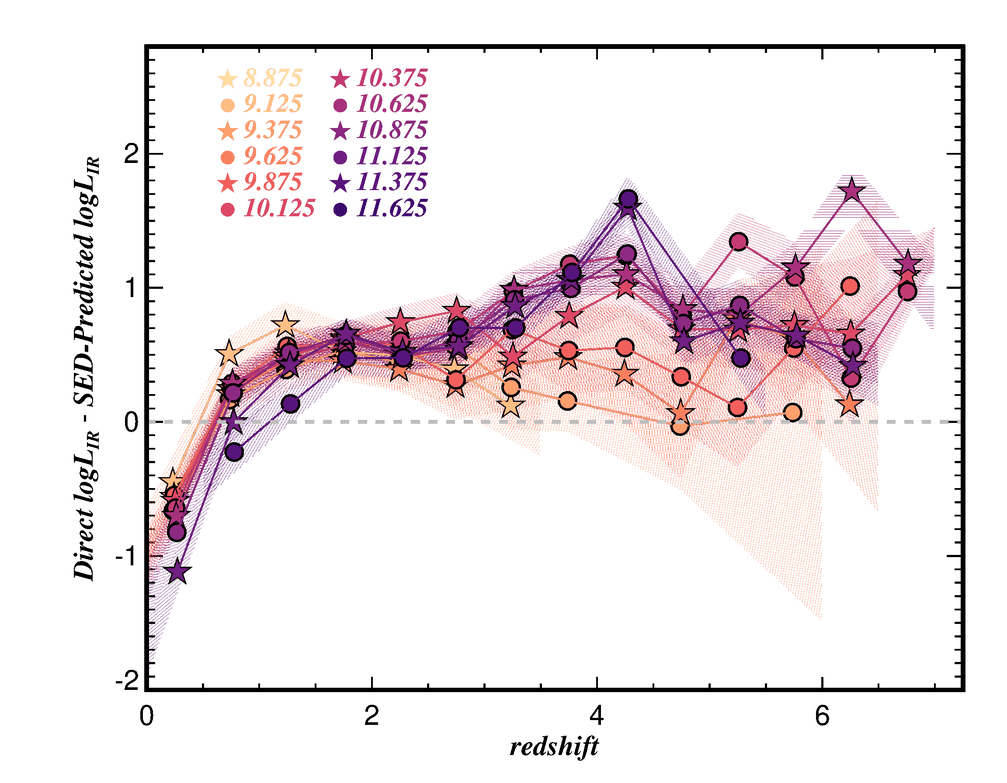}
\caption{At top, we show $\mathcal{G}_{\rm UV}$, the star/dust
  geometry prefactor that relates the magnitudes of attenuation using
  SED-based techniques, \auv, anchored to rest-frame UV/optical
  constraints, to \auvd, a {\it directly}-inferred \auv\ measured from
  IRX as given in Eq.~\ref{eq:auvdirect}.  $\mathcal{G}_{\rm UV}=1$
  corresponds to a foreground dust screen and $\mathcal{G}_{\rm UV}<1$
  corresponds to mixed star/dust geometry that serves to {\it lessen}
  the perceived attenuation for a given known dust column density.  At
  bottom, we show the same quantity a different way: the difference in
  \lir\ measured directly with \lir\ as inferred from \auv\ in the
  UV/optical assuming energy balance.  Here we see that the
  underprediction of \auv\ leads to the
  $\sim$0.5\,dex$\,\approx\,$3$\times$ underestimation of \lir, with
  more substantial variation at high masses and high redshift.}
\label{fig:auvcompare}
\end{figure*}

To answer this question, we can compare the
UV/optical-inferred attenuation, as measured with {\sc Bagpipes}, with
the {\it direct} attenuation inferred from \funobs\ or IRX.
Revisiting our framework from \S~\ref{sec:framework},
Equation~\ref{eq:auvdirect} describes such a \auvd, which is
representative of the \auv\ you would expect to see if all of the dust
in the galaxy presented as a foreground dust screen (which is often
taken as the presumed geometry in the interpretation of \auv).  We
note that Eq.~\ref{eq:auvdirect} does have a coefficient not yet
described, $B^\prime$, which is the ratio of bolometric corrections
between the UV and the FIR.  Conceptually, it translates our
non-bolometric proxies (\luv\ measured at rest-frame 1600\AA\ and
\lir\ integrated between 8--1000\,\um) for absorbed light (UV/optical)
and emitted light (FIR) to their bolometric ideals: the integrated
light lost due to attenuation from the UV through the near-IR and the
total light emitted over all wavelengths in the dust's modified
blackbody emission.
The value\footnote{\citet{meurer99a} and \citet{mclure18a} refer to
this as a bolometric correction ratio simply as $B$, but here we name
it $B^\prime$ to avoid confusion with the modeled bump strength, $B$.
It is defined such that $B^\prime\equiv BC(1600\AA)/BC({\rm FIR})$.}
of $B^\prime$ has been discussed as a constant 1.19$\pm$0.20
\citep{meurer99a}, 1.71$\pm$0.05 \citep{mclure18a}, 1.75
\citep{cullen17a}, or BC(1600${\rm \AA}$)=1.66$\pm$0.15
\citep{meurer99a,seibert05a,overzier11a}.  \citet{hao11a} fit a value
of 0.46$\pm$0.12 for 1/$B^\prime$, which would imply
$B^\prime\approx2.17$.  While prior works have largely treated
$B^\prime$ as a scalar, $B^\prime$ can be derived precisely for a
given modeled SED.  For example, BC(1600${\rm \AA}$) has a predictable
dependence on the shape of the intrinsic UV spectrum (like the
observed rest-frame UV slope $\beta_{\rm UV}$), the total attenuation
(like \av), properties of the attenuation law ($\delta$ and $B$);
similarly BC(FIR) should have some dependence on the FIR SED shape
(primarily driven by \tdust).  We used a wide array of {\sc
  Bagpipes}-generated SEDs and FIR SEDs to derive the following
approximation for $B^\prime$ on \av, $\beta_{\rm UV}$, and $\delta$,
noting that bump strength $B$ and \tdust\ (assuming
\tdust$\lesssim$100\,K) have $<$1\%\ impact.  We find this
approximation is able to infer $B^\prime$ to within
$\sim$5-10\%\ accuracy based on our model SEDs:
\begin{equation}
B^\prime(A_V,\beta_{\rm UV},\delta)=10^{c_0+c_1 A_V + c_2 \beta_{\rm UV} + c_3 \delta}
\end{equation}
where $c_{0}=0.2665\pm0.0006$, $c_{1}=-0.0451\pm0.0003$,
$c_{2}=-0.02087\pm0.00016$, and $c_{3}=0.9774\pm0.0014$.  The scatter
about this relation (owing to other critical details driving the SED
like the star-formation history and ionization parameter) is of order
$\sigma_{B^\prime}\approx0.08$.  Galaxies with low attenuation and
blue $\beta_{\rm UV}$ have higher $B^\prime\approx1.5$ than those with
more attenuation or redder UV slopes ($B^\prime\approx1$).  The
dependence on $\delta$ is such that shallower attenuation results in
higher $B^\prime$.  For the {\sc Bagpipes} SEDs fit directly to the
COSMOS-Web sample, the median $\langle B^\prime\rangle=1.16\pm0.32$.
For each mass and redshift bin of our stack, we calculate a median
value of $B^\prime$ to use in the comparison of \auv\ to \auvd.

Figure~\ref{fig:auvcompare} shows the ratio of SED-inferred \auv\ and
the directly inferred \auvd\ as a function of stellar mass and
redshift.  As a reminder, this ratio is a direct proxy for the impact
of star/dust geometry, $\mathcal{G}_{\rm UV}$, as noted in
Equation~\ref{eq:geometry}.  The figure also shows a direct comparison
between the measured \lir\ luminosity and a SED-{\it predicted}
$L_{\rm IR}$, based on inverting Eq.~\ref{eq:auvdirect} to solve for
\lir\ (via IRX) and using the SED-derived \auv\ instead of \auvd; in
other words, this would be the observed \lir\ if the presumption of a
foreground dust screen were correct.

We note a few key observations from these comparisons, first the
\auv\ comparison.  Recalling that $\mathcal{G}_{\rm UV}$ should be
bounded by $0<\mathcal{G}\le1$, we note that our data are broadly
consistent with that limit.  One notable exception again is the $z=0.25$
bin, which overall sits off the plot around $\mathcal{G}_{\rm
  UV}\approx2-3$.  Values of $\mathcal{G}_{\rm UV}>1$ are technically
unphysical; but this can arise if \auv\ from SED fitting {\it
  overpredicts} the dust content.  This may be the case for the
$z=0.25$ bin as discussed in \S~\ref{sec:lowz}.  Above $z>0.5$, we
note that all measurements of $\mathcal{G}_{\rm UV}$ are statistically
consistent with values $\le1$.  Lower values of $\mathcal{G}_{\rm UV}$
correspond to a larger mismatch between SED-measured \auv\ and
directly inferred \auvd, implying that the ISM is more optically thick
and/or clumpy.  While the scatter on $\mathcal{G}_{\rm UV}$ is
large, because the range of \auv\ in each bin is somewhat broad,
we note the general trend that it increases slightly with
stellar mass (approaching the `screen' approximation) but decreases
with increasing redshift, perhaps signaling that at fixed stellar
mass, the ISM is more optically thick to UV photons at higher
redshifts.  This is not entirely surprising when we also consider how
higher redshift galaxies are more compact at fixed stellar mass
\citep{yang25a}.

\begin{figure}
\includegraphics[width=0.99\columnwidth]{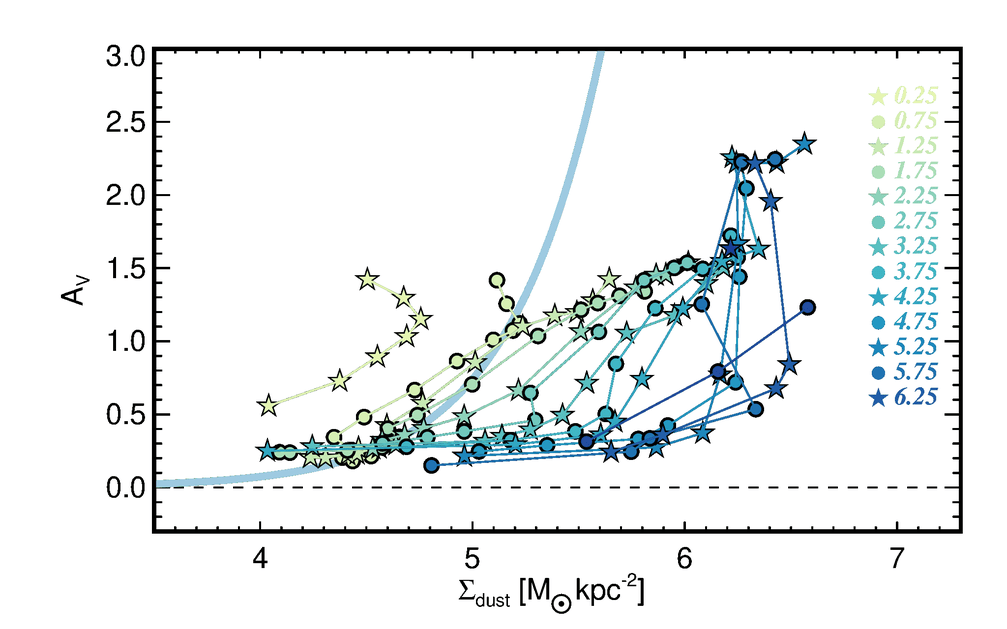}
\caption{The relationship between \av\ and \sdust\ as a function of
  redshift, with color indicating different redshift bins. The thick
  light blue line corresponds to a foreground screen of dust
  calibrated to the properties of Milky Way dust, with \cv=0.74, where
  \cv\ is defined in Eq.~\ref{eq:clambda}.  With increasing redshift,
  a clear trend is seen where fixed \av\ corresponds to a higher and
  higher dust mass surface density \sdust.}
\label{fig:avsigmadust}
\end{figure}

Looking at the \lir\ comparisons in Figure~\ref{fig:auvcompare}, which
is an alternative way of digesting the same information, we note that
overall, \lir\ from UV/optical SEDs underpredicts the true IR
luminosity.  From $0.5<z<3$ across all mass bins, this seems to be by
roughly a factor of $\sim$0.5\,dex $\approx 3\times$.  For
intermediate masses $\sim10^{10}$\,\msun\ this factor of 3
underprediction holds out to $z\sim6-7$, while for the highest mass
bins the underprediction grows to $\sim$1\,dex or a factor of
$10\times$.

Are these results surprising?  If the reader reminds themselves that
the results of Figure~\ref{fig:auvcompare} are not for DSFGs, but
rather representative of the average galaxy at these masses and
redshifts, the answer is likely yes.  It has often been thought that
there is a great chasm between the attenuation properties of `normal'
galaxies and DSFGs \citep{swinbank04a}, with the latter an extreme
caricature of the former.  It has long been thought that rare,
extreme phenomena, like gas rich mergers \citep{hopkins10a}, drive
these extreme characteristics (star formation rates, dust production,
disturbed star/dust geometries) making them so different from
galaxies' normal mode.

What our data suggest is that this chasm does not exist.  Rather,
DSFGs are simply the tip of the iceberg, or the light-post that was
always there giving us early hints that the dust in all galaxies' ISM
is transformative across a wide mass and redshift range.  These
results reinforce the idea that galaxies simply are moderately to
significantly optically thick through the rest-frame UV, and that, as
a result, UV diagnostics -- not just continuum but corresponding line
diagnostics -- can only tell a partial story of galaxies' ISM,
permeating one optical depth path length into a much more complex
ecosystem that can be revealed through its dust emission. Indeed,
other recent results from JWST reinforce this result, that optically
luminous galaxies also have optically thick dust \citep{cheng25a}.
Furthermore, radiative transfer models applied to cosmological
simulations have long emphasized these global trends, that UV can
underestimate the total SFR across a wide range of masses
\citep{sommovigo20a,liang21a,parente24a}.

A related consequence of these comparisons is in the practical
application of energy balance SED fitting.  Energy balance itself {\it
  should} hold, but in the case where the UV portion of the SED is
optically thick, there is no guarantee that the attenuation measured
in the UV should match the emission seen in the FIR.

\subsection{The A$_V$ - $\Sigma_{\rm dust}$ relation}\label{sec:avsdust}

While direct comparisons of \lir\ and \auv\ tell us about dust
geometry, the \av-\sdust\ relation tells us about both dust geometry
and the inherent properties of the dust.  This is because of the
reliance on the dust mass absorption coefficient, $\kappa$, in
calculating dust mass from flux density on the Rayleigh-Jeans
tail. Beyond our earlier calculation of \mdust\ for our stacked SEDs,
we take some steps to convert to $\Sigma_{\rm dust}$.  Rather than
applying any scaling relations on galaxies' sizes, we directly compute
median and inner 68\%\ confidence intervals on sizes for the stacked
galaxy samples in each bin, and propagate the uncertainties on both
\mdust\ and $R_{e}$ via Monte Carlo sampling to generate a realistic
distribution of \sdust\ per mass and redshift bin.
Figure~\ref{fig:avsigmadust} shows where the stacks fall in
\av\ vs. \sdust\ directly in redshift bins.  Higher stellar masses at
all redshifts are at higher \av, following from \S~\ref{sec:auvmstar}
and Fig.~\ref{fig:avmass2}.  It is striking how the stacks evolve
sharply with redshift toward higher dust mass surface densities.  This
is especially interesting considering our earlier discussion of the
decreasing measured attenuation (\av) with increasing redshift; in
other words, at higher redshifts, \av\ decreases (mildly) while
\sdust\ increases by over an order of magnitude.

While the evolution in Figure~\ref{fig:avsigmadust} is striking, we
can be a bit more precise in our calculations if we instead show the
mass dependence and evolution of \cuv\ and \cv, measured via
Eq~\ref{eq:clambda}.  Because the IR SEDs are stacked, we cannot undo
the `averaging' that is inherent to the stack, i.e. we have a single
measurement of \mdust\ per mass and redshift bin.  However, we have
complete information on \auv\ (\av) and $R_e$ for each source contributing to
that stack.  Thus, to compute the most precise bin-averaged \cuv\ (\cv), we
compute the median and inner 68$^{th}$ quartile on \auv$R_{e}^2$, such
that \sdust = (\auv$R_{e}^2$)\,$\pi$/\mdust.

\begin{figure*}
  \centering
  \includegraphics[width=0.99\columnwidth]{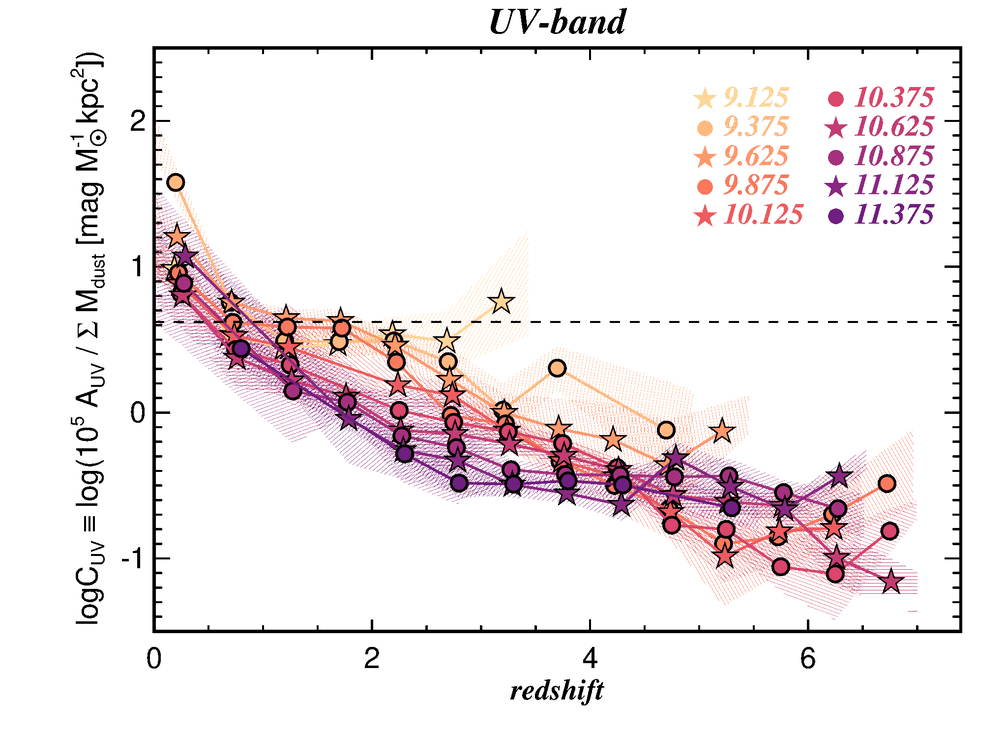}
  \includegraphics[width=0.99\columnwidth]{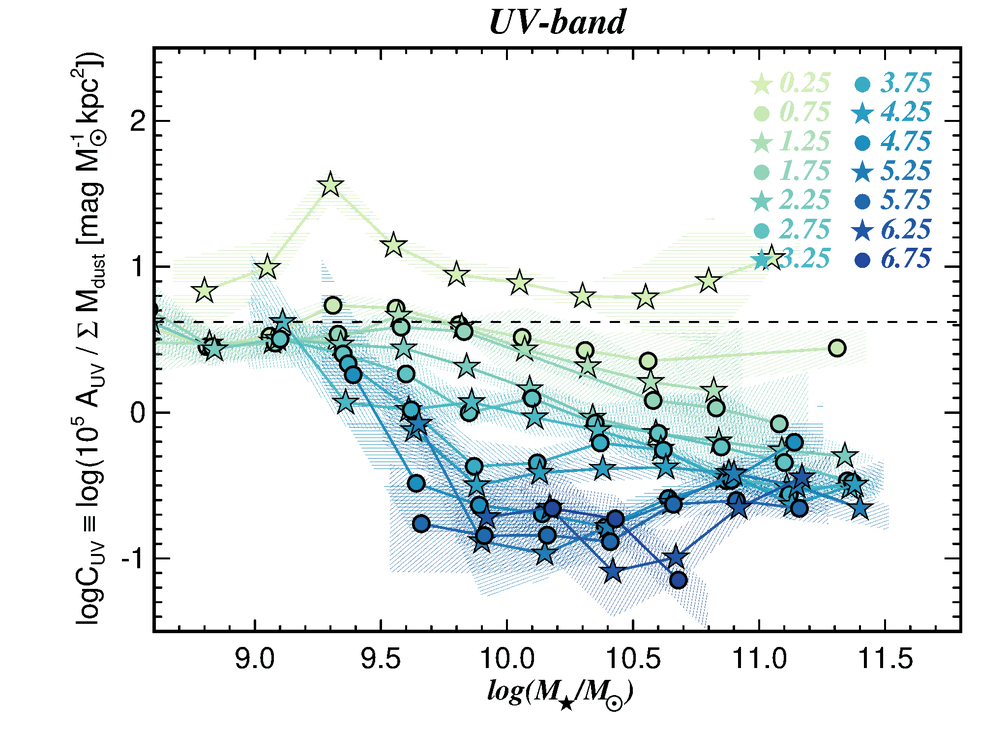}\\
  \includegraphics[width=0.99\columnwidth]{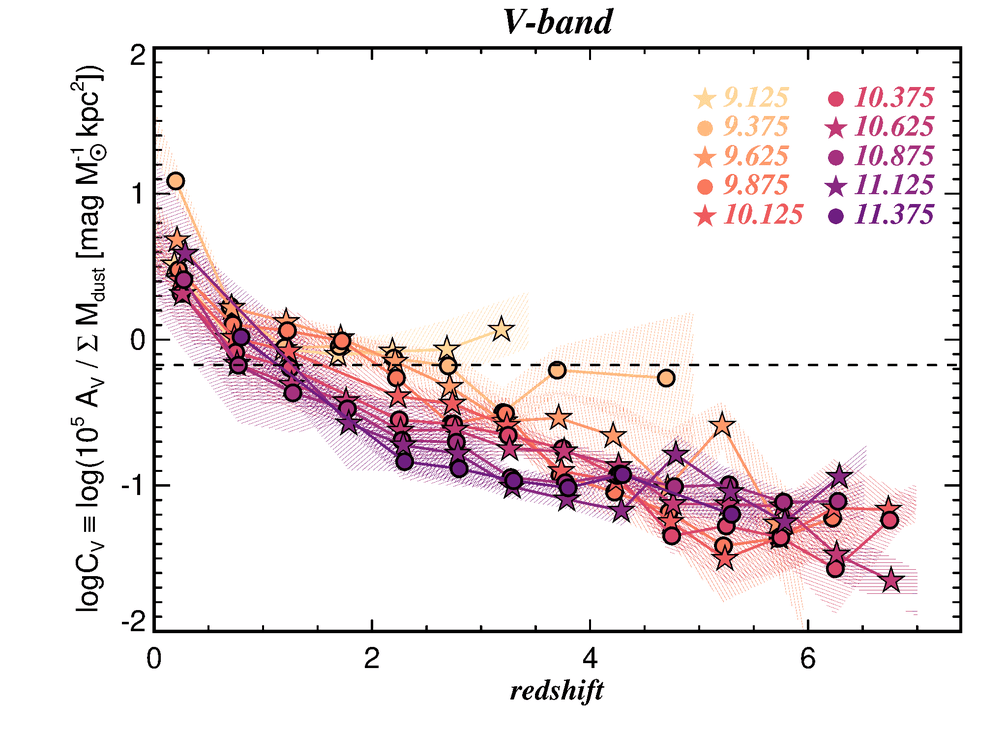}
  \includegraphics[width=0.99\columnwidth]{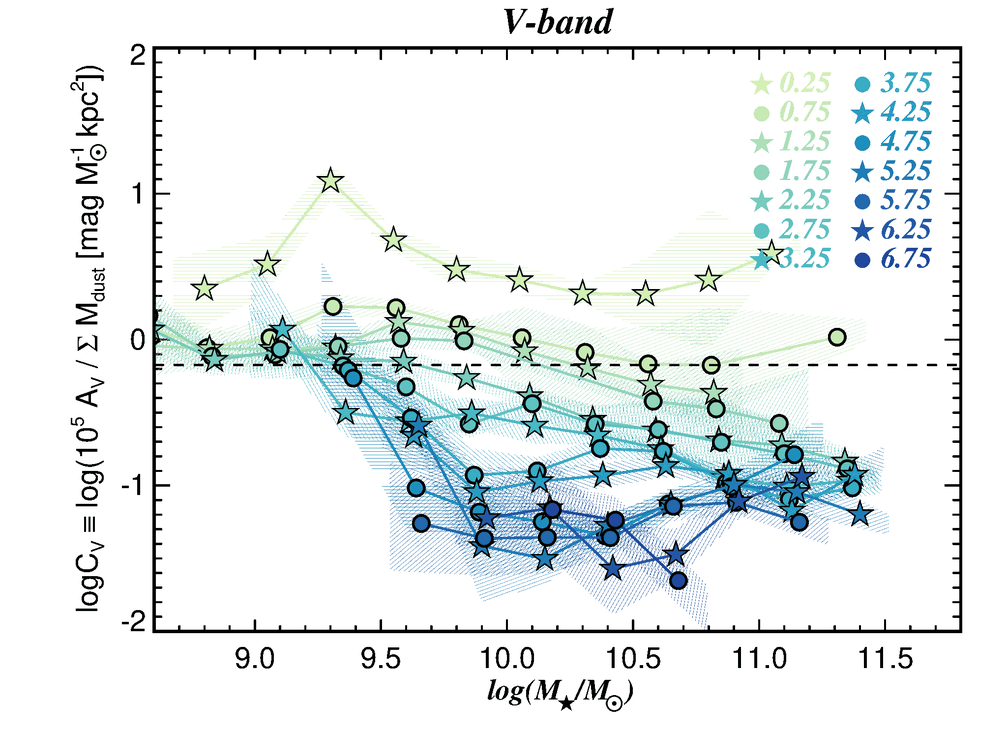}
  \caption{The redshift and stellar mass dependence of \cuv\ (top) and
    \cv\ (bottom); $C_\lambda$ is the ratio relating $A_\lambda$ to
    \sdust, i.e. it captures the evolution seen
    Figure~\ref{fig:avsigmadust} in one coefficient, defined through
    Eq.~\ref{eq:clambda}.  $C_{\lambda}$ captures the unknowns of both
    star/dust geometry and dust opacity. The values applicable to the
    Milky Way are marked in a dashed horizontal line at \cuv=4.18\ and
    \cv=0.67.  Both \cuv\ and \cv\ show a strong redshift dependence
    that is nearly identical, mirroring the similarities in
    Figure~\ref{fig:avmass2}.  To a lesser extent, \cuv\ and \cv\ also
    show a stellar mass dependence, whereby higher stellar masses have
    lower $C_\lambda$ from $0<z<4$; above $z>4$, there may be an
    inversion such that higher masses may have higher $C_\lambda$.
    The fact that \cuv\ and \cv\ show very similar behavior means that
    the evolution is not driven by the evolution in attenuation curve
    slope (as shown in Figure~\ref{fig:z_delta}).}
  \label{fig:cevol}
\end{figure*}

Figure~\ref{fig:cevol} shows the mass dependence and evolution of
\cv\ and \cuv\ with redshift.  Higher \cv\ (or \cuv) corresponds to
higher \av\ (\auv) per unit dust mass surface density.  As a reminder,
\cv\ groups the effect of both geometry ($\mathcal{G}_V$) and dust
physics ($\kappa_V$/$\kappa_{\rm FIR}$) into one factor and reveals
how it evolves.  This figure shows us a strong redshift
dependence in \cuv\ and \cv, where both are over an order of magnitude
lower at $z\sim6$ compared to $z\sim1$.  The stellar mass variance
shows interesting second order effects: at the highest stellar masses
the redshift evolution is more subtle than at lower masses where the
change in \cv\ (or \cuv) is more extreme. What drives the striking
redshift evolution in particular, given that dust masses \mdust\ is
relatively constant with redshift, and \av\ (\auv) only drops by a
factor of $\sim$2$\times$ towards high redshifts?  It comes from the
evolution in galaxy sizes, where, because dust masses (and the DTS
ratio) seem relatively constant but higher redshift galaxies are
smaller, the dust mass surface density of those high-$z$ systems are
substantially higher.

How do we interpret this drop in \cv\ (or \cuv) at high-$z$?  Is this
related to the graying of the attenuation curves at higher-$z$ seen in
Figure~\ref{fig:z_delta}?  Both the attenuation curve and \cv\ (\cuv)
carry the uncertainty of star/dust geometry as well as grain size
distribution and dust grain properties.  How much of it is geometry
versus the fundamental characteristics of the dust grains themselves?
We begin to disentangle the two in the next subsection.

\subsection{An evolving grain size distribution?}\label{sec:kratio}

\begin{figure*}
  \centering
  \includegraphics[width=0.99\columnwidth]{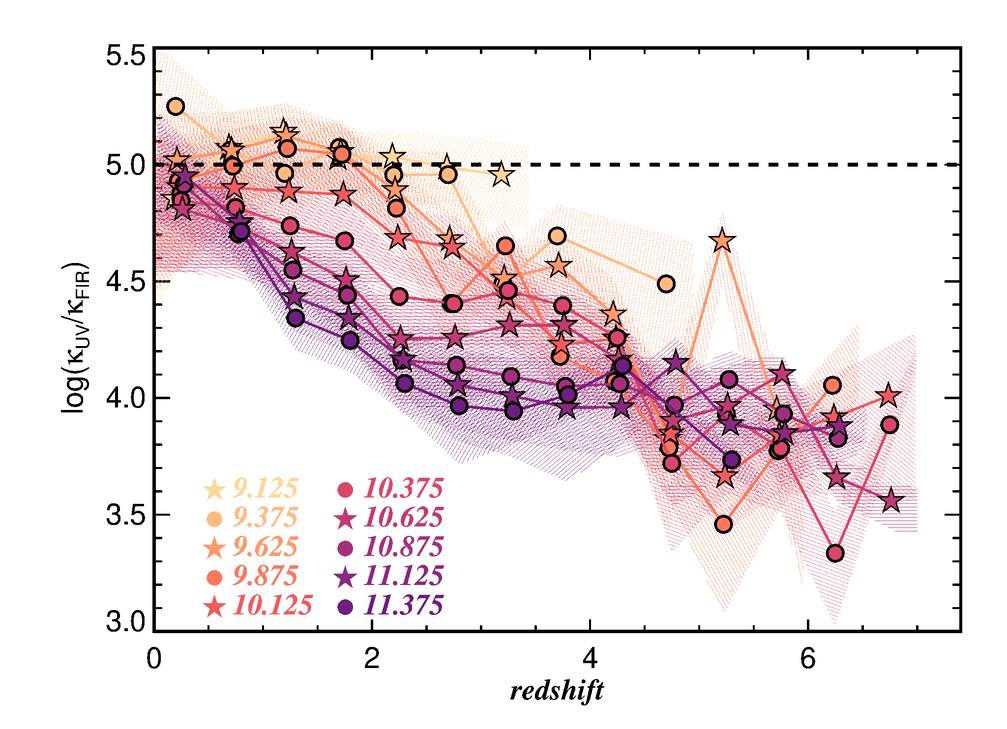}
  \includegraphics[width=0.99\columnwidth]{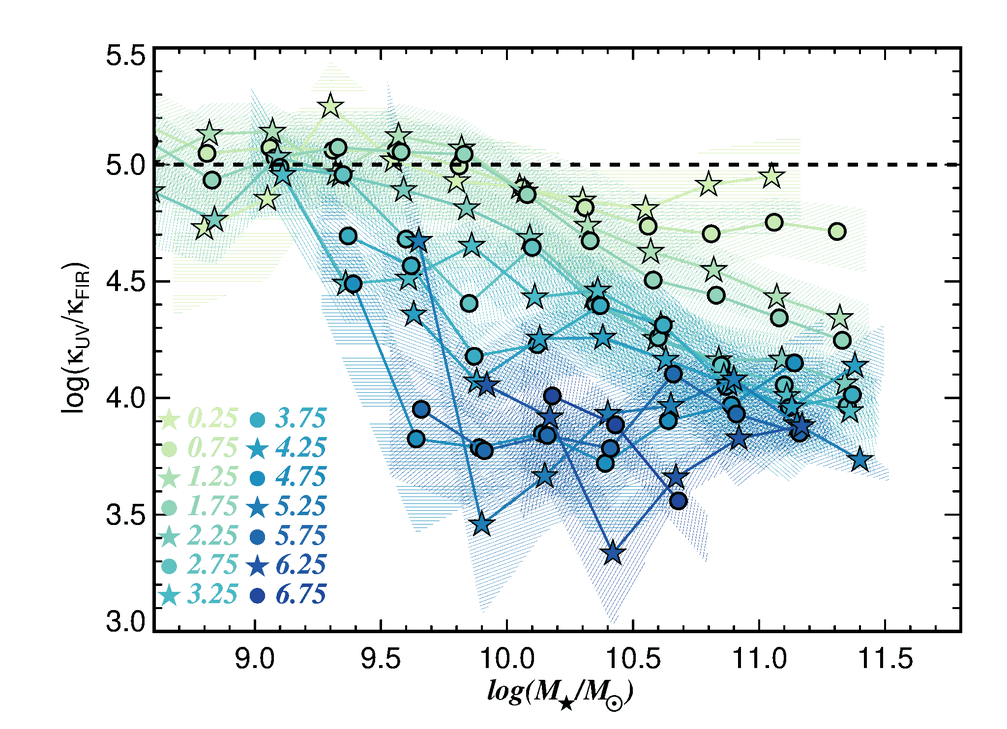}
  \caption{Measurements of the variation in \kratio\ (see
    Eq.~\ref{eq:kratio}) in redshift and stellar mass.  \kratio\ is a
    galaxy-integrated indicator of dust grain properties in the ISM --
    the grain size distribution, composition, and the morphology of
    dust grains. {\it First-order effects of star/dust geometry do not
      impact the measurement of \kratio.} The nominal
      Milky Way value of \kratio$=10^5$ originates from \citep{li01a}
      models.
    There is a precipitous fall in \kratio\ with increasing redshift,
    over an order of magnitude from $z=0$ to $z\sim4-5$.  From $0<z<4$
    there is substantial mass dependence as well, with lower
    \kratio\,$\approx$\,10$^{4}$ in massive galaxies compared to lower
    mass galaxies.  Beyond $z>4$, all mass bins (with data) have low
    \kratio. $\kappa_{\rm UV}$ is most sensitive to the small grain
    distribution, thus lower \kratio\ at high-redshift indicates a
    fundamental depletion of small grains, potentially caused by
    intense radiation fields at early cosmic times; at later times,
    those small grains are only destroyed in the still intense
    radiation fields of high-mass star-forming galaxies with less
    destruction in lower mass galaxies, and at low redshifts the
    radiation fields are significantly less intense across all mass
    ranges.}
  \label{fig:kratio}
\end{figure*}

The wealth of our stacks capture enough detail in the IR SEDs relative
to the rest-frame UV and optical to inform a direct calibration of the
relative opacity of dust grains, from the UV to the FIR, as they
evolve to high redshifts.  This relative opacity is captured in our
dataset by the ratio of dust mass absorption coefficients in the UV to
the FIR, \kratio; this quantity traces the integrated effects of the
grain size distribution (relative mass ratio of small to large
grains), the dust composition, and the morphology of dust grains.  As
a reminder, it is not a true measure of dust microphysics, given that
it is a galaxy-integrated quantity.  Equation~\ref{eq:kratio} sets the
measurement of \kratio\ that we use, set by a measurement of the
relative luminosities emitted in the UV and IR, as well as shape of
the IR SED and galaxy size.  The nominal value of the ratio for Milky
Way dust is $\approx10^{5}$ \citep{draine07a}.

Figure~\ref{fig:kratio} shows the ratio's stellar mass dependence and
evolution.  The ratio clearly falls by over {\it an order of
  magnitude} from $0<z<7$ across all stellar masses.  High stellar
mass galaxies have low \kratio\ at all redshifts beyond $z>2$.  From
$1<z<4$ there is substantial stellar mass dependence.  What physical
phenomena explain this complex evolution in \kratio?

The \citet{draine03a} grain model indicates that $\kappa_{\rm UV}$ is
dominated by the small grain population, including PAHs and very small
carbonaceous and silicate grains.  On the other hand, $\kappa_{\rm
  FIR}$ depends on the total grain cross-section dominated by the
larger grain population.  The factor of 10 drop in \kratio\ from
$0<z\lesssim4$ signals a significant depletion of the small grain
population at high redshifts relative to larger grains.  This may be
expected for a number of reasons.  First, small grains tend to be
destroyed in intense radiation fields
\citep{engelbracht05a,calzetti07a}; galaxies are more compact and have
higher SFRs at high redshifts, driven by their higher gas fractions
\citep{tacconi18a}.  Intense radiation fields leads to sputtering,
Coulomb destruction, and photo-destruction of PAHs and other very
small grains \citep{draine07a,remy-ruyer14a}. Second, high redshift
galaxies' ISMs are generally more metal-poor at fixed mass
\citep{maiolino19a}; at lower metallicities, the dominant dust
production channels shift \citep{galliano18a}.  AGB stars -- while
efficient producers of small carbonaceous grains -- have not had time
to contribute significantly to the ISM at $z>4$ \citep{schneider24a}
in galaxies that are metal poor and did not have significant star
formation at early times.  Instead, dust production may be dominated
by core-collapse supernovae and ISM grain growth, which is thought to
preferentially produce larger grains
\citep{asano13a,asano13b,zhukovska16a}.  There is additional
complexity introduced by the concept of shattering and coagulation.
Grain-grain collisions (shattering) in turbulent media can replenish
the small grain population \citep{hirashita09a,jones17a}, but at
sufficient densities, dust coagulation may counterbalance shattering
by turning smaller grains into larger grains, shifting the balance of
\kratio.  Last, aside from intense star formation, AGN can also
produce the radiation fields that lead to the destruction of small
grains \citep{laor93a,maiolino01a,maiolino01b} further complicating
the physical drivers of an evolving \kratio.

\begin{figure*}
  \centering
  \includegraphics[width=1.6\columnwidth]{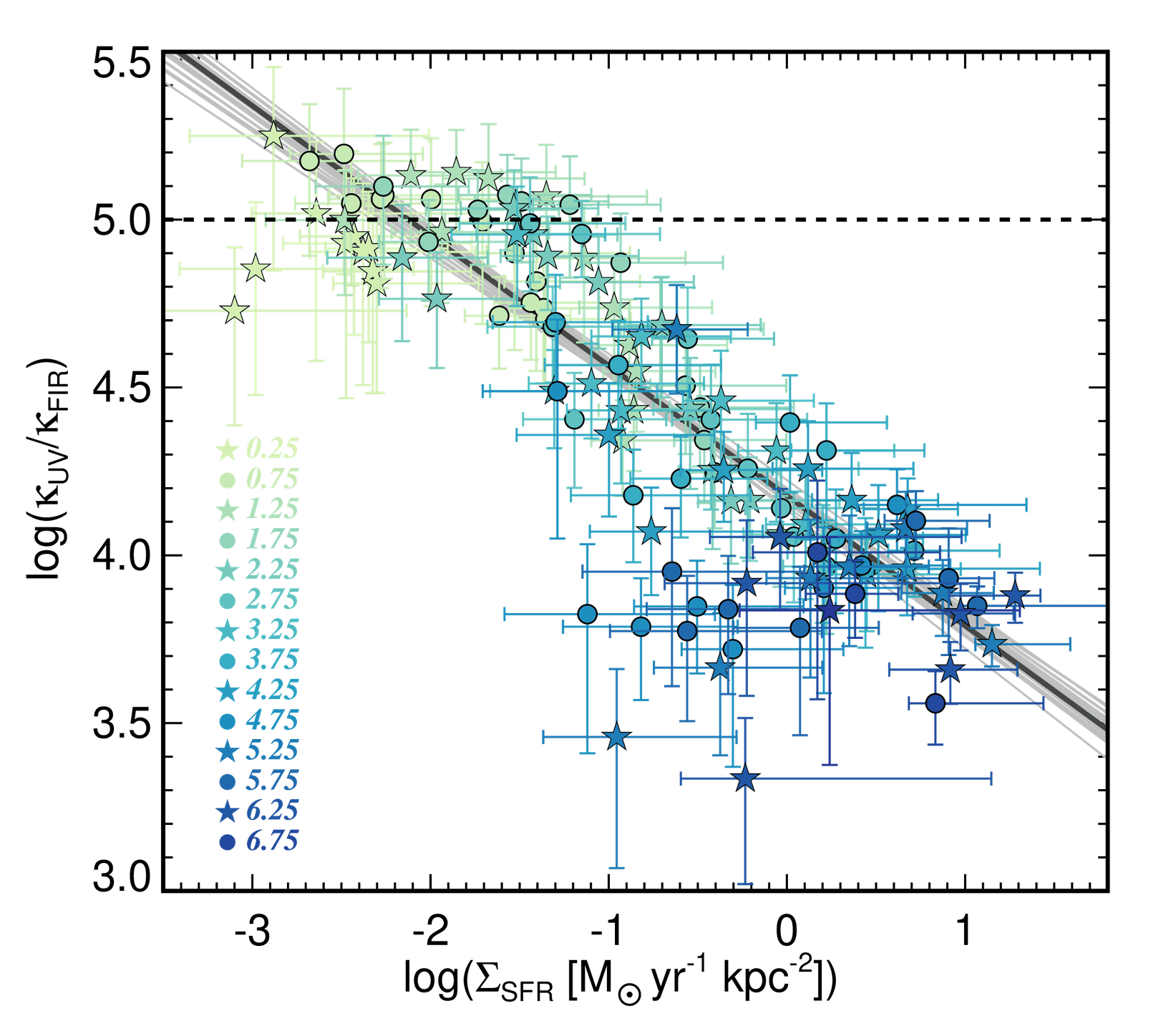}
  \caption{The relationship between SFR surface density and
    \kratio\ for all redshift and stellar mass bins shown on
    Figure~\ref{fig:kratio}.  The horizontal line marks the \kratio\ of
    the Milky Way.  Here we find evidence that the underlying physical
    driver of variation in \kratio\ may be the SFR surface density.
    In the most intense radiation fields where $\Sigma_{\rm SFR}$ is
    high, our results are consistent with a depletion of small grains
    relative to the large dust grain population.  The black line
    indicates the best-fit relation as given in
    Eq.~\ref{eq:kratiosfrd} with gray lines representing 25 random
    draws from the fit's posterior distribution; scatter about the
    relation is 0.69\,dex.}
  \label{fig:kratiosfrd}
\end{figure*}

Because we see a distinctly different behavior in the evolution of
\kratio\ in different mass bins, and because an underlying physical
hypothesis as to the driver of its evolution is the intensity of the
stellar radiation field, we plot \kratio\ against $\Sigma_{\rm SFR}$,
the star-formation rate surface density, in
Figure~\ref{fig:kratiosfrd}.  What we see is clear: that the more
intense radiation fields that are thought to accompany high SFR
surface densities have a much lower \kratio, consistent with a
depletion of small dust grains.  We fit the relationship between
$\Sigma_{\rm SFR}$ and \kratio\ as:
\begin{equation}
  \log\bigg(\frac{\kappa_{\rm UV}}{\kappa_{\rm FIR}}\bigg) = (10.79\pm0.50) - (2.58\pm0.11) \log(\Sigma_{\rm SFR})
  \label{eq:kratiosfrd}
\end{equation}
where the fit has an overall scatter of 0.69\,dex.  This fit is
summarized in our appendix, Table~\ref{tab:allrelations}.

Complementary work on cosmological simulations
\citep[e.g.][]{aoyama17a,hou19a} suggests a similar trend, that a
deficit of small grains at high-$z$ produces flatter (grayer)
intrinsic extinction curves \citep{narayanan26a}.  And while, in this
work, we cannot disentangle the relative impact of grain size
distribution and geometry on the attenuation curve, we do note that
the suggested depletion of small grains at high-$z$ would result in
both a grayer attenuation curve (higher $\delta$ at higher-$z$) and
decrease in \kratio.  Similarly, $\Sigma_{\rm SFR}$ is found to drive
significant variation in line-of-sight attenuation
\citep{sommovigo25a}.

Indeed, the evolution in \kratio, and the evolution in $C_\lambda$
(the scaling between A$_\lambda$ and \sdust) and $\delta$ (slope of
the attenuation law) are all intertwined.  Both $C_{\lambda}$ and
$\delta$ are dependent on {\it both} star/dust geometry and inherent
dust properties, while \kratio\ only informs the latter.  Empirically,
we have measured steep evolution in \kratio\ and $C_\lambda$ (both
changing by over an order of magnitude across $0<z<7$), while the
evolution in $\delta$ is more subtle.  The fact that \cuv\ and
\cv\ evolve in very similar ways to one another relates closely to the
subtle evolution of $\delta$ (a steeper evolution in $\delta$ would
mean that \cuv\ and \cv\ evolve very differently from one another).
It is difficult to take the interpretation of the magnitude of
evolution of these measurements -- i.e. steep evolution in \cuv\ and
\kratio\ relative to $\delta$ -- further without making some
significant assumptions (on the relative ratio of geometries between
UV and V, and between the unanchored evolution of $\kappa_{\rm
  UV}/\kappa_{\rm V}$).  Another cautionary note is that we
fundamentally restricted the dynamic range of $\delta$ to
$-0.5<\delta<0.1$ to avoid overfitting the UV/optical portion of
galaxies' SEDs.  Future work with large spectroscopic samples may be
able to disentangle the relationship between $\delta$ and fundamental
dust properties, as well as geometry, further with more precise
constraints on hand.

\subsection{What drives dust temperature?}\label{sec:tdustdiscuss}

The physical drivers of a galaxies' average dust temperature have been
extensively discussed in the literature
\citep[e.g.][]{chapman03b,symeonidis13a,magnelli14a,simpson17a} with
many works, especially those focusing on nearby galaxies where SEDs
are often more extensively studied
\citep{lehnert96a,chanial07a,lutz16a,diaz-santos10a} drawing the
conclusion that star formation surface density, or IR luminosity
surface density, $\Sigma_{\rm IR}$, correlates most tightly with
\tdust.  This conclusion is also drawn in \citet{burnham21a} for
high-$z$ galaxies studied with high-resolution ALMA imaging and from
radiative transfer simulations \citep{parente25a}.  The framework from
\citeauthor{burnham21a} is useful for parameterizing a predictive
behavior of \tdust\ as a function of both redshift and stellar mass.
While \citet{drew22a} recognized that galaxies' dust temperatures at
fixed stellar masses would evolve primarily between $0<z<2$ from the
evolution of the SFR-\mstar\ relation alone, it did not account for
the potential increase in temperature due to galaxies' smaller sizes
and thus increase in $\Sigma_{\rm IR}$.  To accurately model the
evolution of $\Sigma_{\rm IR}$ and its impact on \tdust, the size
evolution of galaxies forms a critical piece.  Following the work of
\citet{yang25a} using the COSMOS-Web data, we fit a global average
relation to the size-mass evolution of galaxies' half light radii in
the stacked sample we use in the analysis:
\begin{equation}
R_{\rm e}(M_\star,z) = C M_\star^\alpha (1+z)^{-\beta}.
\label{eq:sizeevol}
\end{equation}
We measure $C=0.0178^{+0.0018}_{-0.0011}$, $\alpha=0.287\pm0.004$, and
$\beta=1.23\pm0.02$.  We note that \citeauthor{yang25a} measure
$\alpha=0.20$ and $\beta=1.21$ which roughly agrees with our fit to
the `stackable' sample here.

From a galaxy's stellar mass and redshift, one can estimate the star
formation rate using the \citet{speagle14a} parameterization.
\citet{burnham21a} presents, in their Eq.~7, a tight empirical
relationship between $\lambda_{\rm peak}$ ($\propto$\,\tdust$^{-1}$)
and $\Sigma_{\rm IR}$.  The translation of that relation using SFR and
$R_{\rm e}$ directly is:
\begin{equation}
\begin{aligned}
\log(\lambda_{\rm peak}) ={}&
\frac{1}{(3.80 \pm 0.16)}
\Big[
(28.61 \pm 0.33) \\
&\quad - \log(\mathrm{SFR}) + \log\!\left(\pi R_{\rm e}^2\right)
\Big]
\end{aligned}
\label{eq:sfrtdust}
\end{equation}
With SFR in \sfr\ and $R_{\rm e}$ in kpc. The translation between
$\log(\lambda_{\rm peak})$ and \tdust\ for the type of SED
fitting used in this work is given in Eq~\ref{eq:tdustlpeak}.

We refer to Equation~\ref{eq:sfrtdust}, or the expected relationship
between SFR surface density and $\lambda_{\rm peak}$, as the
``Stefan-Boltzmann'' expectation.  This is because the backbone of
this inference comes from an application of the Stefan-Boltzmann law
to galaxy scales (see the \citealt{burnham21a} discussion \S~5.1 and
measurement in 5.2.1).  To summarize, Stefan-Boltzmann informs the
relationship between the emergent luminosity and temperature from an
optically thick body; galaxies are likely optically thick in some
regimes but certainly not others.  In the rest-frame millimeter, we
explicitly presume they are optically thin (allowing the calibration
of the dust masses and dust mass surface densities), but the dominant
IR emission is at significantly shorter wavelengths, at rest-frame
$\approx$100\,\um.  Despite complex geometry, by making a simplified
assumption (that galaxies are optically thick disks rather than
spheroids), Equation~\ref{eq:sfrtdust} emerges.  Does it hold with our
data?

\begin{figure}
\centering
\includegraphics[width=0.99\columnwidth]{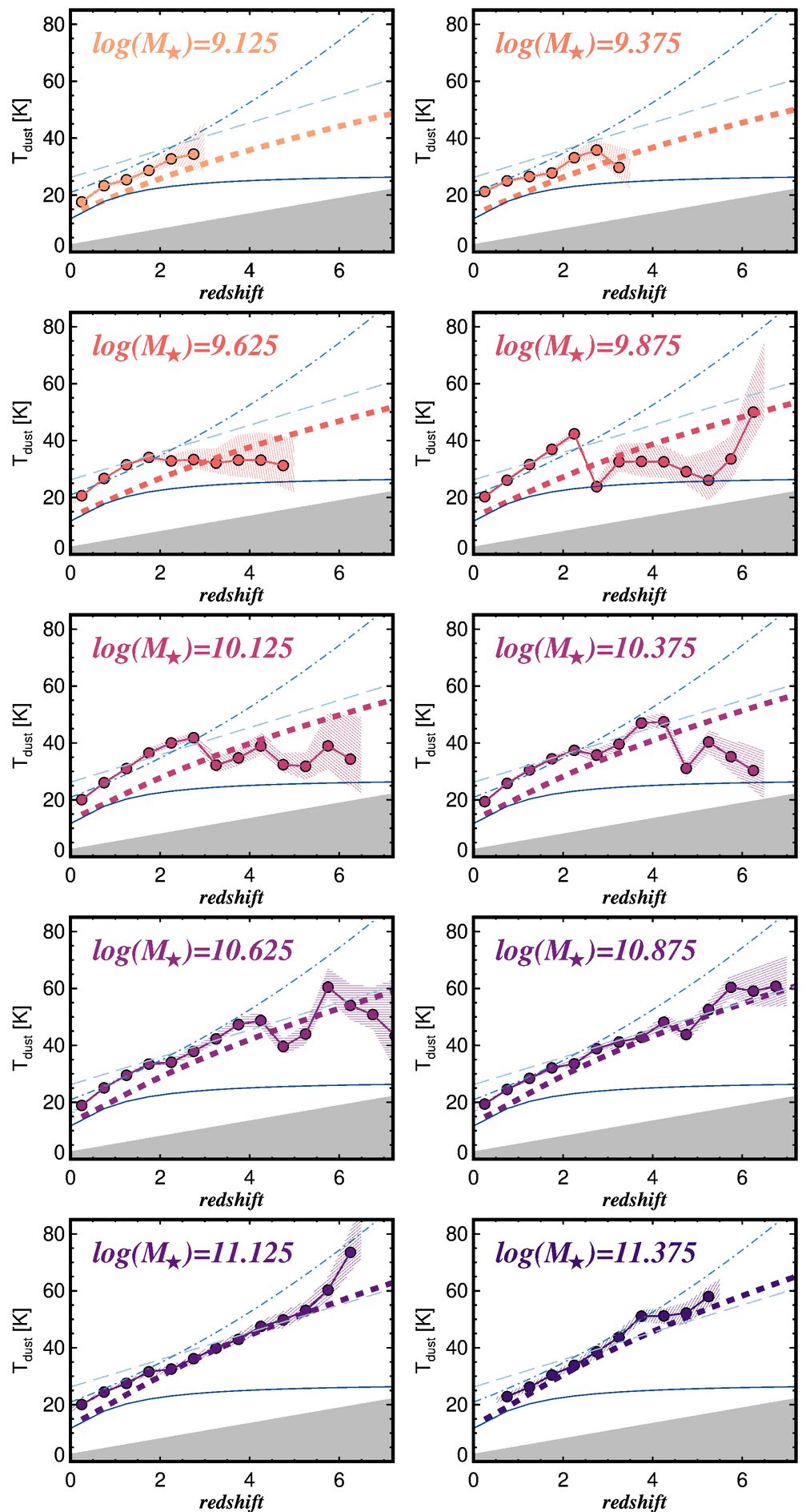}
\caption{The evolution of dust temperature, broken into different
  stellar mass bins, compared against expectation for an
  optically thick-through-100\um\ disk with SFR and size determined by
  evolution in the SFR-\mstar\ and the size-mass relationships (dashed
  lines matched in color to the data). The expected temperature given
  an SFR and size is given by Equation~\ref{eq:sfrtdust}. The model
  fits the data well at high stellar masses
  (\mstar\,$>10^{10.5}$\,\msun), suggesting the optically thick
  assumption may be an appropriate approximation in that mass regime
  across all redshifts. At more modest masses, temperatures are flat
  beyond $z>2$, perhaps indicating that the model is less appropriate
  in that mass regime. The three blue lines are the same literature
  comparisons shown in Figure~\ref{fig:tdust}.}
\label{fig:tdust2}
\end{figure}

Figure~\ref{fig:tdust2} compares measurements to this Stefan-Boltzmann
expectation at fixed stellar mass intervals.  All mass bins show
increasing temperatures over the $0<z<2$ mass range, consistent with
the dominant evolution in the SFRs of galaxies on the main sequence
between those epochs.  At higher redshifts, we note that the high mass
bins ($>10^{10.5}$\,\msun) show remarkably good agreement with the
Stefan-Boltzmann model; those bins predict the steepest evolution in
dust temperature due to their relatively compact sizes and high SFRs.
The intermediate mass bins ($\approx10^{9.5-10.5}$\,\msun) are
slightly discrepant from the model, with some indication of flatter
temperature evolution than expected beyond $z>2$.  That may indicate
that the ISM is less optically thick near the $\approx100$\,\um\ peak.
The lowest mass bins do not have sufficient constraints at high
redshifts to draw conclusions.

We note that \citet{dudzeviciute21a} compare dust-mass matched samples
at $z=1-2$ and $z=3-4$ and find dust continuum sizes roughly half as
large at higher redshifts, which is consistent with increased star
formation surface densities and higher dust temperatures.

\subsection{Dust-to-Stellar Ratio in the context of evolving metallicity and gas fractions}\label{sec:dts_discuss}

In Figure~\ref{fig:mdustmstar} we have measured the evolution and mass
dependence of the dust-to-stellar ratio finding a
$\propto(1+z)^{1.36}$ evolution and $\propto M_\star^{-0.3}$ stellar
mass dependence.  How does this measured DTS evolution stack up
against other literature measurements?

\begin{figure*}
  \centering
  \includegraphics[width=1.8\columnwidth]{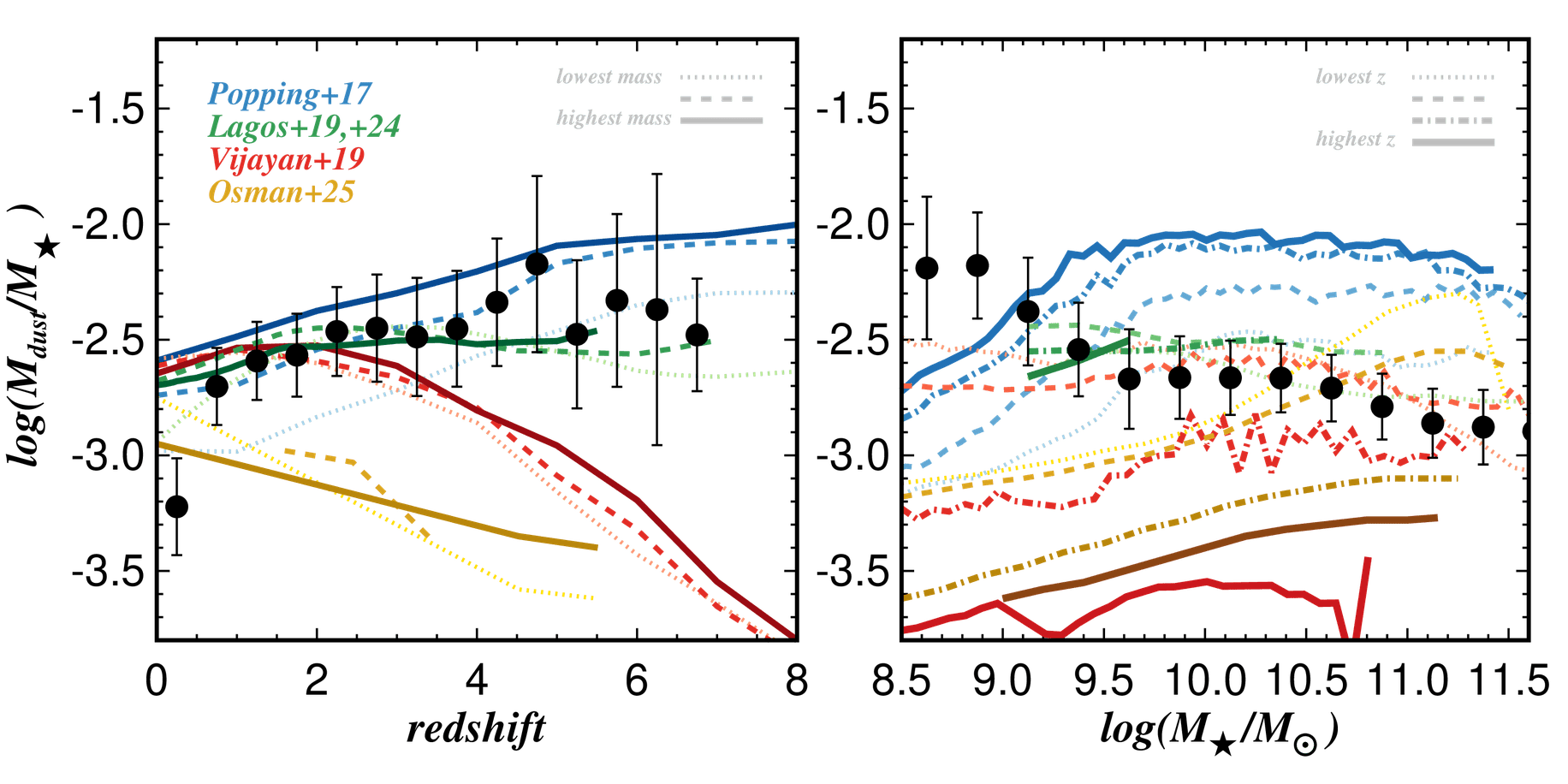}
  \caption{A comparison of the averaged binned dependence of the DTS
    from our data (black points, taken from
    Figure~\ref{fig:mdustmstar}) against expectations from
    semi-analytic models with dust prescriptions.  Blue lines show
    expectation from the Santa Cruz model \citet{popping17a}; mass
    bins taken at $10^{9.125}$\,\msun, $10^{9.675}$\,\msun, and
    $10^{10.125}$\,\msun\ and redshift bins at $z=1$, $z=3$, $z=5$,
    and $z=7$. Green lines are from {\sc SHARK}
    \citep{lagos19a,lagos24a} with the same mass and redshift
    bins. Red lines are from {\sc L-Galaxies} \citep{vijayan19a}; mass
    bins are $10^{9}$\,\msun, $10^{9.5}$\,\msun, and $10^{10}$\,\msun
    and the same redshift bins.  Yellow/golden lines are from {\sc
      Dusty-Gaea} \citep{osman25a}, with the same mass bins as {\sc
      L-Galaxies} but shown at $z=1.6$, $z=2.45$, $z=3.48$ and
    $z=5.51$.  On the left panel, darker colors and
      progressively more solid lines correspond to higher stellar
    mass bins.  On the right panel, darker colors and
      progressively more solid lines correspond to higher redshift
    bins.}
  \label{fig:dts_models}
\end{figure*}

Theoretical predictions for the DTS are largely inferred through
semi-analytic models that employ scaling relations determined using
hydrodynamic simulations on smaller cosmological volumes, with
radiative transfer applied. Our work here does not perform detailed
comparisons with simulations predictions (given the scope required for
such a comparison), though we do note the general trends expected for
the DTS in a suite of semi-analytic models including the Santa Cruz
model \citep{popping17a}, {\sc SHARK} \citep{lagos19a,lagos24a}, {\sc
  L-Galaxies} \citep{vijayan19a}, and {\sc Gaea} \citep{osman25a}.
The mean model expectations are shown in Figure~\ref{fig:dts_models}
relative to the averaged binned relations in our work.
\citet{popping17a} generally observe a DTS that falls with cosmic
time, in line with our observations, though their stellar mass
dependence is substantially flatter than what our data suggest.  {\sc
  SHARK} \citep{lagos18a,lagos19a,lagos24a} shows similar trends as
the Santa Cruz model, though with generally better agreement with our
measurements over the mass and redshift ranges where there is
constraining power.  The {\sc L-Galaxies} model \citep{vijayan19a}
predicts a trend in the opposite direction, such that higher redshift
galaxies have lower DTS ratios, discrepant with our data, though the
stellar mass dependence in {\sc L-Galaxies}, like SHARK, reflects what
we see: higher mass galaxies have lower DTS ratios.  {\sc Dusty-Gaea}
\citep{osman25a} show lower DTS ratios at higher redshifts and higher
DTS ratios with increased mass.  From existing models, it seems clear
there is little convergence on the behavior of the DTS with stellar
mass and redshift.

Observational measurements of the DTS at high-$z$ are somewhat limited,
though \citet{bethermin15b}, \citet{magnelli20a}, \citet{donevski20a},
and recently \citet{jolly25a} and \citet{eales26a} offer direct
measurements for high-redshift samples.  The \citeauthor{bethermin15b}
is analogous to our measurements here, in that they stack massive
galaxies in the \citet{ilbert13a} iteration of the COSMOS catalog and
fit full dust SEDs (a notably shallower catalog than used herein,
though the best available at the time).  Their analysis is limited to
$z<4$.  While they did not have the resolution to infer a stellar mass
dependence of the DTS, they do see some redshift evolution
$\propto(1+z)^{0.4-0.7}$, shallower than seen in our dataset.  This
could be driven by the limitation of their analysis to the most
massive stellar reservoirs.

The \citeauthor{magnelli20a} work does stack galaxies by stellar mass
and redshift in ASPECS 1.2\,mm maps, and infers dust mass with adopted
$\kappa_{\rm FIR}$, as we do in this work (though we stack multiple
FIR datasets).  They anchor the DTS measurement in an integrated sense
by comparing the dust mass density to the stellar mass function
\citep{mortlock15a} and constrain it to $z\sim3$ as having a steeper
redshift dependence than we find here, $\propto (1+z)^{2.6}$ and
shallow stellar mass dependence, $\propto M_\star^{0.1}$.  Broadly
speaking, our results are consistent in order of magnitude: most of
the change in the DTS is in its redshift evolution, with a more minor
variation dependent on the underlying galaxy mass.

The \citeauthor{donevski20a} work presents the DTS for galaxies
directly detected in the IR; they find a much shallower redshift
evolution in this sample $\propto (1+z)^{0.5}$ with a stellar mass
dependence $\propto M_\star^{-(0.2-0.6)}$.  While our stacking results
indicate steeper redshift evolution of the DTS, we are consistent with
\citeauthor{donevski20a} in its finding of a weak, but negative,
stellar mass dependence.

\citet{jolly25a} stacks 1.2\,mm continuum data from the ALMA Lensing
Cluster Survey and find mild evidence for lower dust-to-stellar mass
ratios at higher redshifts than at $z\sim1$; their assumptions
regarding $\kappa_{\rm FIR}^{\rm fix}$ are different than our work by
a factor of two and high-$z$ samples are limited by small sample
sizes; the work does see lower DTS ratios in higher mass galaxies.

\citet{euclid-collaboration26a} stack {\it Euclid}-detected
main-sequence galaxies on both {\it Herschel} and SCUBA-2 imaging,
fitting modified blackbodies in a similar fashion as our analysis,
stepping out to $z=3$.  They find higher DTS ratios at higher
redshifts, with a flattening past $z>1$ with lower DTS ratios at
higher stellar masses, which mirrors our data directly as shown in
Figure~\ref{fig:mdustmstar_lit}.

Finally, recent work in \citet{eales26a} presents a similar stacking
analysis as our work, using the COSMOS2025 catalog to stack the
SCUBA-2 imaging and infer dust mass evolution with stellar mass and
Hubble Sequence. Even stacking on a single band and varying their
assumptions with respect to dust temperatures, \citet{eales26a} find
an evolution of the DTS that is consistent with what we find:
increasing DTS at higher redshift and lower stellar masses.

These literature sources set the stage for contextualizing the (lack
of) precision with which the DTS is understood.  It is helpful to
pedagogically break the DTS into its components:
\begin{align}
  \frac{M_{\rm dust}}{M_\star} &= \left[\frac{M_{\rm dust}}{M_{\rm gas}}\right]\left[\frac{M_{\rm gas}}{M_\star}\right] = \left[\frac{M_{\rm dust}}{M_{\rm gas}}\right]\mu_{\rm gas} \nonumber \\
  {\rm DTS} &= {\rm DTG(Z(M_\star,z))} \times {\rm GTS} 
  \label{eq:dts2}
\end{align}
In other words, the DTS ratio can be expressed as the product of the
dust-to-gas (DTG) ratio and the gas-to-stellar (GTS) ratio, or
$\mu_{\rm gas}$.  Does our earlier fit to the DTS ratio agree with
literature measurements of the DTG ratio and $\mu_{\rm gas}$?

Our understanding of the dust-to-gas and its metallicity dependence is
anchored to works like \citet{remy-ruyer14a} and \citet{galliano21a}.
The DTG ratio is substantially lower at lower metallicity.  Further,
the relationship between galaxies' metallicities and their stellar
masses is extensively studied
\citep{maiolino08a,mannucci10a,zahid13a,zahid14a,maiolino19a,curti20a,sanders21a,sanders23a,jain26a}.
While some work argue for a fundamental metallicity relationship,
where $Z$ is exclusively determined from \mstar\ and SFR alone, we
instead adopt an evolving mass-metallicity relationship (MZR) in the
form of Eq.~2 of \citet{curti20a} with redshift evolution described by
Eq.~4 of \citet{jain26a}.

\begin{figure}
  \includegraphics[width=0.99\columnwidth]{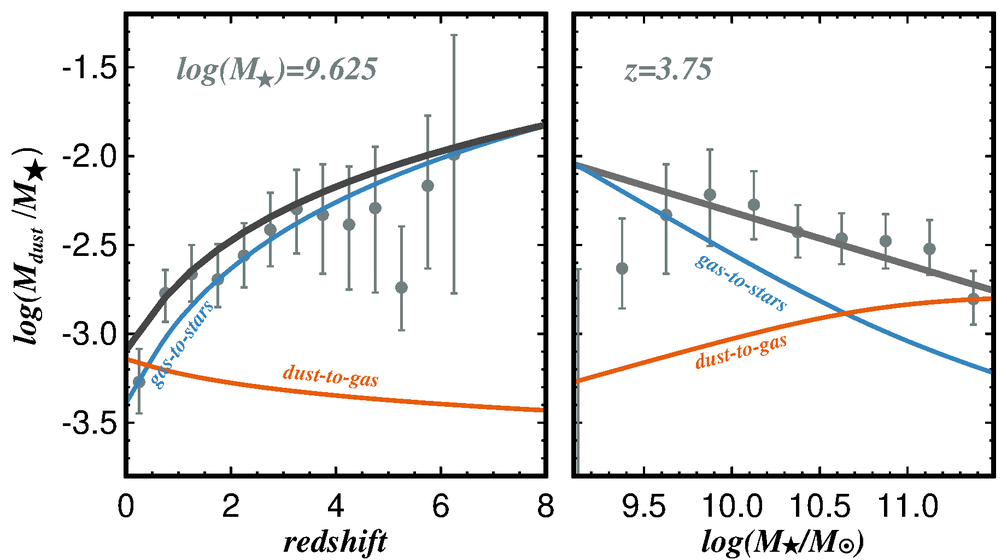}
  \caption{An illustration of the estimated relative contributions to
    the evolution of the dust-to-stellar (DTS) mass ratio.  Here we
    show a single mass bin and its evolution in redshift
    ($\log(M_\star)=9.625$, left) and a single redshift bin and the
    stellar mass dependence of the DTS ratio ($z=3.75$, right).  By
    anchoring to literature measurements of the mass metallicity
    relation \citep{zahid14a,jain26a} and the metallicity dependence
    of the dust-to-gas ratio \citep{remy-ruyer14a}, we infer the
    evolutionary shape and mass dependence of the dust-to-gas (DTG)
    ratio (orange curves).  It deviates substantially from the
    evolution seen in the DTS ratio (gray curves, Eq~\ref{eq:dts1}).
    The difference can be accounted for by the steep evolution (and
    mass dependence) of the gas-to-stellar ratio (GTS; blue curves).
    While \citet{tacconi18a} and \citet{liu19a} measure an even
    steeper evolution in the GTS ($\propto(1+z)^{2.5}$ and
    $\propto(1+z)^{2.9}$ vs. our inferred $\propto(1+z)^{1.6}$), there
    is sufficient uncertainty in the possible evolution of
    $\kappa_{\rm FIR}$ to make up the difference.}
  \label{fig:cartoon_dts}
\end{figure}

Combining the MZR with the GTS ratio in \citet{remy-ruyer14a}, then
gives us a concrete prediction of the evolution in the DTG ratio with
redshift and stellar mass.  We show the implied DTG evolution,
renormalized to the measured DTS ratio, in
Figure~\ref{fig:cartoon_dts} and parameterize it as a 2$^{nd}$ order
polynomial of the form
\begin{equation}
\log(\mathrm{DTG}) = A + b\,\Delta M + c\,\Delta M^2
    + d\,\ell_z + e\,\Delta M\,\ell_z + f\,\ell_z^2
\label{eq:dtg}
\end{equation}
Where $\Delta M \equiv \log(M_\star/10^{10}\,M_\odot)$ and $\ell_z
\equiv \log(1+z)$.  We fit the coefficients such that
$(A,\,b,\,c,\,d,\,e,\,f) =
(-2.191,\,+0.143,\,-0.056,\,-0.178,\,+0.135,\,-0.069)$.  The DTG ratio
increases for higher stellar masses (due to their higher
metallicities) but decreases at high redshift (due to their higher
star formation rates and overall lower metallicities).

Given the expectation for the evolution of the DTG ratio in
Eq.~\ref{eq:dtg}, and the earlier fit to the DTS ratio in
Eq.~\ref{eq:dts1}, we explicitly predict that the gas-to-stellar ratio
would evolve as $\mu_{\rm gas}\propto M_\star^{-0.49} (1+z)^{1.57}$.
In other words, this would imply that higher gas-to-stellar ratios
exist at higher redshifts but that at fixed redshift, higher mass
systems have lower relative gas-to-stellar ratios.  This implied GTS
evolution is also shown in Figure~\ref{fig:cartoon_dts}; again, it
represents the difference in evolution between the measured DTS and
the implied DTG given our understanding of metallicity evolution and
dependence of DTG on metallicity.

We can then compare this directly to measurements of $\mu_{\rm gas}$
from \citet{tacconi18a} and \citet{liu19a} and reflect on
discrepancies.  Both \citeauthor{tacconi18a} and \citeauthor{liu19a}
find a steeper redshift dependence and shallower mass dependence than
is implied in our data, such that $\mu_{\rm
  gas}\propto(1+z)^{2.5}M_\star^{-0.36}$ and $\mu_{\rm
  gas}\propto(1+z)^{2.9}M_\star^{-0.69}$, respectively.  What causes
this discrepancy?  There may be wiggle room in this comparison that
alleviates sufficient tension, or it may hint at some breakdown of
assumptions.

A major assumption on our part in the measurement of the DTS is that
$\kappa_{\rm FIR}$ {\it does not evolve}; all dust masses in this work
use a fixed value.  That would place all of the evolution we observe
in \kratio\ on the evolution of $\kappa_{\rm UV}$ and the lack of
small grains, with little to no change in the large grain population
to which $\kappa_{\rm FIR}$ is most sensitive.  If, instead, there is
some redshift evolution in $\kappa_{\rm FIR}$ due to different
characteristics of the large grain population, that would change the
redshift evolution of the DTS measured in Eq.~\ref{eq:dts1}.
Importantly, simulations suggest that the dust-to-stellar ratio is
also directly related to the dust size distribution and thus
$\kappa_{\rm FIR}$: SN-produced dust would preferentially form larger
grains dictating lower DTS ratios \citep{schneider24a}.
An evolving $\kappa_{\rm FIR}$ would not only impact the measured DTS,
but indeed it would also percolate to an evolution in DTG, because the
DTG-$Z$ relationship derived in \citet{remy-ruyer14a} would have an
additional redshift dependence with a variable $\kappa_{\rm
  FIR}$. Furthermore, a changing $\kappa_{\rm FIR}$ would {\it also}
change the GTS ratio; this is not because gas or stars are directly
linked to $\kappa_{\rm FIR}$, but because many of the gas measurements
used to anchor measurements of GTS rely on dust masses and a presumed
DTG ratio!

Beyond this concern regarding the evolution of $\kappa_{\rm FIR}$ is
the additional dependence of dust mass on dust temperature where, for
example, \citet{tacconi18a} presume a mass-weighted dust temperature
of 25\,K; if that differs substantially at higher redshifts where dust
temperatures may be higher, then dust masses could be lower than
currently inferred, drawing down the GTS ratio.

The complex relationship between $\kappa_{\rm FIR}$ and the evolution
of DTS, as well as DTG and GTS is beyond the scope of what our data
can constrain directly, but with more detailed sets of observations to
calibrate these relationships at high-$z$, we may learn quite a bit
about the relative role that dust and gas play in shaping our
interpretation of the growth of galaxies alongside their well-studied
stellar components.

\section{Conclusions}\label{sec:conclusions}

In this work we have presented a detailed comparison of galaxies'
attenuation characteristics -- inferred from the rest-frame UV/optical
-- to their average dust emission characteristics -- inferred from the
rest-frame FIR.  We take $>$500,000 galaxies from the latest JWST
catalog in the COSMOS field and pair that catalog with some of the
deepest, wide-field submm extragalactic maps to reach sensitivities
down to stellar masses $\sim10^{9}$\,\msun\ and redshifts $z\sim7$ in
a stacking analysis.  Our main conclusions are as follows:
\begin{enumerate}
\item {\bf Attenuation inferred from the UV/optical is a strong
  function of mass and falls slightly at increased redshift
  (\S~\ref{sec:auvmstar}, Eq.~\ref{eq:avmstar},
  Figs.~\ref{fig:avmass1}, \ref{fig:avmass2}).}  Specifically the UV
  attenuation to stellar mass relation (\auv-\mstar) shows mild
  redshift evolution out to $z\sim6$ with decreasing \auv\ (and \av)
  towards high-$z$, followed by a reversal and upturn beyond $z\sim6$
  whose origin is uncertain.
\item {\bf Attenuation curves are grayer at higher redshifts
  (\S~\ref{sec:flexiblecurve}, Eq.~\ref{eq:z_delta},
  Fig.~\ref{fig:z_delta}).} The slope of the attenuation curve, as
  inferred from UV/optical tracers, evolves with redshift, steepening
  monotonically toward lower redshift with the slope evolving as
  $\langle \delta(z)\rangle = -0.36+0.03z$, transitioning from
  near-Calzetti at high-$z$ to SMC-like slopes locally.  This points
  to a deficit of small grains in the early Universe.
\item {\bf Dust temperatures rise steeply with redshift, tracking
  expectations from a toy Stefan-Boltzmann model at
  $>10^{10}$\,\msun\ (\S~\ref{sec:tdust}, \ref{sec:tdustdiscuss},
  Eq~\ref{eq:tdustlpeak}, \ref{eq:sfrtdust}
  Fig.~\ref{fig:tdust},~\ref{fig:tdust2}).}  We postulate this is
  driven by the evolution of star formation surface densities.  Lower
  mass galaxies show less evolution in dust temperature beyond $z>2$
  than high mass galaxies, possibly hinting that FIR SEDs for
  lower-mass systems are less optically thick around their peak at
  rest-frame 100\,\um.
\item {\bf Star formation is overwhelmingly obscured ($>90\%$) for
  galaxies above $z>0.5$ and stellar masses
  $>10^{10}$\,\msun\ (\S~\ref{sec:funobsmstar}, Eq.~\ref{eq:funobs}
  Fig.~\ref{fig:funobs}).}  Higher mass galaxies are {\it more}
  obscured at increasing redshifts, with unobscured SFR fractions
  $\lesssim$1\%\ at $>10^{10.5}$\,\msun.  Galaxies with stellar masses
  $\sim10^{9.5}$\,\msun\ are $\sim$10\%\ unobscured at all redshifts
  $z>1$.
\item {\bf UV/optical SED fitting systematically underestimates
  attenuation and obscured star formation by a factor of
  $\boldmath{\sim3\times}$ at $\boldmath{z>0.5}$
  (\S~\ref{sec:auvcompare}, Fig.~\ref{fig:auvcompare})} for all mass
  bins, and by up to an order of magnitude for the most massive
  galaxies at high redshifts.   Put another way, energy
    balance fits driven by OIR constraints will underestimate the dust
    SED, and nearly all galaxies are effectively optically thick to
  rest-frame UV light, meaning UV diagnostics alone give a
  fundamentally incomplete picture of star formation and the ISM.
\item {\bf The relationship between dust mass surface density and
  UV/optical attenuation evolves by over an order of magnitude between
  $\boldmath{0<z<7}$ (\S~\ref{sec:avsdust},
  Figs.~\ref{fig:avsigmadust}, \ref{fig:cevol}).} The
  \auv-\sdust\ (\av-\sdust) relation is parameterized via the
  evolution of \cuv\ (\cv), a coefficient relating the two.  The
  strong evolution seen in \cuv\ and \cv\ is attributable to both an
  evolving star/dust geometry and dust grain size distribution and
  composition.  By combining constraints that relate \auv\ to
  \lir\ (independent of dust properties) and $\Sigma_{\rm dust}$-\auv,
  we loosely constrain the evolution in dust geometry,
  $\mathcal{G_{\rm UV}}$.  We find that the impact of geometry is
  weakly evolving though largely inconsistent with a foreground dust
  screen across all masses and redshifts $z>0.5$.
\item {\bf Significant evolution in the grain size distribution is
  inferred via the measurement of \kratio\ (\S~\ref{sec:kratio},
  Eq~\ref{eq:kratiosfrd}
  Figs.~\ref{fig:kratio},~\ref{fig:kratiosfrd}).}  \kratio\ is
  measured independent of first-order geometric effects and is
  inferred to fall by over an order of magnitude from $z\sim0$ to
  $z\sim7$.  This is consistent with preferential destruction of small
  grains by intense radiation fields at early cosmic times and appears
  to be primarily driven by a redshift-invariant inverse relationship
  to SFR surface density.
\item {\bf The dust-to-stellar mass ratio (DTS) evolves as
  $\sim(1+z)^{1.36}$ with slight mass dependence $\sim M_\star^{-0.3}$,
  with the most rapid evolution seen between $z\sim0$ and $z\sim0.5$
  (\S~\ref{sec:DTS},~\ref{sec:dts_discuss}, Eq.~\ref{eq:dts2},
  Figs.~\ref{fig:mdustmstar_lit}, \ref{fig:mdustmstar},
  \ref{fig:cartoon_dts}).}  The evolution in the dust-to-stellar ratio
  is dominated by the steep redshift dependence of the gas-to-stellar
  mass ratio which counterbalances the shallow (and negative)
  evolution in the dust-to-gas ratio.  The mass dependence originates
  from the higher gas fractions of lower mass galaxies.
\end{enumerate}
We have found that the standard framework for inferring dust
properties from the rest-frame UV/optical -- attenuation curves, a
presumed foreground dust screen geometry, and fixed opacity -- is
insufficient across most masses and redshifts.  The combination of
JWST photometry and deep ALMA/(sub)mm stacking reveals that the ISM of
typical star-forming galaxies is moderately to significantly opaque
through the rest-frame UV across a wide range of masses and redshifts,
not just the extreme, dusty star-forming galaxy population as
previously thought.  This holds even with high quality constraints
from JWST.  As a result, FIR measurements remain irreplaceable for
accurate star formation rate budgets across cosmic time: UV/optical
SEDs are not simply noisy proxies for the total SFR, but
systematically biased ones whose systematics worsen in precisely the
mass and redshift regimes where the most star formation is occurring.

The measurement of an evolving \kratio, constrained independent of
first-order star/dust geometry effects, suggests a significant shift
in typical ISM dust grain properties in the early Universe relative to
now.  This may serve as direct observational evidence that small
grains are being preferentially destroyed in the intense radiation
fields of early star-forming galaxies, but we caution that we are
constraining the galaxy-integrated ratio, \kratio, not the UV and FIR
evolution of $\kappa$ independently, and certainly not on microscopic
scales; it is possible that, aside from small grain depletion, the
evolution of large grains and dust composition could impact this ratio
significantly.  The implications of such an evolution, whatever the
cause, are significant.  It implies that interpretation and
calibration of both dust attenuation inferences and dust mass
measurements may be moving targets.  Any high-$z$ work relying on
fixed dust prescriptions calibrated in the local universe should
proceed with caution.

\begin{acknowledgements}

   We thank the anonymous referee for helpful comments that improved
   the manuscript. CMC thanks the University of California Santa
   Barbara's Division of Mathematical, Life and Physical Sciences for
   support of this work, as well as the National Science Foundation
   for support through grants AST-2009577 and AST-2307006 and to NASA
   through grant JWST-GO-01727 awarded by the Space Telescope Science
   Institute, which is operated by the Association of Universities for
   Research in Astronomy, Inc., under NASA contract NAS 5-26555.
  HBA acknowledges support from the Harrington Graduate Fellowship at
  UT Austin as well as the National Science Foundation Graduate
  Research Fellowship.
  ET acknowledges support from the ANID CATA-BASAL program FB210003,
  and FONDECYT Regular 1241005 and 1250821.
  This project has received funding from the European Union's Horizon
2020 research and innovation program under the Marie
Sk\l{}odowska-Curie grant agreement No. 101148925.
MA is supported by FONDECYT grant number 1252054, and gratefully
acknowledges support from ANID Basal Project FB210003, ANID MILENIO
NCN2024\_112 and ANID + Vinculaci\'on Internacional + FOVI250261.
The NIKA2 N2CLS is based on observations carried out under project
192-16 with the IRAM 30m telescope, and follow-up projects W16EE,
E16AI, W21CV, W23CX, W23CJ, and S24CF with NOEMA.
Much of this work was conducted on land traditionally occupied by the
Chumash people in what is now known as the Santa Barbara area of
California.  We pay our respects to the Chumash elders, past, present,
and future, who call this place-—the land that the university sits
upon—-their home.

\end{acknowledgements}

\appendix

\section{Best-Fit Relations Derived in this Work}

Here we provide a summary of the fitted relations in this work and
show those fitted relations explicitly binned by stellar mass and
redshift.  These include attenuation as a function of redshift and
mass, both in the UV, \auv($z$,$M$), as well as optical, \av($z$,$M$);
the fraction of star formation that is unobscured, \funobs($z$,$M$);
the dust-to-stellar mass ratio, DTS($z$,$M$); the rest-frame peak
wavelength of the dust SED, \llp($z$,$M$) ($\lambda_{\rm peak}$ is
inversely proportional to temperature); and the relationship between
\kratio\ and $\Sigma_{\rm SFR}$.  Table~\ref{tab:allrelations}
presents a summary of the derived relationships and their best-fit
parameters.

\begin{figure*}
  \includegraphics[width=\columnwidth]{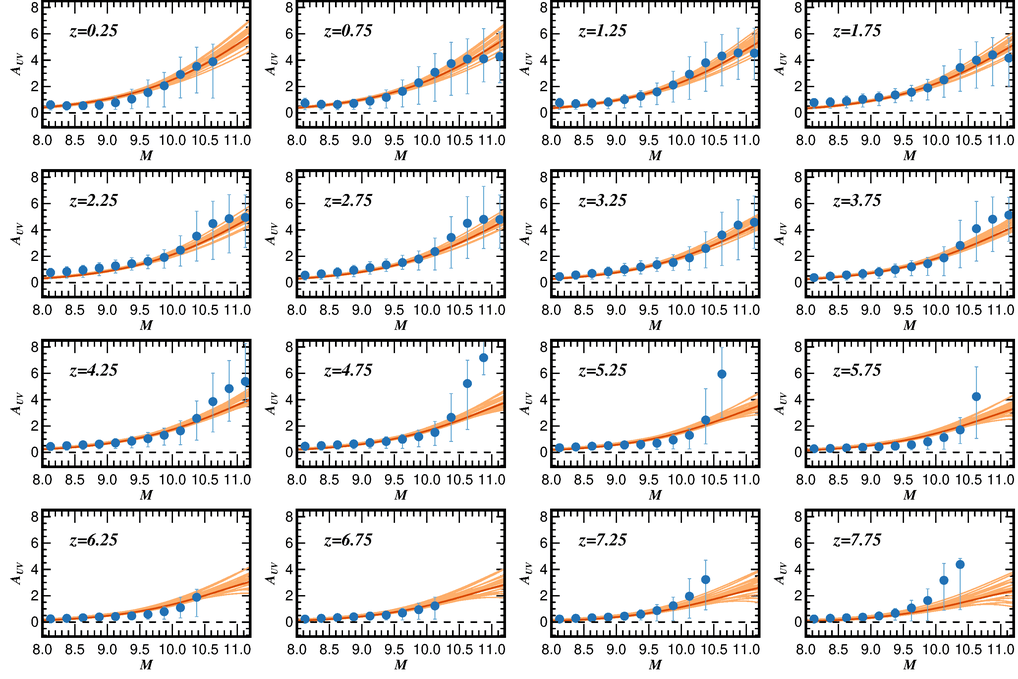}
  \caption{Redshift evolution of the \auv-\mstar\ relation as measured
    in our dataset (blue points with 68\%\ confidence intervals on the
    sample distribution).  Orange curves show the best-fit derived
    relation for \auv($z$,$M$) (Eq.~\ref{eq:avmstar}) fit jointly to
    all data points shown. }
  \label{fig:panels_auvmstar}
\end{figure*}

\begin{figure*}
  \includegraphics[width=\columnwidth]{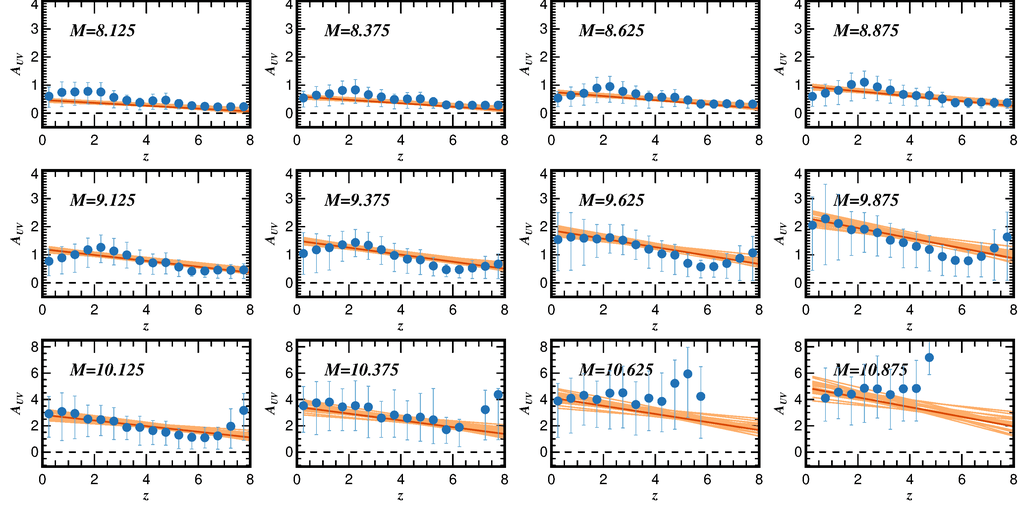}
  \caption{Stellar mass bins showing the redshift evolution of
    \auv\ (blue points). Orange curves show best-fit model as given in
    Eq~\ref{eq:avmstar}.}
  \label{fig:panels_auvz}
\end{figure*}

\begin{figure*}
  \includegraphics[width=\columnwidth]{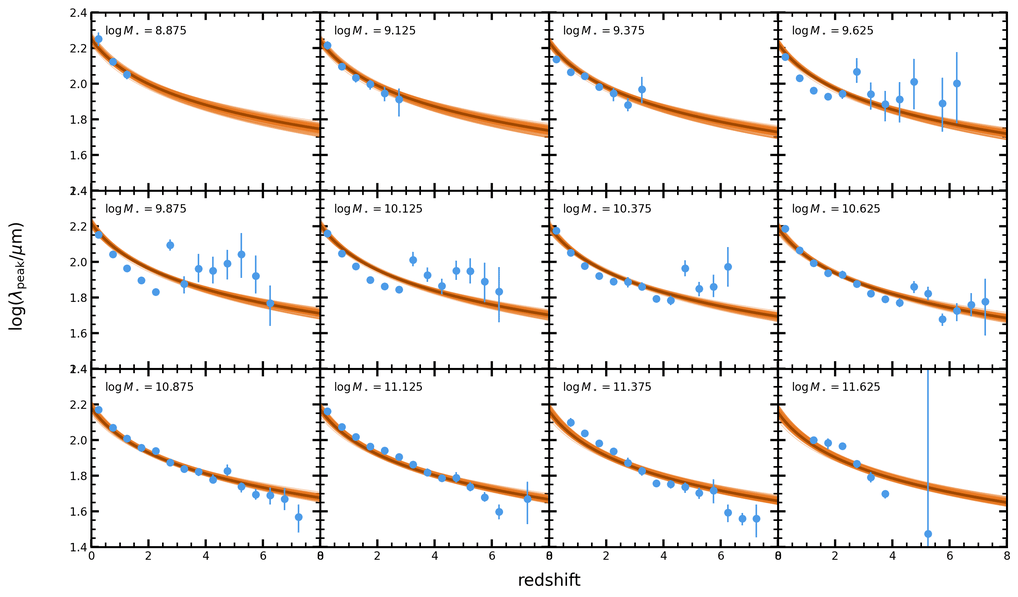}
  \caption{The evolution in dust temperature, as given by \llp, in
    stellar mass bins.  Blue points show our data while orange shows
    the best-fit model from Eq.~\ref{eq:tdustfit}.}
  \label{fig:panels_llp}
\end{figure*}

\begin{figure*}
  \includegraphics[width=\columnwidth]{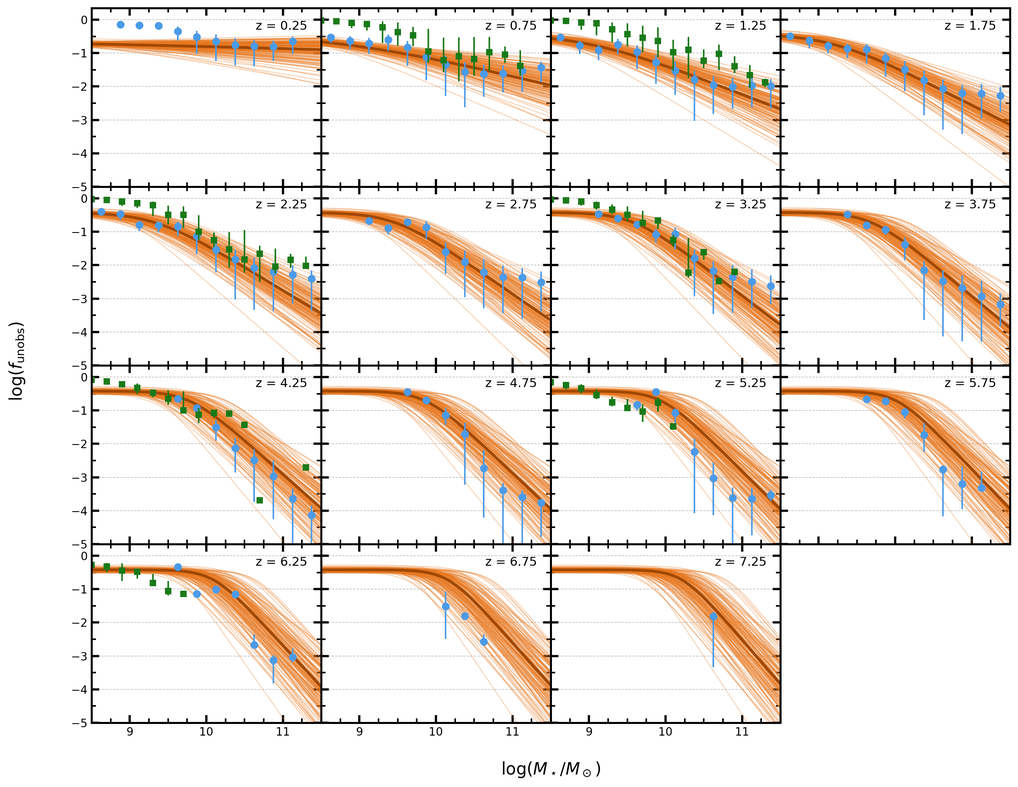}
  \caption{The fraction of star formation that is unobscured as a
    function of \mstar\ in different redshift bins.  Blue show our
    data and orange show the model fit jointly across all $z$, $M$
    bins simultaneously.  Green points are drawn from hydrodynamic
    simulations in \citet{zimmerman24a}, showing clear redshift
    evolution that is more pronounced in higher redshift bins than
    lower redshift bins, and in agreement with the broad trend we
    observe in this dataset.}
  \label{fig:panels_funobsmstar}
\end{figure*}

\begin{figure*}
  \includegraphics[width=\columnwidth]{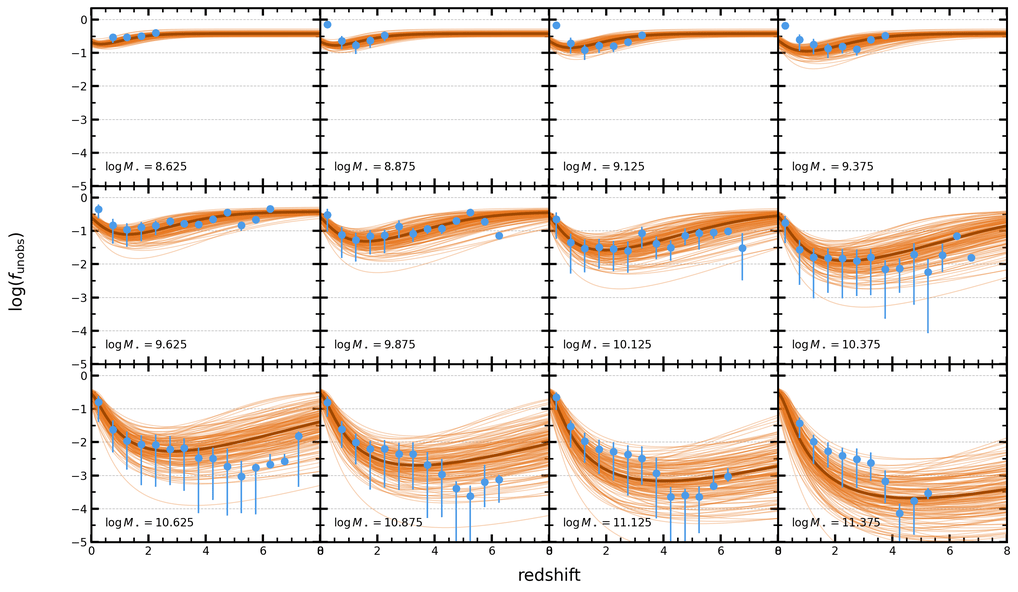}
  \caption{The fraction of star formation that is unobscured as a
    function of redshift.  Blue show our data and orange show the
    best-fit model; one notes that the general trend is, at fixed
    stellar mass, that \funobs\ drops precipitously from $0<z<2$ but
    then rises modestly at higher redshifts.}
  \label{fig:panels_funobsz}
\end{figure*}

\begin{figure*}
  \includegraphics[width=\columnwidth]{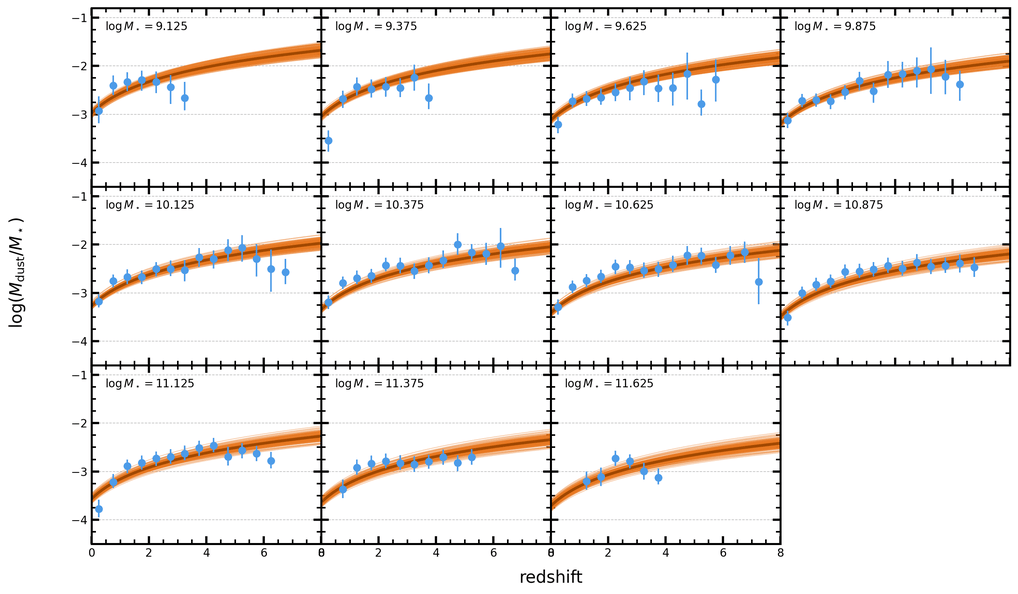}
  \caption{The evolution of the dust-to-stellar mass ratio assuming a
    fixed $\kappa_{\rm FIR}$.  Blue points show our data while orange
    shows the best-fit model from Eq.~\ref{eq:dts1}.}
  \label{fig:panels_dtsz}
\end{figure*}

\begin{figure*}
  \includegraphics[width=\columnwidth]{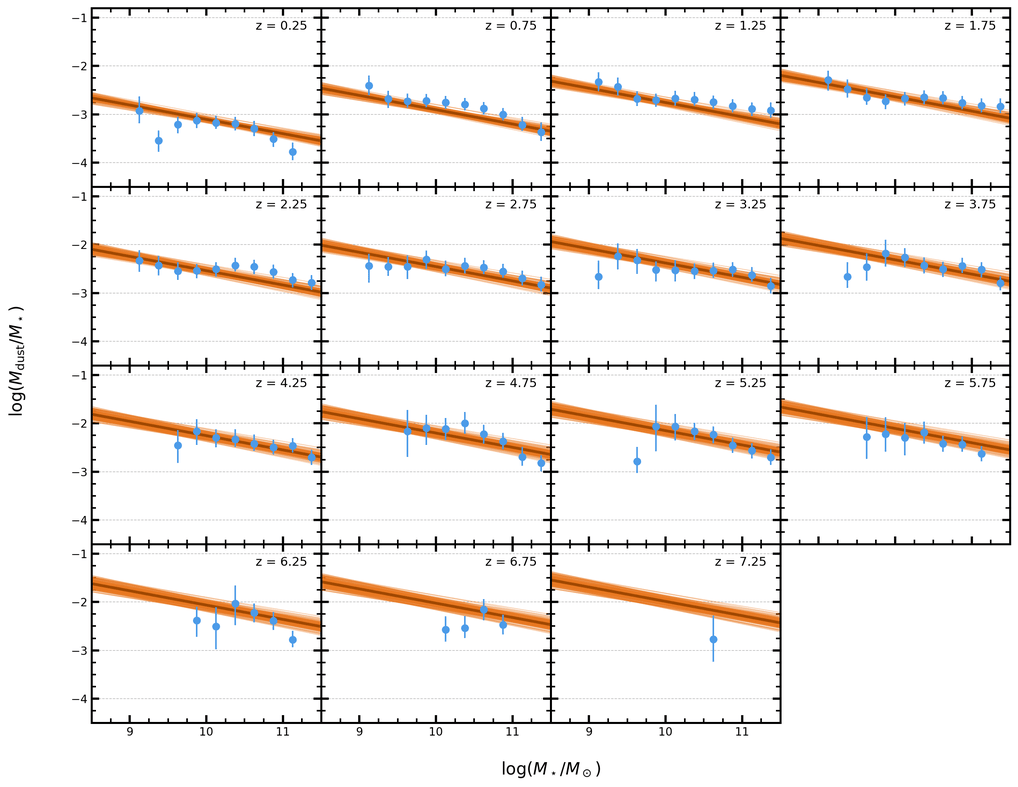}
  \caption{The stellar mass dependence of the dust-to-stellar mass
    ratio assuming a fixed $\kappa_{\rm FIR}$.}
  \label{fig:panels_dtsmstar}
\end{figure*}

Figure~\ref{fig:panels_auvmstar} and Figure~\ref{fig:panels_auvz} show
the stellar mass dependence in redshift bins, and redshift evolution
in stellar mass bins, respectively for \auv.  A similar behavior is
seen in the evolution of \av.  Figure~\ref{fig:panels_llp} show the
redshift evolution of dust temperature via the measured quantity,
\llp, in different stellar mass bins.
Figure~\ref{fig:panels_funobsmstar} and
Figure~\ref{fig:panels_funobsz} show the behavior of \funobs\ with
redshift and stellar mass.  Figure~\ref{fig:panels_dtsz} and
Figure~\ref{fig:panels_dtsmstar} show the evolution of the
dust-to-stellar mass ratio as well as its stellar mass dependence;
note that the DTS calculations presume that there is no redshift
evolution or mass dependence of $\kappa_{\rm FIR}$ as discussed in the
main text.


\begin{deluxetable}{ll}
\tablecaption{Summary of empirical relations derived in this work.\label{tab:allrelations}}
\tablewidth{\textwidth}
\tablehead{
  \colhead{Relationship} &
  \colhead{Description \&\ Parameters}
}
\startdata

\multicolumn{2}{c}{\bf Eq.~\ref{eq:avmstar}: Attenuation}\\[6pt]

\begin{minipage}{0.4\textwidth}
  $A_{\lambda}(z,M) = I(M) + S(M)(1+z)$\\
  $I(M) = \frac{A_{\rm max}}{1+\exp(-k(M-M_0))}$\\
  $S(M) = -\exp(b_{0}+b_{1}M)$\\
\end{minipage}
&
\begin{minipage}{0.6\textwidth}
  Applies to both \auv($z$,$M$) and \av($z$,$M$).\\ The best-fit
  parameters for \auv: $A_{\rm max,UV}=16.33^{+3.87}_{-3.00}$,
  $k=0.970_{-0.074}^{+0.073}$, \\$M_0=11.652_{-0.189}^{+0.191}$,
  $b_{0}=-8.84_{-1.37}^{+1.31}$, and
  $b_1=0.722_{-0.143}^{+0.141}$.\\ The best-fit parameters for \av:
  $A_{\rm max,V}=4.37^{+1.32}_{-0.96}$, $k=0.932_{-0.089}^{+0.085}$,\\
  $M_0=11.675_{-0.196}^{+0.201}$, $b_0=-9.94_{-1.57}^{+1.47}$, and
  $b_1=0.693_{-0.163}^{+0.166}$.
\end{minipage}\\

\tableline
\multicolumn{2}{c}{\bf Eq.~\ref{eq:z_delta}: Attenuation Slope}\\[6pt]

\begin{minipage}{0.4\textwidth}
$\langle \delta(z)\rangle = a + b z$ \\
\end{minipage}
&
\begin{minipage}{0.6\textwidth}
  Best-fit parameters are: $a=-0.36\pm0.07$, and $b=0.030\pm0.014$.\\
\end{minipage}
\\

\tableline

\multicolumn{2}{c}{\bf Eq.~\ref{eq:tdustfit}: Dust Temperature (Rest-Frame Peak Wavelength of Dust SED)}\\[6pt]

\begin{minipage}{0.4\textwidth}
  $\log(\lambda_{\rm peak}/\mu{\rm m}) = a + b\,(M-10) + c\,\log(1+z)$\\
\end{minipage}
&
\begin{minipage}{0.6\textwidth}
  Best-fit parameters are: $a=2.213\pm0.012$,
$b=-0.035_{-0.006}^{+0.007}$, and\\ $c=-0.531\pm0.022$.
\end{minipage}
\\[6pt]

\multicolumn{2}{c}{\bf Rest-Frame Peak Wavelength to Dust Temperature Conversion}\\[6pt]

\begin{minipage}{0.4\textwidth}
  {\bf Optically-thin dust:}\\
  $\log(T_{\rm dust}/K)=3.462-\log(\lambda_{\rm peak/\mu\!m})$\\
  {\bf Generic Opacity (this work):}\\
  $\log(T_{\rm dust}/K)=3.756-1.048\log(\lambda_{\rm peak/\mu\!m})$\\
\end{minipage}
&
\begin{minipage}{0.6\textwidth}
  When a dust SED is measured, the observable quantity is \llp, the
  rest-frame peak wavelength. $\lambda_{\rm peak}\propto T_{\rm dust}^{-1}$ via
  Wien's Displacement Law. However, the exact mapping of modeled dust
  temperature to \llp\ depends on the opacity of the dust throughout
  the IR.  Caution should be used in comparing dust temperatures from
  different works that use different opacity assumptions; it is safest
  to use \llp\ and understand how $T_{\rm dust}$ relates to the
  observable.
\end{minipage}\\
\tableline

\multicolumn{2}{c}{\bf Eq.~\ref{eq:funobs}: Fraction of Star Formation that is Unobscured}\\[6pt]

\begin{minipage}{0.4\textwidth}
  $\log(f_{\rm unobs})(z,M) = \mathcal{F}_0 -\log\!\left[1+10^{\,\alpha(z)(M-M_t(z))}\right]$\\
  $\alpha(z)=\alpha_0+\alpha_1\log(1+z)$\\
  $M_t(z)=M_{t,0}+M_{t,1}\log(1+z)$\\
\end{minipage}
&
\begin{minipage}{0.6\textwidth}
  Best-fit parameters
are: $\mathcal{F}_0=-0.420_{-0.047}^{+0.050}$,
$\alpha_0=-0.206_{-0.146}^{+0.128}$,\\
$\alpha_1=3.118_{-0.288}^{+0.380}$,
$M_{t,0}=8.198_{-0.161}^{+0.159}$, and
$M_{t,1}=2.197_{-0.157}^{+0.154}$.
  \end{minipage}\\

\tableline

\multicolumn{2}{c}{\bf Eq.~\ref{eq:dts1}: Dust-to-Stellar Mass Ratio}\\[6pt]

\begin{minipage}{0.4\textwidth}
  $\log(M_{\rm dust}/M_\star)(z,M) = A + \eta_0\,(M-10) + \eta_1\,\log(1+z)$\\
\end{minipage}
  &
\begin{minipage}{0.6\textwidth}
    Best-fit parameters are:
  $A=-3.237\pm0.036$, $\eta_0=-0.296\pm0.023$, and\\
  $\eta_1=1.364\pm0.066$.
\end{minipage}\\
\tableline


\multicolumn{2}{c}{\bf Eq.~\ref{eq:kratiosfrd}: Ratio of Dust Mass Absorption Coefficients UV-to-FIR}\\[6pt]

\begin{minipage}{0.4\textwidth}
  $\log(\kappa_{\rm UV}/\kappa_{\rm FIR})(\Sigma_{\rm SFR}) = c_0 + c_1\,\log\Sigma_{\rm SFR}$
\end{minipage}
&
\begin{minipage}{0.6\textwidth}
  While this ratio shows clear redshift evolution and mass dependence,
  it shows a more direct relation to the inferred star formation
  surface density, $\Sigma_{\rm SFR}$. Best-fit parameters are:
  $c_0=10.79\pm0.50$, and $c_1=-2.58\pm0.11$.
\end{minipage}
\\
\enddata
\tablecomments{$M\equiv\log(M_\star/M_\odot)$ throughout.
  Parameter uncertainties represent 68\% credible intervals from MCMC sampling.}
\end{deluxetable}


\section{Stacking Procedure Details}

There are several possible approaches that have been used to stack
IR/submm data in the literature. We specifically explore direct mean
and median stacking \citep{schreiber15a}, and a clustering-based
approach to modeling the stacking signal
\citep{kurczynski10a,viero13a,viero22a,bethermin15a,bethermin17a}.
The stacking methodology will change the results of the stacking in
regimes where maps are dominated by confusion noise: large beamsizes
and high source densities.  In practice this is most significant in
the {\it Herschel}/SPIRE bands, as detailed in many of the references
above.

To fully test stacking techniques and their sensitivity to confusion
noise and clustering we generate two sets of mock submillimeter
images.  The first set of maps use random positions to populate maps
and are thus unclustered, while the second set of maps is built on the
framework of a cosmological light cone and thus contains clustering.

The unclustered model maps are built on the framework presented in
\citet{casey18a} to generate the infrared luminosity function; the
exact IRLF parameters are specified in \citet{zavala21a} which also
uses the same framework.  Stellar masses are assigned to sources of a
given IR luminosity working backward from the SFR-\mstar\ relation
\citep[adopting the][model]{speagle14a} following similar methodology
to \citet{long23a}; scatter is added as inferred from the width of the
SFR-\mstar\ relation.

The clustered model builds on the Simulated Infrared
  Dusty Extragalactic Sky (SIDES) framework \citep{bethermin17a} and
uses the same 2\,deg$^2$ light cone produced from the Bolshoi-Planck
cosmological simulation \citep{klypin16a,rodriguez-puebla16a}.
\citet{bethermin17a} populated dark-matter halos using abundance
matching \citep{vale04a} and assigned various physical (luminous)
properties to galaxies in the halos using the 2SFM model
\citep{sargent12a,bethermin12a,bethermin13a}.  While the SIDES model
itself has associated submillimeter maps, we note that SIDES has
adopted a different IR SED convention than we have in this work.  In
order to directly test the impact of clustering alone, and not
clustering {\it plus} the differences in IR SEDs, we use a uniform
procedure for assigning flux densities to simulated sources.  Thus we
use the stellar masses and positions from the light cone, but reassign
flux densities.  We do note that the clustered model
(473,500 sources) has about 50\%\ more sources as the
  unclustered model (318,900 sources).

The IR emission associated with sources in the unclustered maps is
generated using a tweak of the prescription given in
\citeauthor{casey18a}; mainly, we introduce some redshift evolution of
dust temperature towards higher redshifts following the
\citet{jones23a} meta-analysis with a characteristic scatter about the
mean temperature of $\sim$20\%\ (in line with observations of large
samples at $z<2$, \citealt{drew22a}).  We note that this model is
roughly in agreement with our final findings of the evolution of dust
temperature with redshift in \S~\ref{sec:tdust}.  Monochromatic flux
densities are then inferred from SEDs and injected into maps at each
wavelength, then convolved with the observed beam as noted in Table~1
of \citet{casey18a}.  Instrumental noise is added to match the
observed dataset.  In the {\it Herschel}/SPIRE bands, there is a
strong positive skew to the flux distribution in the map from
confusion noise; we then subtract the median value of the map to reset
the pixel distribution about zero.  This has minimal impact for highly
significant emission and allows stacking to proceed with the expected
results: a null result for stacking on blank sky.  We note that SPIRE
maps, by default, have the {\it mean} set to zero (not the median) so
we shift the map to the median using offsets of 1.262\,mJy, 1.202\,mJy
and 0.946\,mJy at 250\um, 350\um, and 500\um, respectively.

\begin{figure*}
\includegraphics[width=0.99\textwidth]{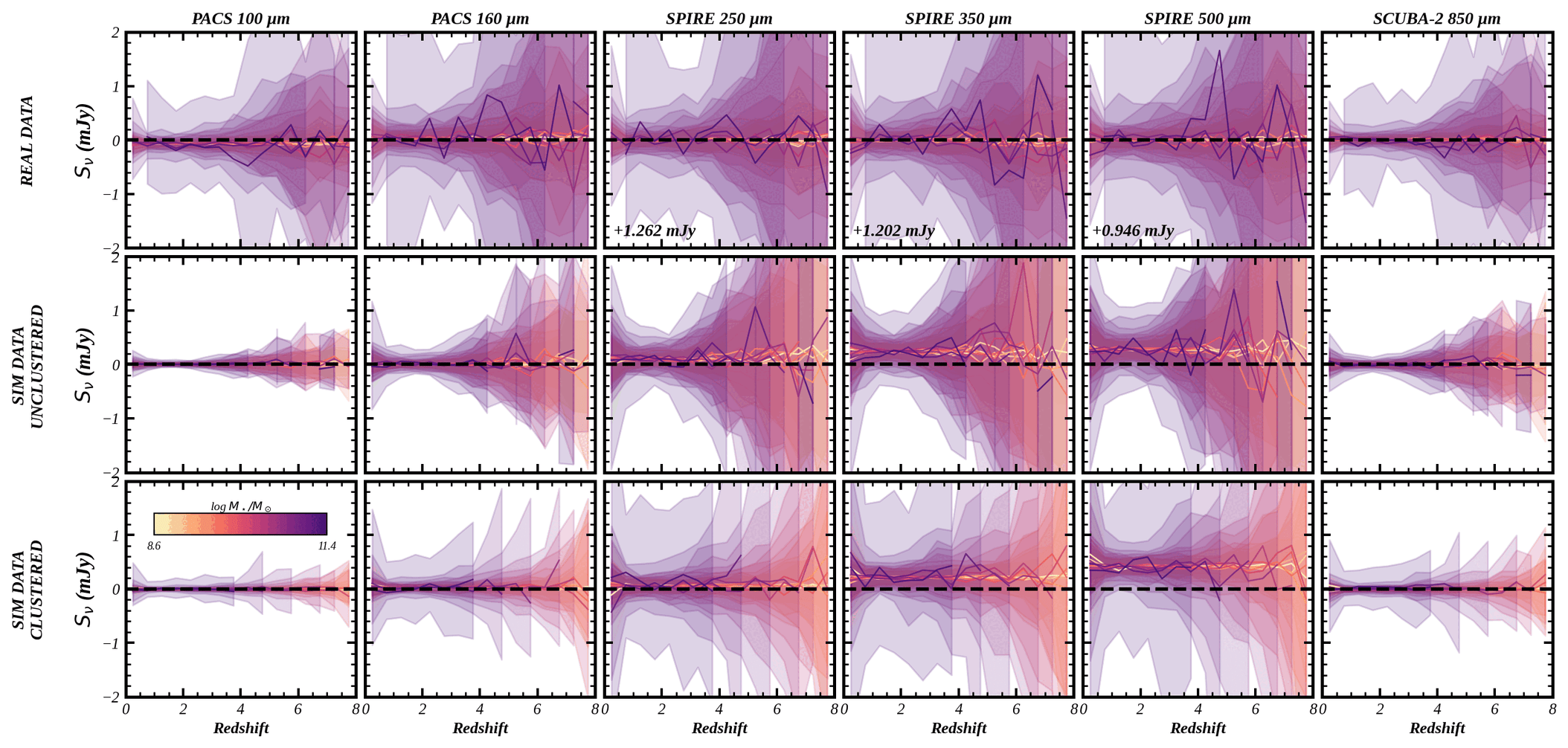}
\caption{The measured stacked flux density in the real maps, simulated
  unclustered, and simulated clustered maps when sampling the same
  number of sources in the real data with scrambled, random positions.
  Median stacking on a median-subtracted map results in median flux
  density measurements of zero across all flux densities, redshifts
  and mass bins.  What is important to gather from this plot is that
  the random stacking on simulated maps is analogous to the random
  stacking on real maps.}
\label{fig:randomstacks}
\end{figure*}

We first present the results of random position stacks on the real
maps, simulated unclustered maps, and simulated clustered maps in
Figure~\ref{fig:randomstacks}.  Within each mass and redshift bin we
have stacked the same number of sources that exist in the real data,
but scrambled their positions in the map such that they should probe
blank portions of sky. Thus, the random stacks give us the intrinsic
map background.  As should be the case for random positions, there is
no significant deviation from zero across all redshift and stellar
mass bins at any wavelength.  We note that SPIRE bands, due to their
significant source confusion, do show slight deviations in the
positive direction, even after median subtraction, though these
deviations are well within the measurement uncertainty inferred
through 100 different realization of stacking measurements over the
same map.

\begin{figure*}
\includegraphics[width=0.99\textwidth]{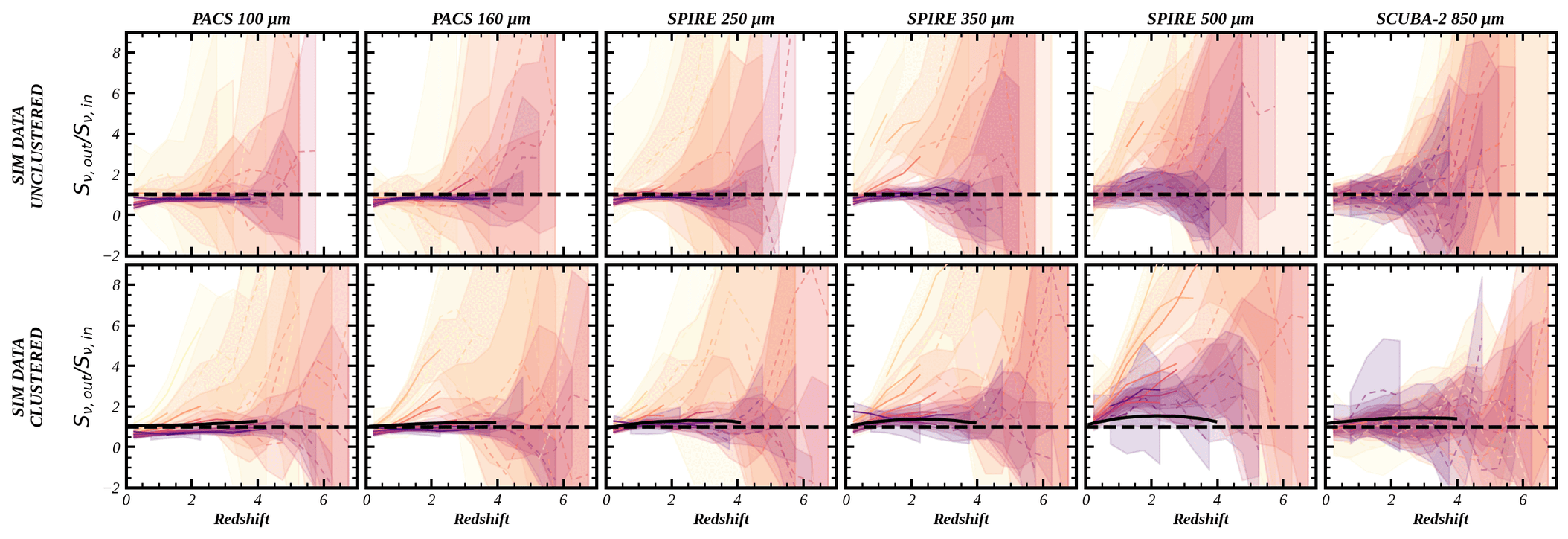}
\caption{A comparison of output stacked flux density in redshift and
  mass bins relative to input flux density (median of the population
  in the same bin) across the unclustered (top row) and clustered
  (bottom row) simulations.  Solid lines denote a regime (redshift and
  stellar masses) where the significance of the stacked result in the
  real maps is $>$3$\sigma$ and dashed lines indicate stacked flux
  densities $<$3$\sigma$. Within measurement uncertainties for
  $>$3$\sigma$ significant measurements, ratios are broadly consistent
  with unity, with some exception for the SPIRE bands at low mass
  (unclustered maps) or SPIRE and PACS 160\um\ maps in the clustered
  analysis. The solid black lines show the estimates of
    clustering bias from \citet{bethermin15a} for galaxies above
    stellar masses $>3\times10^{10}$\,\msun.}
\label{fig:sims_inandout}  
\end{figure*}

We then proceed to analyze the input and output of median stacking
measurements on the simulated maps to understand the impact of both
confusion noise and clustering.  We first investigate the ratio of
measured output flux density to input flux density across all
redshifts and stellar masses.  Figure~\ref{fig:sims_inandout} shows
the results.  Within uncertainties, we see broad agreement with a flux
ratio of unity, especially where the significance of real stacked
detections is high ($>$3$\sigma$, solid lines).  However, there are
some regimes of greater concern.  In the unclustered simulation, some
of the lower mass bins in SPIRE 350\um\ or 500\um\ show significantly
elevated output flux density ($\sim$2-4$\times$ input) from $0<z<2$.
This is more significant in the clustered simulation, but extends to
PACS 160\um\ and all of the SPIRE bands, with SPIRE 500\um\ the most
severe, with 4$\times$ elevated flux density at
$\sim10^{9.3}$\,\msun\ and even higher, reaching $\sim$10$\times$ at
10$^{8.5}$\,\msun.  We note that uncertainties on this ratio are quite
large in the higher redshift bins: in this regime source numbers
dwindle.  We note that, despite large implied ratios of stacked flux
to true input flux in this regime, {\it none} of the reported flux
densities in the corresponding mass, redshift, wavelength bins are
deemed statistically significant.  In other words, only flux density
limits are used to constrain the SED in this regime.

The deviation towards higher stacked flux densities than input is a
result of how confusion-limited the maps are in the given mass regime
-- when stacking low mass galaxies with a very high source density and
low expected flux density, the contribution of flux from other
(brighter) sources within the beamsize becomes significant.  If we are
to recover true flux densities in this regime, we should divide the
measured stacked flux by the measured ratio implied in
Figure~\ref{fig:sims_inandout}, in particular using the results from
the clustered simulation.  Indeed, the difference between
non-clustered and clustered simulations should be caused by the
relative clustering of low mass galaxies around higher mass galaxies
which does not exist in the unclustered maps.  This is the approach
that \citet{bethermin15b} take; we overplot their measured clustering
bias for galaxies (in their samples for galaxies with
\mstar$>$3$\times10^{10}$\,\msun) as a solid black line on the bottom
row of Figure~\ref{fig:sims_inandout}; our clustered map results agree
for the matched mass bins with the \citeauthor{bethermin15b} estimate
of clustering bias.  We infer a much larger clustering correction for
lower mass galaxies, below the mass threshold fit in previous stacking
analyses.  However, an important note is that we infer much larger
uncertainties on the ratio of output to input flux density than previous
work suggests.  In other words, we find the same clustering bias but at
far less statistical significance.  We attribute the difference in the
significance of the clustering signal to be due to the differences
between our SED model and that of SIDES: with greater dynamic range of
SED shapes (ie. input dust temperatures thus input flux density
distributions in a given mass and redshift bin), there is naturally
far larger scatter in the resulting output stack measurement and
associated uncertainty, as well as greater range in the input flux
density.  While it is a less tidy outcome, we feel this may better
reflect the true range of SEDs present in the real universe.

Given the low significance of implied clustering bias inferred from
our analysis, it is not entirely clear if the correction of clustering
bias should be applied to our measured flux densities or not.  To test
the relative impact, we did a direct comparison of
clustering-corrected flux densities to non-corrected flux densities,
the impact on the best-fit SEDs and the scientific conclusions in this
work overall.  As per Figure~\ref{fig:sims_inandout}, the most
clustering-impacted regime is at low mass (\mstar$<10^{9.5}$\,\msun);
beyond $z>2$, the maximum SNR in this mass regime is $\approx3\sigma$
across all PACS and SPIRE bands, but the stacks are primarily
dominated by non-detections.  We found that, given the inherently low
SNR of the stacks and the uncertainty of the clustering correction
factor, the clustering-corrected flux densities are statistically
consistent with the original flux density measurements and have no
resulting impact on the derived SED properties.  One might think that
the additional uncertainty introduced by the correction factor would
change at least the uncertainty on the derived dust temperature, for
example; however, the uncertainties on luminosity and dust temperature
remain about the same with and without the correction because the
clustering correction downweights the flux density while inflating the
uncertainty, so the $\sim$2$\sigma$ upper limits on non-detections is
largely unchanged.  Because we find that the clustering effect is, in
the end, not impacting our conclusions and simultaneously is dependent
on model assumptions for SEDs and HOD modeling, we decline to
implement it in our flux density measurements to provide a more
straightforward reporting of our direct measurements.

\begin{figure*}
\centering
\includegraphics[width=0.89\textwidth]{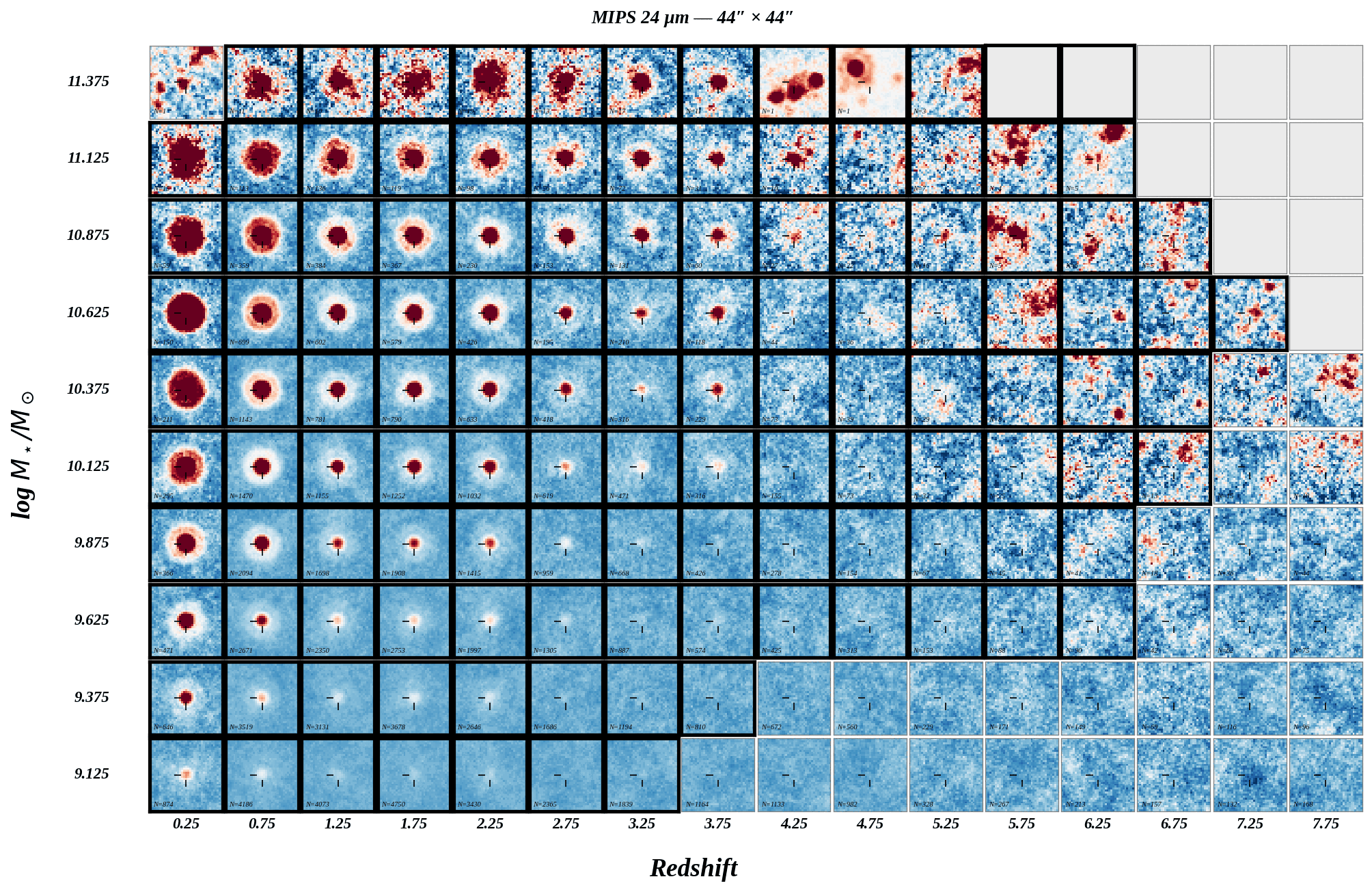}
\caption{Two-dimensional stacked MIPS 24\,\um\ cutouts.}
\label{fig:stack2d_24}
\end{figure*}

\begin{figure*}
\centering
\includegraphics[width=0.89\textwidth]{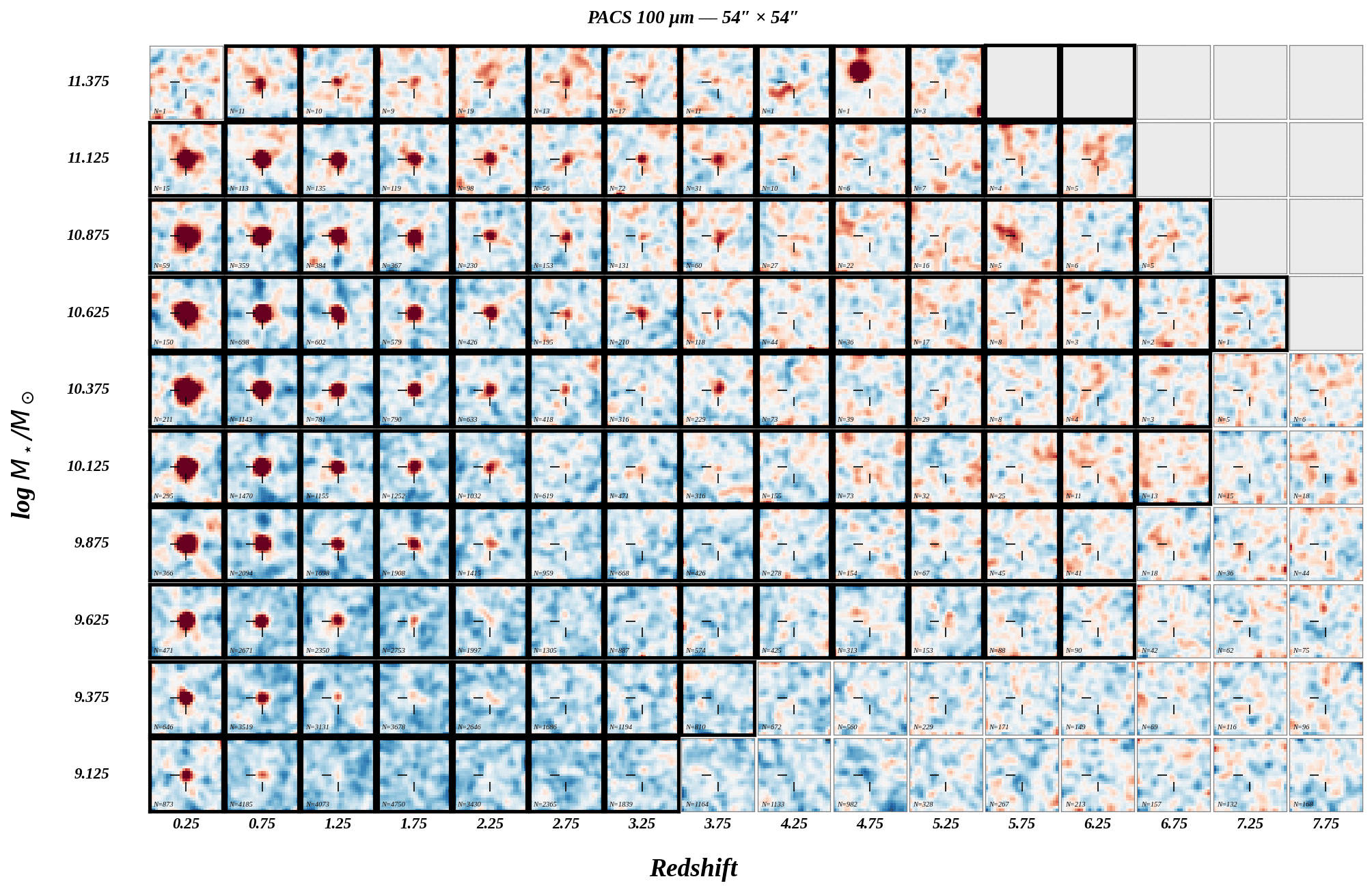}
\caption{Two-dimensional stacked PACS 100\,\um\ cutouts.}
\label{fig:stack2d_100}
\end{figure*}

\begin{figure*}
\centering
\includegraphics[width=0.89\textwidth]{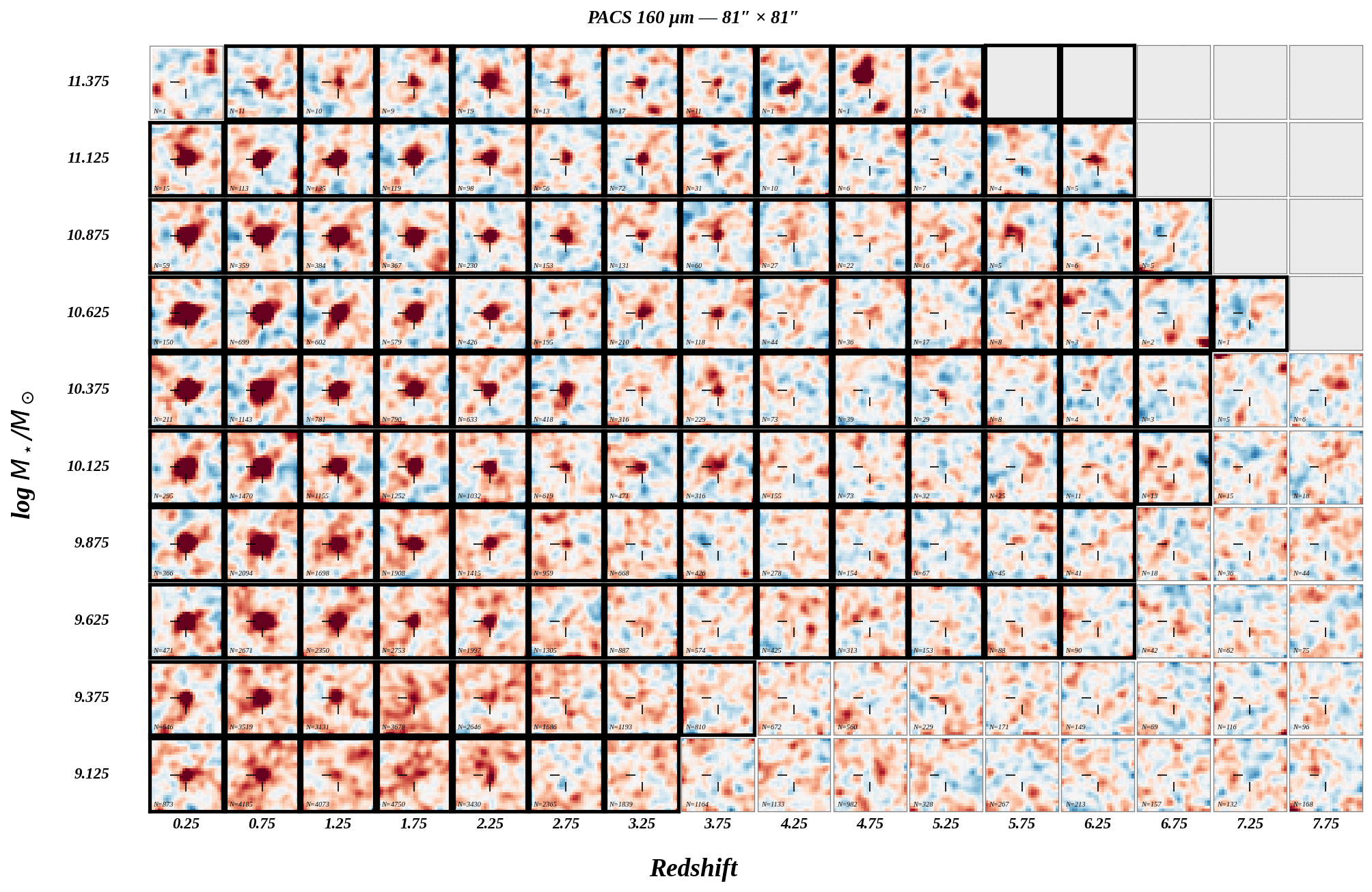}
\caption{Two-dimensional stacked PACS 160\,\um\ cutouts.}
\label{fig:stack2d_160}
\end{figure*}

\begin{figure*}
\centering
\includegraphics[width=0.89\textwidth]{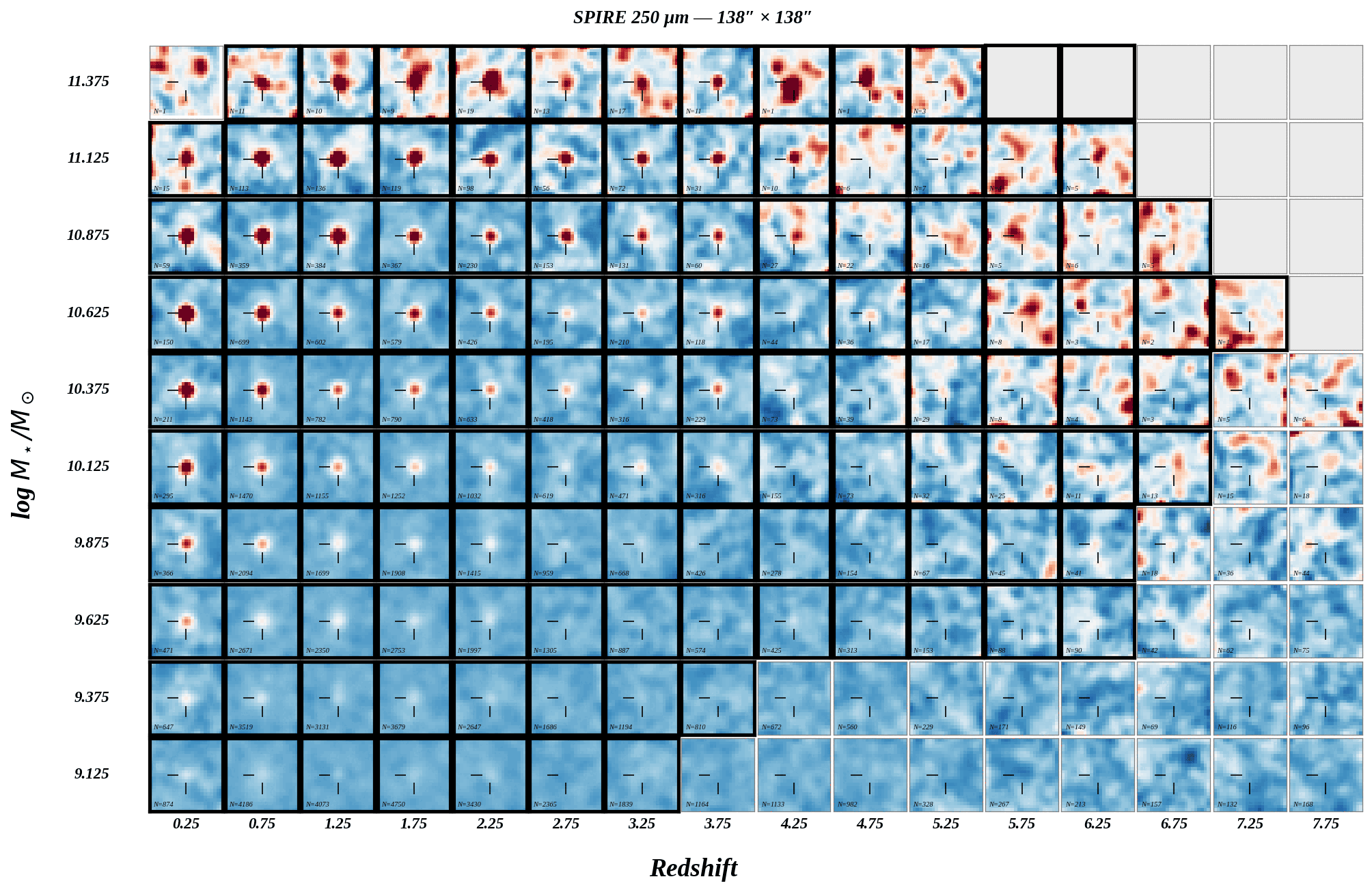}
\caption{Two-dimensional stacked SPIRE 250\,\um\ cutouts.}
\label{fig:stack2d_250}
\end{figure*}

\begin{figure*}
\centering
\includegraphics[width=0.89\textwidth]{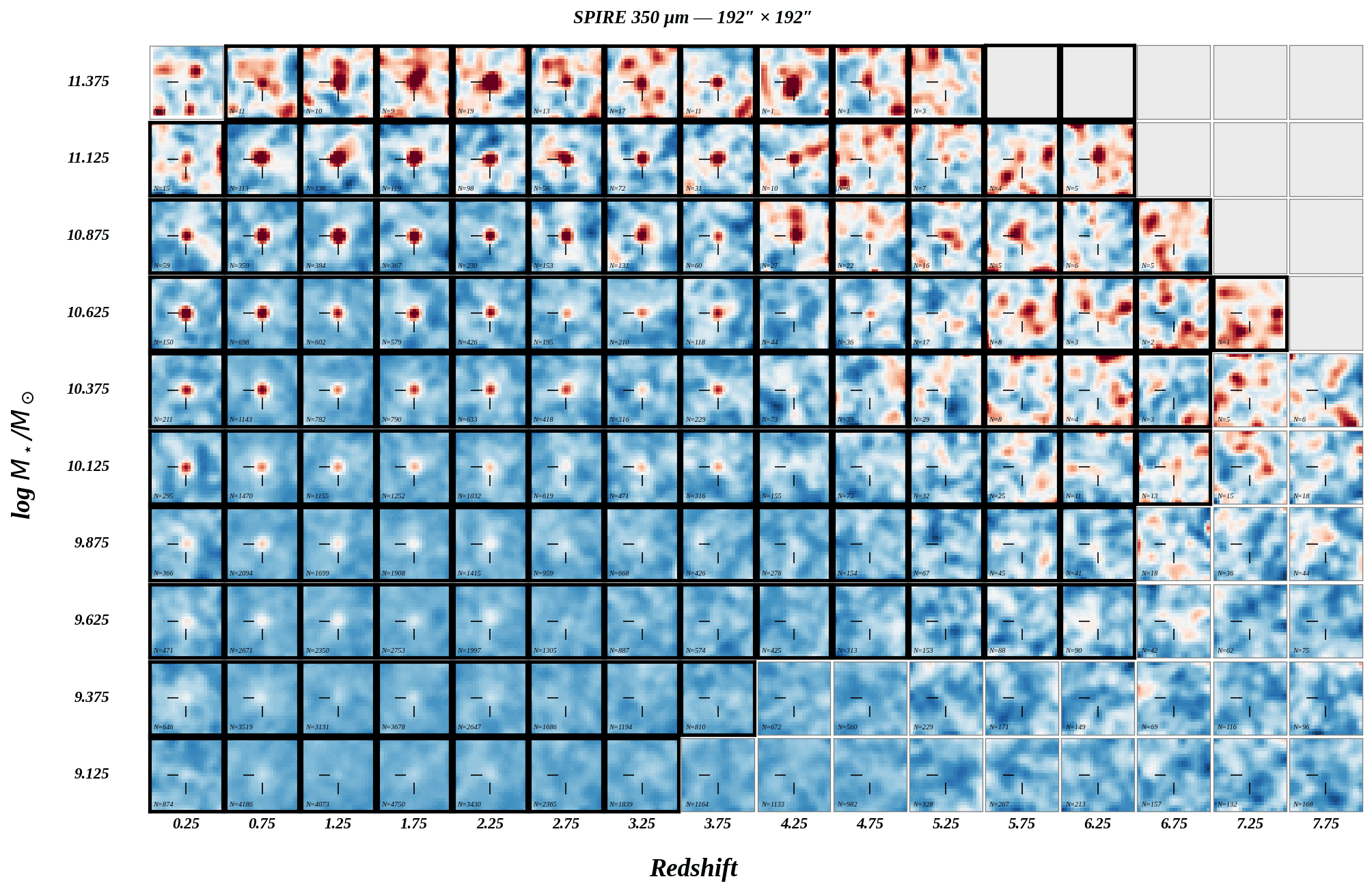}
\caption{Two-dimensional stacked SPIRE 350\,\um\ cutouts.}
\label{fig:stack2d_350}
\end{figure*}

\begin{figure*}
\centering
\includegraphics[width=0.89\textwidth]{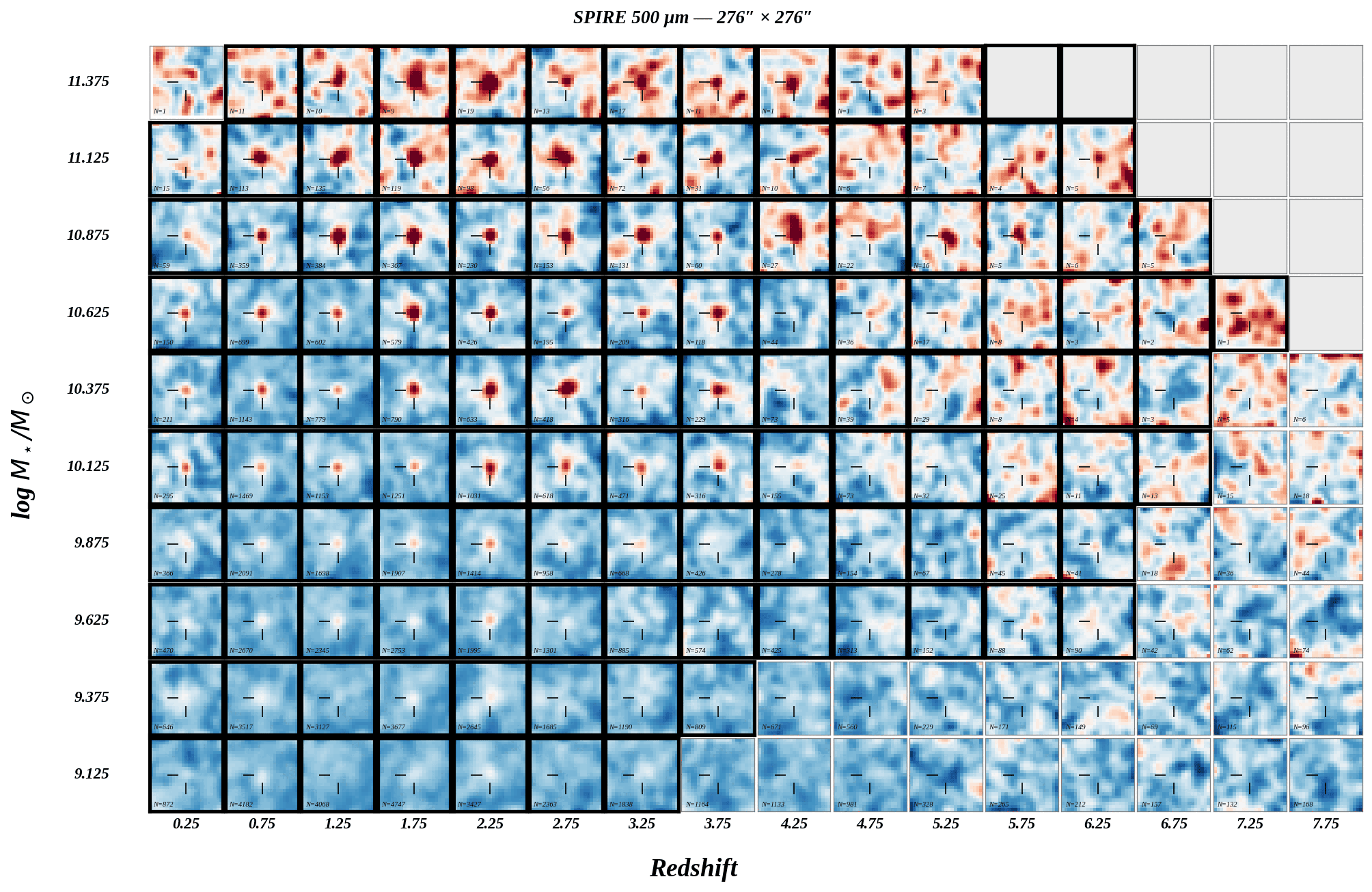}
\caption{Two-dimensional stacked SPIRE 500\,\um\ cutouts.}
\label{fig:stack2d_500}
\end{figure*}

\begin{figure*}
\centering
\includegraphics[width=0.89\textwidth]{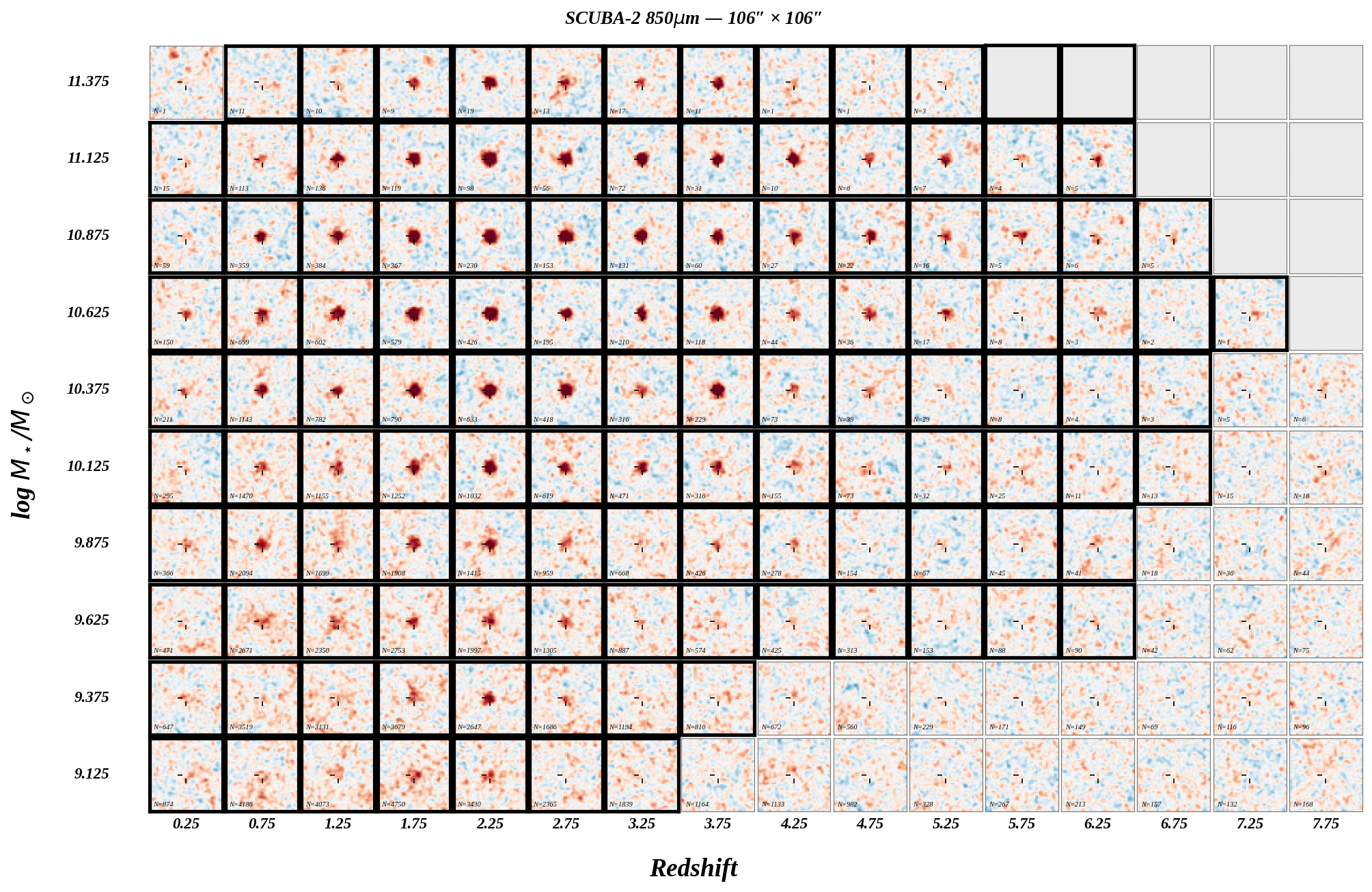}
\caption{Two-dimensional stacked SCUBA-2 850\,\um\ cutouts.}
\label{fig:stack2d_850}
\end{figure*}

\begin{figure*}
\centering
\includegraphics[width=0.89\textwidth]{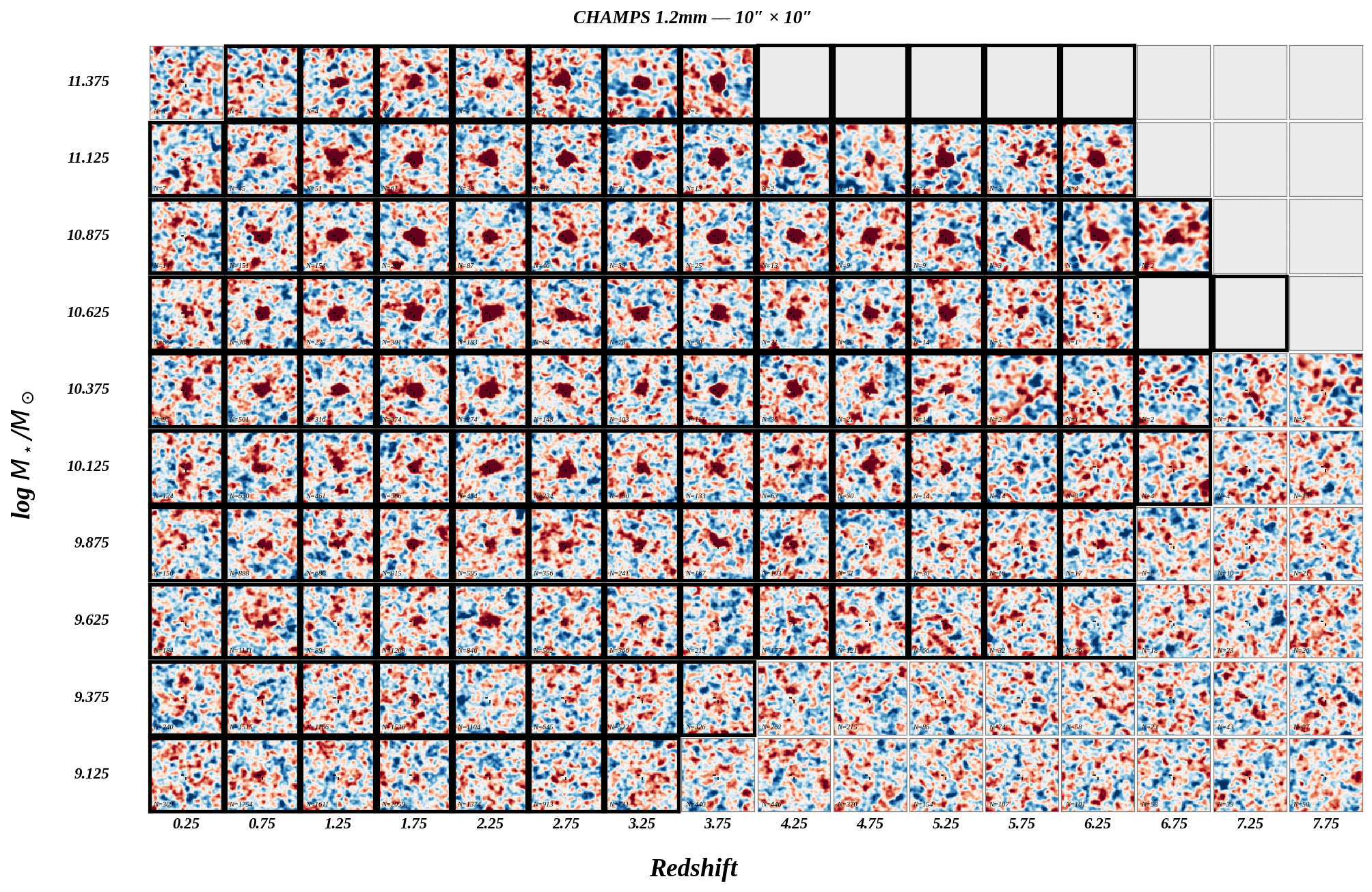}
\caption{Two-dimensional stacked CHAMPS 1.2\,mm cutouts.}
\label{fig:stack2d_1200}
\end{figure*}

\begin{figure*}
\centering
\includegraphics[width=0.89\textwidth]{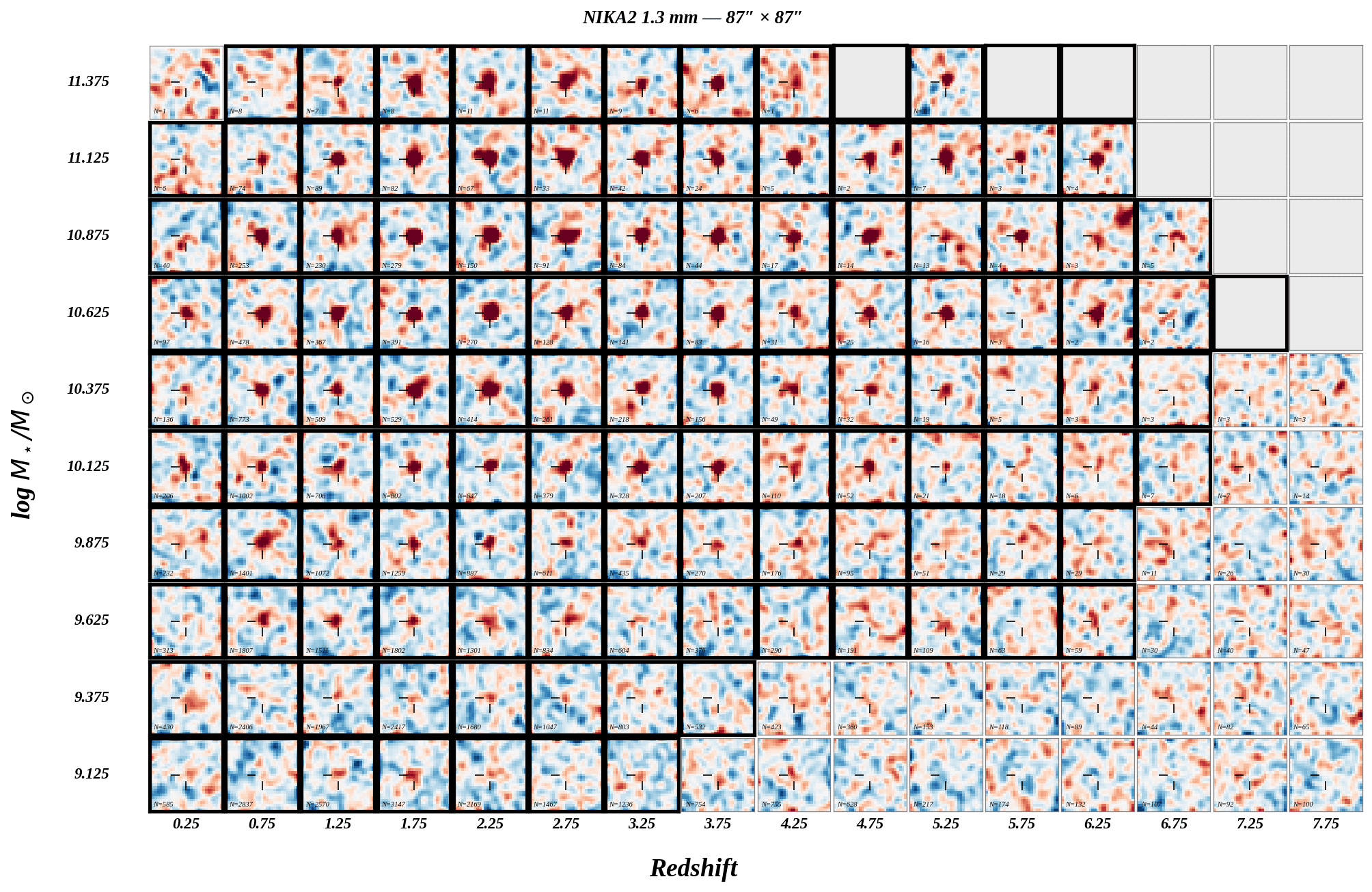}
\caption{Two-dimensional stacked NIKA-2 1.2\,mm cutouts.}
\label{fig:stack2d_nika1}
\end{figure*}

\begin{figure*}
\centering
\includegraphics[width=0.89\textwidth]{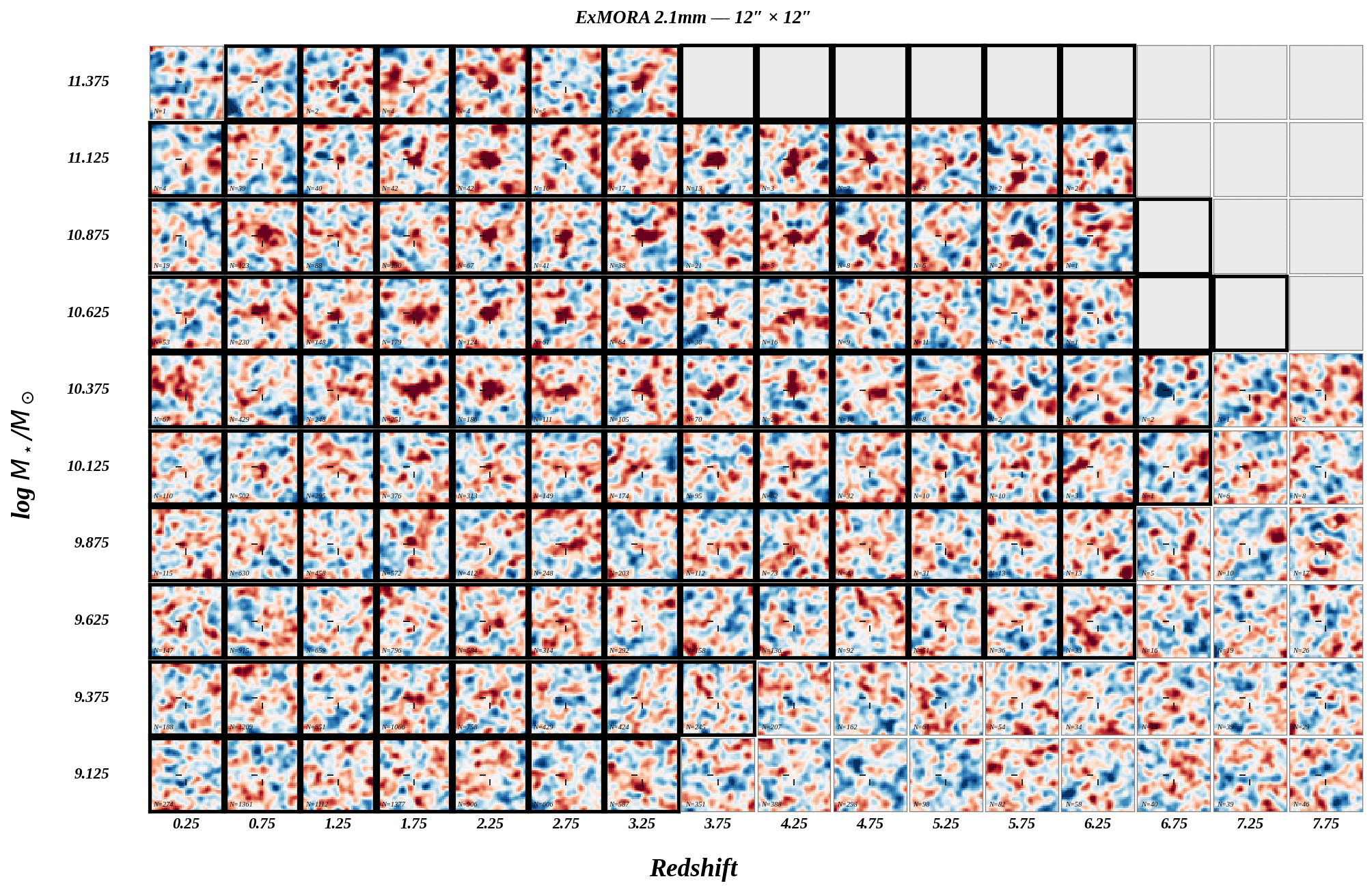}
\caption{Two-dimensional stacked Ex-MORA 2.1\,mm cutouts.}
\label{fig:stack2d_2100}
\end{figure*}

\begin{figure*}
\centering
\includegraphics[width=0.89\textwidth]{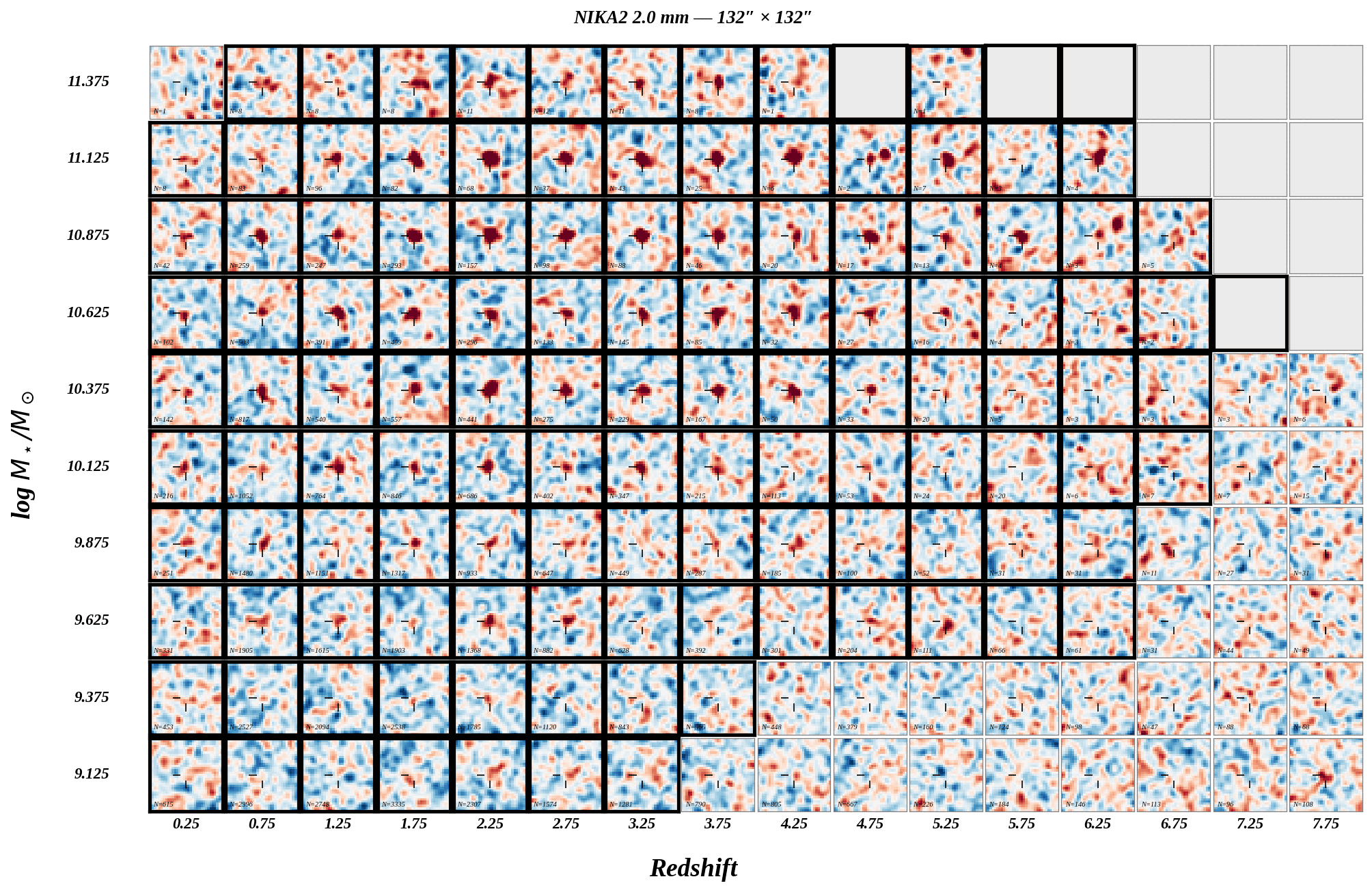}
\caption{Two-dimensional stacked NIKA-2 2\,mm cutouts.}
\label{fig:stack2d_nika2}
\end{figure*}

A full table of our measured flux density constraints in every
redshift and mass bin shown in Figure~\ref{fig:masterseds} is given in
Table~\ref{tab:masterflux}.  Note we only report flux densities for
bins where dust SED fits have been measured: those with at least two
$>$3$\sigma$ significant detections or one $>$5$\sigma$ detection.
For illustrative purposes, we also show stacked cutouts in all mass
and redshift bins at each wavelength in the following:
Figure~\ref{fig:stack2d_24} (24\um), Figure~\ref{fig:stack2d_100}
(100\um), Figure~\ref{fig:stack2d_160} (160\um),
Figure~\ref{fig:stack2d_250} (250\um), Figure~\ref{fig:stack2d_350}
(350\um), Figure~\ref{fig:stack2d_500} (500\um),
Figure~\ref{fig:stack2d_850} (850\um), Figures~\ref{fig:stack2d_1200}
and ~\ref{fig:stack2d_nika1} (1200\um\ from CHAMPS and
1300\um\ NIKA2, respectively), and Figures~\ref{fig:stack2d_2100} and
~\ref{fig:stack2d_nika2} (2100\um\ from Ex-MORA and 2000\um\ from
NIKA2, respectively).  Stacked 2D images are generated by shifting
imaging on sub-pixel scales to center sources before median stacking.
Bins with SED fits are outlined in thick black.

\begin{longrotatetable}

\setlength{\tabcolsep}{0.3pt}
\renewcommand{\arraystretch}{1}
\tabletypesize{\fontsize{5}{7}\selectfont}
\begin{deluxetable*}{cc rrrrr ccccccccccc}
\tablewidth{0pt}
\tablecaption{Master stacked flux densities.\label{tab:masterflux}}
\tablehead{
\colhead{$z$} & \colhead{$\log M_*$} & \colhead{$N_\mathrm{F}$} & \colhead{$N_\mathrm{1}$} & \colhead{$N_\mathrm{2}$} &
\colhead{$N_\mathrm{N1}$} & \colhead{$N_\mathrm{N2}$} &
\colhead{$S_{24}$} & \colhead{$S_{100}$} & \colhead{$S_{160}$} & \colhead{$S_{250}$} &
\colhead{$S_{350}$} & \colhead{$S_{500}$} & \colhead{$S_{850}$} &
\colhead{$S^{\rm CHAMPS}_{1.2\,\rm mm}$} & \colhead{$S^{\rm NIKA2}_{1.3\,\rm mm}$} &
\colhead{$S^{\rm NIKA2}_{2.0\,\rm mm}$} & \colhead{$S^{\rm ExMORA}_{2.1\,\rm mm}$} \\[0.8ex]
\colhead{} & \colhead{$(M_\odot)$} & \colhead{} & \colhead{} & \colhead{} &
\colhead{} & \colhead{} &
\colhead{(mJy)\rule{0pt}{2.2ex}} & \colhead{(mJy)\rule{0pt}{2.2ex}} & \colhead{(mJy)\rule{0pt}{2.2ex}} & \colhead{(mJy)\rule{0pt}{2.2ex}} &
\colhead{(mJy)\rule{0pt}{2.2ex}} & \colhead{(mJy)\rule{0pt}{2.2ex}} & \colhead{(mJy)\rule{0pt}{2.2ex}} &
\colhead{(mJy)\rule{0pt}{2.2ex}} & \colhead{(mJy)\rule{0pt}{2.2ex}} & \colhead{(mJy)\rule{0pt}{2.2ex}} & \colhead{(mJy)\rule{0pt}{2.2ex}} \\[0.6ex]
}
\startdata
0.25 & 9.125 & 4838 & 1873 & 1565 & 615 & 610 & $(0.039\pm0.032)$ & $0.282\pm0.057$ & $0.76\pm0.12$ & $(-0.55\pm0.10)$ & $(-0.38\pm0.11)$ & $(-0.293\pm0.089)$ & $0.066\pm0.013$ & $(0.0025\pm0.0031)$ & $(0.020\pm0.021)$ & $(0.0009\pm0.0079)$ & $(-0.0028\pm0.0018)$ \\
0.25 & 9.375 & 3800 & 1472 & 1262 & 454 & 451 & $0.239\pm0.046$ & $0.672\pm0.071$ & $1.31\pm0.14$ & $(-0.11\pm0.10)$ & $(-0.05\pm0.13)$ & $(-0.11\pm0.11)$ & $0.074\pm0.014$ & $(0.0036\pm0.0034)$ & $(0.036\pm0.021)$ & $(0.001\pm0.011)$ & $(0.0016\pm0.0019)$ \\
0.25 & 9.625 & 2843 & 1130 & 937 & 337 & 332 & $0.673\pm0.072$ & $1.28\pm0.11$ & $2.31\pm0.18$ & $0.98\pm0.15$ & $0.43\pm0.14$ & $(0.04\pm0.12)$ & $0.076\pm0.016$ & $0.0126\pm0.0039$ & $(0.041\pm0.034)$ & $(0.019\pm0.011)$ & $(0.0053\pm0.0023)$ \\
0.25 & 9.875 & 2210 & 877 & 737 & 249 & 248 & $1.38\pm0.16$ & $1.99\pm0.19$ & $3.93\pm0.29$ & $2.40\pm0.17$ & $1.28\pm0.17$ & $0.45\pm0.14$ & $0.144\pm0.019$ & $0.0194\pm0.0045$ & $(0.042\pm0.031)$ & $(0.0156\pm0.0089)$ & $(0.0074\pm0.0025)$ \\
0.25 & 10.125 & 1691 & 656 & 571 & 211 & 212 & $2.28\pm0.24$ & $2.83\pm0.30$ & $5.40\pm0.36$ & $3.85\pm0.23$ & $2.26\pm0.17$ & $0.98\pm0.18$ & $0.204\pm0.021$ & $0.0295\pm0.0052$ & $(0.154\pm0.053)$ & $(0.022\pm0.010)$ & $0.0106\pm0.0029$ \\
0.25 & 10.375 & 1241 & 503 & 428 & 143 & 142 & $3.20\pm0.28$ & $3.63\pm0.40$ & $7.13\pm0.54$ & $6.10\pm0.32$ & $3.76\pm0.24$ & $1.76\pm0.19$ & $0.257\pm0.025$ & $0.0494\pm0.0059$ & $(0.083\pm0.037)$ & $(0.026\pm0.017)$ & $(0.0097\pm0.0033)$ \\
0.25 & 10.625 & 788 & 319 & 288 & 99 & 98 & $4.23\pm0.39$ & $4.66\pm0.62$ & $9.41\pm0.74$ & $8.43\pm0.46$ & $5.57\pm0.35$ & $2.39\pm0.30$ & $0.390\pm0.031$ & $0.0543\pm0.0074$ & $(0.141\pm0.060)$ & $(0.040\pm0.016)$ & $0.0184\pm0.0040$ \\
0.25 & 10.875 & 369 & 119 & 131 & 43 & 42 & $5.22\pm0.94$ & $5.62\pm0.83$ & $12.1\pm1.1$ & $10.23\pm0.68$ & $5.68\pm0.51$ & $2.55\pm0.41$ & $0.487\pm0.046$ & $0.060\pm0.012$ & $0.208\pm0.065$ & $(0.032\pm0.028)$ & $0.0249\pm0.0061$ \\
\enddata
\tablecomments{Median (sub)mm stacked flux densities (mJy) per redshift and stellar-mass bin in this work.
Values in parentheses are $<3\sigma$ detections.
$N_\mathrm{F}$: number of sources stacked maps that cover the entirety of COSMOS-Web;
$N_\mathrm{1}$: number of sources covered in CHAMPS (1.2\,mm); $N_\mathrm{2}$: ExMORA (2.1\,mm); $N_\mathrm{N1}$: NIKA2 (1.3\,mm); $N_\mathrm{N2}$: NIKA2 (2.0\,mm). This table is available online in fits format at https://github.com/caitlinmcasey/duststacks.}
\end{deluxetable*}
\renewcommand{\arraystretch}{1}
\tabletypesize{\footnotesize}
\setlength{\tabcolsep}{6pt}

\end{longrotatetable}

\begin{longrotatetable}

\begin{deluxetable*}{ccccccccccccc}
\tabletypesize{\tiny}
\tablewidth{0pt}
\tablecaption{Derived SED Physical Properties.\label{tab:mastersed}}
\tablehead{
\colhead{$z$} & \colhead{$\log M_*$} &
\colhead{$L_{\rm IR}$} & \colhead{$\lambda_{\rm peak}$} &
\colhead{$\alpha_{\rm MIR}$} & \colhead{$\beta_{\rm IR}$} &
\colhead{$M_{\rm dust}$} &
\colhead{${\rm SFR}_{\rm UV}$} & \colhead{${\rm SFR}_{\rm IR}$} &
\colhead{$f_{\rm unobs}$} & \colhead{$B^{\prime}$} & \colhead{$A_{\rm UV}$} & \colhead{$A_V$} \\[0.8ex]
\colhead{} & \colhead{$(M_\odot)$} &
\colhead{$(L_\odot)$} & \colhead{$(\mu{\rm m})$} &
\colhead{} & \colhead{} &
\colhead{$(M_\odot)$} &
\colhead{$(M_\odot\,{\rm yr}^{-1})$} & \colhead{$(M_\odot\,{\rm yr}^{-1})$} &
\colhead{} & \colhead{} & \colhead{(mag)} & \colhead{(mag)}
}
\startdata
0.25 & 9.125 & $(4.21^{+1.27}_{-0.76})\times10^{8}$ & $164.2^{+8.6}_{-8.6}$ & $4.90^{+0.10}_{-0.06}$ & $-$ & $(1.58^{+0.92}_{-0.48})\times10^{6}$ & $0.13^{+0.13}_{-0.07}$ & $0.062^{+0.015}_{-0.010}$ & $(6.75^{+1.35}_{-1.08})\times10^{-1}$ & $1.43^{+0.22}_{-0.24}$ & $0.80^{+0.41}_{-0.55}$ & $0.26^{+0.15}_{-0.18}$ \\
0.25 & 9.375 & $(5.80^{+1.12}_{-0.81})\times10^{8}$ & $136.9^{+4.5}_{-4.2}$ & $4.92^{+0.08}_{-0.05}$ & $-$ & $(6.80^{+1.90}_{-1.80})\times10^{5}$ & $0.17^{+0.27}_{-0.09}$ & $0.086^{+0.013}_{-0.012}$ & $(6.69^{+1.67}_{-1.28})\times10^{-1}$ & $1.47^{+0.29}_{-0.31}$ & $1.11^{+0.67}_{-0.86}$ & $0.38^{+0.24}_{-0.30}$ \\
0.25 & 9.625 & $(1.77^{+0.13}_{-0.11})\times10^{9}$ & $141.1^{+1.4}_{-3.7}$ & $-$ & $1.97^{+0.10}_{-0.12}$ & $(2.59^{+0.37}_{-0.44})\times10^{6}$ & $0.21^{+0.49}_{-0.11}$ & $0.26^{+0.02}_{-0.02}$ & $(4.46^{+1.74}_{-2.48})\times10^{-1}$ & $1.49^{+0.43}_{-0.28}$ & $1.64^{+0.94}_{-1.27}$ & $0.56^{+0.36}_{-0.44}$ \\
0.25 & 9.875 & $(3.61^{+0.27}_{-0.28})\times10^{9}$ & $142.3^{+2.0}_{-3.7}$ & $4.93^{+0.07}_{-0.04}$ & $1.96^{+0.07}_{-0.11}$ & $(5.55^{+0.80}_{-0.64})\times10^{6}$ & $0.23^{+0.44}_{-0.16}$ & $0.52^{+0.03}_{-0.02}$ & $(3.08^{+1.47}_{-2.43})\times10^{-1}$ & $1.45^{+0.56}_{-0.30}$ & $2.2^{+1.0}_{-1.7}$ & $0.73^{+0.37}_{-0.60}$ \\
0.25 & 10.125 & $(5.57^{+0.20}_{-0.45})\times10^{9}$ & $144.1^{+2.5}_{-2.8}$ & $4.90^{+0.10}_{-0.05}$ & $2.01^{+0.07}_{-0.12}$ & $(8.94^{+0.94}_{-0.53})\times10^{6}$ & $0.24^{+0.56}_{-0.14}$ & $0.81^{+0.06}_{-0.04}$ & $(2.26^{+1.47}_{-1.66})\times10^{-1}$ & $1.33^{+0.56}_{-0.45}$ & $2.8^{+1.2}_{-1.8}$ & $0.89^{+0.47}_{-0.58}$ \\
0.25 & 10.375 & $(7.96^{+0.47}_{-0.59})\times10^{9}$ & $150.1^{+2.9}_{-4.2}$ & $4.80^{+0.14}_{-0.10}$ & $2.08^{+0.08}_{-0.08}$ & $(1.52^{+0.16}_{-0.13})\times10^{7}$ & $0.25^{+0.55}_{-0.15}$ & $1.2^{+0.1}_{-0.1}$ & $(1.72^{+0.97}_{-1.50})\times10^{-1}$ & $1.20^{+0.43}_{-0.57}$ & $3.3^{+1.1}_{-1.9}$ & $1.03^{+0.46}_{-0.65}$ \\
\enddata
\tablecomments{Derived SED and physical properties per redshift and stellar-mass bin from this work.
$L_{\rm IR}$ and $M_{\rm dust}$ are derived from modified-blackbody SED fits to stacked photometry.
$\alpha_{\rm MIR}$ and $\beta_{\rm IR}$ are the mid-infrared power-law slope and Rayleigh--Jeans emissivity index; entries with $\beta_{\rm IR}$ fixed to~2 by the fit prior and $\alpha_{\rm MIR}$ with no meaningful posterior constraint are shown as $-$.
${\rm SFR}_{\rm UV}$ is uncorrected for dust attenuation; ${\rm SFR}_{\rm IR}$ is inferred from $L_{\rm IR}$.
$f_{\rm unobs}$ is the fraction of star formation unobserved in the UV; a lower error of~0 indicates the posterior lower bound is consistent with zero.
$B^{\prime}$ is the ratio of bolometric corrections between the 1600\AA\ to total UV and 8--1000\,\um\ to total IR.  Both \auv\ and \av\ are inferred from rest-frame UV/optical SED fitting where the quoted statistic is representative of the median and 68\%\ confidence intervals on the stackable sample in the corresponding bin (where the number of sources included are $N_{\rm F}$; these are not representative of errors on individual galaxy \auv\ or \av\ measurements).
This table is available online in fits format at https://github.com/caitlinmcasey/duststacks.}
\end{deluxetable*}
\tabletypesize{\footnotesize}
\setlength{\tabcolsep}{6pt}

\end{longrotatetable}


\bibliographystyle{apsrev4-1}
\bibliography{caitlin-bibdesk}


\newpage
{\footnotesize
\noindent\textbf{Affiliations}
\vspace{4pt}

\begin{list}{}{%
  \setlength{\leftmargin}{1.8em}
  \setlength{\itemindent}{-1.8em}
  \setlength{\itemsep}{2pt}
  \setlength{\parsep}{0pt}
}
\item [1] Department of Physics, University of California, Santa Barbara, Santa Barbara, CA 93106, USA
\item [2] Cosmic Dawn Center (DAWN), Denmark
\item [3] Department of Astronomy, The University of Texas at Austin, 2515 Speedway Blvd Stop C1400, Austin, TX 78712, USA
\item [4] International Centre for Radio Astronomy Research, University of Western Australia, 35 Stirling Hwy, Crawley, WA 6009, Australia
\item [5] Research School of Astronomy and Astrophysics, Australian National University, Cotter Road, Weston Creek, ACT 2611, Australia
\item [6] Instituto de Alta Investigaci\'on, Universidad de Tarapac\'a, Casilla 7D, Arica, Chile
\item [7] University of Massachusetts Amherst, 710 North Pleasant Street, Amherst, MA 01003-9305, USA
\item [8] Academia Sinica Institute of Astronomy and Astrophysics (ASIAA), No.\ 1, Sec.\ 4, Roosevelt Road, Taipei 10617, Taiwan
\item [9] Instituto de Estudios Astrof\'{\i}cos, Facultad de Ingenier\'{i}a y Ciencias, Universidad Diego Portales, Av.\ Ej\'{e}rcito 441, Santiago, Chile
\item [10] Millenium Nucleus for Galaxies (MINGAL)
\item [11] Department of Astronomy, The University of Washington, Seattle, WA 98195, USA
\item [12] Department of Physics and Astronomy, University of Hawaii, Hilo, 200 W Kawili St, Hilo, HI 96720, USA
\item [13] Caltech/IPAC, MS 314-6, 1200 E.\ California Blvd., Pasadena, CA 91125, USA
\item [14] Universit\'{e} Paris-Saclay, Universit\'{e} Paris Cit\'{e}, CEA, CNRS, AIM, 91191 Gif-sur-Yvette, France
\item [15] David A. Dunlap Department of Astronomy \& Astrophysics, University of Toronto, 50 St.\ George Street, Toronto, ON M5S 3H4, Canada
\item [16] Department of Computer Science, Aalto University, P.O. Box 15400, FI-00076 Espoo, Finland
\item [17] Department of Physics, University of Helsinki, P.O. Box 64, FI-00014 Helsinki, Finland
\item [18] Department of Physics and Astronomy, University of California, Riverside, 900 University Ave, Riverside, CA 92521, USA
\item [19] Space Telescope Science Institute, 3700 San Martin Dr., Baltimore, MD 21218, USA
\item [20] Institute of Physics, GalSpec, Ecole Polytechnique Federale de Lausanne, Observatoire de Sauverny, Chemin Pegasi 51, 1290 Versoix, Switzerland
\item [21] INAF, Astronomical Observatory of Trieste, Via Tiepolo 11, 34131 Trieste, Italy
\item [22] Aix Marseille Univ, CNRS, CNES, LAM, Marseille, France
\item [23] Kavli Institute for Astronomy and Astrophysics, Peking University, Beijing 100871, China
\item [24] Laboratory for Multiwavelength Astrophysics, School of Physics and Astronomy, Rochester Institute of Technology, 84 Lomb Memorial Drive, Rochester, NY 14623, USA
\item [25] National Astronomical Observatory of Japan, 2-21-1 Osawa, Mitaka, Tokyo 181-8588, Japan
\item [26] NASA-Goddard Space Flight Center, Code 662, Greenbelt, MD 20771, USA
\item [27] Purple Mountain Observatory, Chinese Academy of Sciences, 10 Yuanhua Road, Nanjing 210023, China
\item [28] DTU-Space, Technical University of Denmark, Elektrovej 327, 2800, Kgs.\ Lyngby, Denmark
\item [29] Niels Bohr Institute, University of Copenhagen, Jagtvej 128, DK-2200, Copenhagen, Denmark
\item [30] Department of Physics, Centre for Extragalactic Astronomy, Durham University, South Road, Durham DH1 3LE, UK
\item [31] Department of Physics, Northeastern University, 360 Huntington Ave, Boston, MA
\item [32] Institut d'Astrophysique de Paris, CNRS, Sorbonne Universit\'{e}, 98 bis Boulevard Arago, F-75014 Paris, France
\item [33] University of Bologna -- Department of Physics and Astronomy ``Augusto Righi'' (DIFA), Via Gobetti 93/2, I-40129 Bologna, Italy
\item [34] INAF -- Osservatorio di Astrofisica e Scienza dello Spazio, Via Gobetti 93/3, I-40129 Bologna, Italy
\item [35] INFN -- Sezione di Bologna, Viale Berti Pichat 6/2, I-40127 Bologna, Italy
\item [36] Department of Astronomy, University of Florida, 211 Bryant Space Sciences Center, Gainesville, FL 32611, USA
\item [37] Department of Space, Earth and Environment, Chalmers University of Technology, SE-412 96 Gothenburg, Sweden
\item [38] Jet Propulsion Laboratory, California Institute of Technology, 4800 Oak Grove Drive, Pasadena, CA 91109, USA
\item [39] Department of Astronomy and Astrophysics, University of California, Santa Cruz, 1156 High Street, Santa Cruz, CA 95064, USA
\item [40] Department of Astronomy, National Research Institute of Astronomy and Geophysics (NRIAG), Cairo 11421, Egypt
\item [41] Minnesota Institute for Astrophysics, School of Physics and Astronomy, University of Minnesota, 116 Church St.\ SE, Minneapolis, MN 55455, USA
\item [42] University of Geneva, 24 rue du G\'{e}n\'{e}ral-Dufour, 1211 Gen\`{e}ve 4, Switzerland
\item [43] Center for Computational Astrophysics, Flatiron Institute, 162 Fifth Avenue, New York, NY 10010, USA
\item [44] Astronomy Centre, University of Sussex, Falmer, Brighton BN1 9QH, UK
\item [45] School of Astronomy and Space Science, Nanjing University, Nanjing, Jiangsu 210093, China
\item [46] Key Laboratory of Modern Astronomy and Astrophysics, Nanjing University, Ministry of Education, Nanjing 210093, China
\item [47] Kavli Institute for the Physics and Mathematics of the Universe (WPI), The University of Tokyo, Kashiwa, Chiba 277-8583, Japan
\end{list}
\vspace{4pt}
\begin{list}{}{%
  \setlength{\leftmargin}{1.8em}
  \setlength{\itemindent}{-1.8em}
  \setlength{\itemsep}{2pt}
  \setlength{\parsep}{0pt}
}
\item [$\ast$] NSF Graduate Research Fellow
\item [$\dag$] NASA Hubble Fellow
\item [$\ddag$] NPP Fellow
\end{list}
}

\end{document}